\documentstyle[10pt,epsf,epsfig,amssymb,amsfonts,amsmath,amsthm,cite,dp_delphititle]{dp_delphi}
\makeindex
\def\DpTitle{{A Measurement of the Tau Hadronic Branching Ratios}}
\def\DpPaperGroup{PH-EP}
\def\DpPaperRef{2004-046}
\def\DpDate{18 December 2003}
\def\DpAuthors{DELPHI Collaboration}
\def\DpSubmit{( Accepted by Euro. Phys. Journ. C. )}
\def\DpComment{ }
\def\DpEMail{ }
\newcommand{\qcth}{\cos\!\theta}
\newcommand{\AVHC}{E_{hcal}}

\newcommand{\pbeam}{p_{beam}}

\newcommand{\Bone}{B_1}

\def\mtau{m_\tau}
\def\mtau2{m^2_{\tau}}

\def\elec{\rm e}
\def\ee{ {\rm e}^+ {\rm e}^-}
\def\tt{ \tau^+\tau^-}
\def\mm{ \mu^+\mu^-}

\def\qq{ {\rm q} \bar{\rm q}}

\def\gg{\gamma\gamma}
\def\eett{\ee \rightarrow \tt}
\def\eeztt{\ee \rightarrow Z \rightarrow \tt}
\def\eemm{\ee \rightarrow \mm}
\def\bhab{\ee \rightarrow \ee}
\def\eeha{\ee \rightarrow \qq}

\newcommand{\aone}{{\rm a}_1}
\def\a1nu{\aone\nu}

\def\pull{\Pi_{dE/dx}}

\def\s2thw{\sin^{2}\theta^{\mbox{\scriptsize lept}}_{\mbox{\scriptsize eff}}}
\def\kos{K^{0}_{\! \mathrm S}}
\def\kol{K^{0}_{\! \mathrm L}}
\def\ko{K^{0}}

\newcommand{\nut}{\nu_{\tau}}
\newcommand{\num}{\nu_{\mu}}
\newcommand{\nue}{\nu_{\elec}}
\newcommand{\Tauto}{\tau^- \rightarrow}
\def\pio{\pi^0}
\def\ko{K^0}
\def\nt{\nu_\tau}

\newcommand{\HH}{h^-\nut}

\newcommand{\TH}{\Tauto\HH}

\newcommand{\MU}{\mu^- \nut \bar{\nu}_{\mu}}
\newcommand{\TMU}{\Tauto\MU}
\newcommand{\EL}{\elec^- \nut \bar{\nu}_{\mathrm e} }
\newcommand{\TEL}{\Tauto\EL}
\newcommand{\TRO}{\Tauto \rho^- \nut}

\newcommand{\HZ}{h^-\pi^0 \nut}

\newcommand{\THZ}{\Tauto \HZ}

\newcommand{\HZZ}{h^-\pi^0\pi^0 \nut}

\newcommand{\HGZZ}{h^-\geq \! 2\pi^0 \nut}

\newcommand{\HGZZZ}{h^-\geq \! 3\pi^0 \nut}

\newcommand{\PPP}{2\pi^-\pi^+ \nut}
\newcommand{\HHH}{2h^-h^+ \nut}

\newcommand{\THHH}{\Tauto \HHH}
\newcommand{\PPPZ}{2\pi^-\pi^+\pi^0 \nut}

\newcommand{\PPPPP}{3\pi^-2\pi^+ \nut}

\newcommand{\PPPZZ}{3\pi^\pm 2\pi^0 \nut}
\newcommand{\PPPZZZ}{3\pi^\pm 3\pi^0 \nut}

\newcommand{\PPPPPZ}{3\pi^-2\pi^+ \pi^0 \nut}

\def\tpppppno{\tau^- \rightarrow 3 h^- 2 h^+ \pi^0\nu_\tau}
\def\tppppp{\tau^- \rightarrow 3 h^- 2 h^+\nu_\tau}
\begin{document}
%%%%%%%%%%%%%%%%%%%%%%%%%% They are a problem with Coll.Sty ?
\makeatletter
\makeatother
%%%%%%%%%%%%%%%%%%%%%%%%%% ??????????????????????????????????
% Generate the title page
\begin{titlepage}
\pagenumbering{roman}
\CERNpreprint{\DpPaperGroup}{\DpPaperRef} % Reference of the paper
\date{{\small\DpDate}} % Date of the paper
\title{\DpTitle} % Title of the paper
\address{\DpAuthors} % General name of the author(s)
\begin{shortabs} % Start the abstract
\noindent
%   abstract.tex
%
%   abstract.tex
%
     
    The
    exclusive and semi-exclusive branching ratios of the
    $\tau$ lepton hadronic decay modes
    (    $h^- \nut$,
    $h^- \pi^0 \nut$,
    $h^- \pi^0 \pi^0  \nut$,
    $h^- \geq\! 2 \pi^0 \nut$,
    $h^- \geq\! 3 \pi^0 \nut$,
    $2h^- h^+  \nut$,
    $2h^- h^+\pi^0 \nut$,
    $2h^- h^+ \geq\! 2 \pi^0 \nut$,
    $3h^- 2h^+ \nut$ and
    $3h^- 2h^+ \geq 1 \pi^0 \nut$) were measured with data from
    the DELPHI detector at LEP.
\end{shortabs}
\vfill
\begin{center}
\DpSubmit \ \\ % Horrible hack to allow to have DpSubmit empty
\DpComment \ \\
\DpEMail \ \\
\end{center}
\vfill
\clearpage
\headsep 10.0pt
\addtolength{\textheight}{10mm}
\addtolength{\footskip}{-5mm}
\begingroup
% Commands to process the author names
%
\newcommand{\DpName}[2]{\hbox{#1$^{\ref{#2}}$},\hfill}
\newcommand{\DpNameTwo}[3]{\hbox{#1$^{\ref{#2},\ref{#3}}$},\hfill}
\newcommand{\DpNameThree}[4]{\hbox{#1$^{\ref{#2},\ref{#3},\ref{#4}}$},\hfill}
\newskip\Bigfill \Bigfill = 0pt plus 1000fill
\newcommand{\DpNameLast}[2]{\hbox{#1$^{\ref{#2}}$}\hspace{\Bigfill}}
%
%CD\small
\footnotesize
\noindent
\DpName{J.Abdallah}{LPNHE}
\DpName{P.Abreu}{LIP}
\DpName{W.Adam}{VIENNA}
\DpName{P.Adzic}{DEMOKRITOS}
\DpName{T.Albrecht}{KARLSRUHE}
\DpName{T.Alderweireld}{AIM}
\DpName{R.Alemany-Fernandez}{CERN}
\DpName{T.Allmendinger}{KARLSRUHE}
\DpName{P.P.Allport}{LIVERPOOL}
\DpName{U.Amaldi}{MILANO2}
% ...
\normalsize
\endgroup
\titlefoot{Department of Physics and Astronomy, Iowa State
     University, Ames IA 50011-3160, USA
    \label{AMES}}
\titlefoot{Physics Department, Universiteit Antwerpen,
     Universiteitsplein 1, B-2610 Antwerpen, Belgium \\
     \indent~~and IIHE, ULB-VUB,
     Pleinlaan 2, B-1050 Brussels, Belgium \\
     \indent~~and Facult\'e des Sciences,
     Univ. de l'Etat Mons, Av. Maistriau 19, B-7000 Mons, Belgium
    \label{AIM}}
\titlefoot{Physics Laboratory, University of Athens, Solonos Str.
     104, GR-10680 Athens, Greece
    \label{ATHENS}}
\titlefoot{Department of Physics, University of Bergen,
     All\'egaten 55, NO-5007 Bergen, Norway
    \label{BERGEN}}
\titlefoot{Dipartimento di Fisica, Universit\`a di Bologna and INFN,
     Via Irnerio 46, IT-40126 Bologna, Italy
    \label{BOLOGNA}}
\titlefoot{Centro Brasileiro de Pesquisas F\'{\i}sicas, rua Xavier Sigaud 150,
     BR-22290 Rio de Janeiro, Brazil \\
     \indent~~and Depto. de F\'{\i}sica, Pont. Univ. Cat\'olica,
     C.P. 38071 BR-22453 Rio de Janeiro, Brazil \\
     \indent~~and Inst. de F\'{\i}sica, Univ. Estadual do Rio de Janeiro,
     rua S\~{a}o Francisco Xavier 524, Rio de Janeiro, Brazil
    \label{BRASIL}}
\titlefoot{Coll\`ege de France, Lab. de Physique Corpusculaire, IN2P3-CNRS,
     FR-75231 Paris Cedex 05, France
    \label{CDF}}
\titlefoot{CERN, CH-1211 Geneva 23, Switzerland
    \label{CERN}}
\titlefoot{Institut de Recherches Subatomiques, IN2P3 - CNRS/ULP - BP20,
     FR-67037 Strasbourg Cedex, France
    \label{CRN}}
\titlefoot{Now at DESY-Zeuthen, Platanenallee 6, D-15735 Zeuthen, Germany
    \label{DESY}}
\titlefoot{Institute of Nuclear Physics, N.C.S.R. Demokritos,
     P.O. Box 60228, GR-15310 Athens, Greece
    \label{DEMOKRITOS}}
\titlefoot{FZU, Inst. of Phys. of the C.A.S. High Energy Physics Division,
     Na Slovance 2, CZ-180 40, Praha 8, Czech Republic
    \label{FZU}}
\titlefoot{Dipartimento di Fisica, Universit\`a di Genova and INFN,
     Via Dodecaneso 33, IT-16146 Genova, Italy
    \label{GENOVA}}
\titlefoot{Institut des Sciences Nucl\'eaires, IN2P3-CNRS, Universit\'e
     de Grenoble 1, FR-38026 Grenoble Cedex, France
    \label{GRENOBLE}}
\titlefoot{Helsinki Institute of Physics, P.O. Box 64,
     FIN-00014 University of Helsinki, Finland
    \label{HELSINKI}}
\titlefoot{Joint Institute for Nuclear Research, Dubna, Head Post
     Office, P.O. Box 79, RU-101 000 Moscow, Russian Federation
    \label{JINR}}
\titlefoot{Institut f\"ur Experimentelle Kernphysik,
     Universit\"at Karlsruhe, Postfach 6980, DE-76128 Karlsruhe,
     Germany
    \label{KARLSRUHE}}
\titlefoot{Institute of Nuclear Physics PAN,Ul. Radzikowskiego 152,
     PL-31142 Krakow, Poland
    \label{KRAKOW1}}
\titlefoot{Faculty of Physics and Nuclear Techniques, University of Mining
     and Metallurgy, PL-30055 Krakow, Poland
    \label{KRAKOW2}}
\titlefoot{Universit\'e de Paris-Sud, Lab. de l'Acc\'el\'erateur
     Lin\'eaire, IN2P3-CNRS, B\^{a}t. 200, FR-91405 Orsay Cedex, France
    \label{LAL}}
\titlefoot{School of Physics and Chemistry, University of Lancaster,
     Lancaster LA1 4YB, UK
    \label{LANCASTER}}
\titlefoot{LIP, IST, FCUL - Av. Elias Garcia, 14-$1^{o}$,
     PT-1000 Lisboa Codex, Portugal
    \label{LIP}}
\titlefoot{Department of Physics, University of Liverpool, P.O.
     Box 147, Liverpool L69 3BX, UK
    \label{LIVERPOOL}}
\titlefoot{Dept. of Physics and Astronomy, Kelvin Building,
     University of Glasgow, Glasgow G12 8QQ
    \label{GLASGOW}}
\titlefoot{LPNHE, IN2P3-CNRS, Univ.~Paris VI et VII, Tour 33 (RdC),
     4 place Jussieu, FR-75252 Paris Cedex 05, France
    \label{LPNHE}}
\titlefoot{Department of Physics, University of Lund,
     S\"olvegatan 14, SE-223 63 Lund, Sweden
    \label{LUND}}
\titlefoot{Universit\'e Claude Bernard de Lyon, IPNL, IN2P3-CNRS,
     FR-69622 Villeurbanne Cedex, France
    \label{LYON}}
\titlefoot{Dipartimento di Fisica, Universit\`a di Milano and INFN-MILANO,
     Via Celoria 16, IT-20133 Milan, Italy
    \label{MILANO}}
\titlefoot{Dipartimento di Fisica, Univ. di Milano-Bicocca and
     INFN-MILANO, Piazza della Scienza 2, IT-20126 Milan, Italy
    \label{MILANO2}}
\titlefoot{IPNP of MFF, Charles Univ., Areal MFF,
     V Holesovickach 2, CZ-180 00, Praha 8, Czech Republic
    \label{NC}}
\titlefoot{NIKHEF, Postbus 41882, NL-1009 DB
     Amsterdam, The Netherlands
    \label{NIKHEF}}
\titlefoot{National Technical University, Physics Department,
     Zografou Campus, GR-15773 Athens, Greece
    \label{NTU-ATHENS}}
\titlefoot{Physics Department, University of Oslo, Blindern,
     NO-0316 Oslo, Norway
    \label{OSLO}}
\titlefoot{Dpto. Fisica, Univ. Oviedo, Avda. Calvo Sotelo
     s/n, ES-33007 Oviedo, Spain
    \label{OVIEDO}}
\titlefoot{Department of Physics, University of Oxford,
     Keble Road, Oxford OX1 3RH, UK
    \label{OXFORD}}
\titlefoot{Dipartimento di Fisica, Universit\`a di Padova and
     INFN, Via Marzolo 8, IT-35131 Padua, Italy
    \label{PADOVA}}
\titlefoot{Rutherford Appleton Laboratory, Chilton, Didcot
     OX11 OQX, UK
    \label{RAL}}
\titlefoot{Dipartimento di Fisica, Universit\`a di Roma II and
     INFN, Tor Vergata, IT-00173 Rome, Italy
    \label{ROMA2}}
\titlefoot{Dipartimento di Fisica, Universit\`a di Roma III and
     INFN, Via della Vasca Navale 84, IT-00146 Rome, Italy
    \label{ROMA3}}
\titlefoot{DAPNIA/Service de Physique des Particules,
     CEA-Saclay, FR-91191 Gif-sur-Yvette Cedex, France
    \label{SACLAY}}
\titlefoot{Instituto de Fisica de Cantabria (CSIC-UC), Avda.
     los Castros s/n, ES-39006 Santander, Spain
    \label{SANTANDER}}
\titlefoot{Inst. for High Energy Physics, Serpukov
     P.O. Box 35, Protvino, (Moscow Region), Russian Federation
    \label{SERPUKHOV}}
\titlefoot{J. Stefan Institute, Jamova 39, SI-1000 Ljubljana, Slovenia
     and Laboratory for Astroparticle Physics,\\
     \indent~~Nova Gorica Polytechnic, Kostanjeviska 16a, SI-5000 Nova Gorica, Slovenia, \\
     \indent~~and Department of Physics, University of Ljubljana,
     SI-1000 Ljubljana, Slovenia
    \label{SLOVENIJA}}
\titlefoot{Fysikum, Stockholm University,
     Box 6730, SE-113 85 Stockholm, Sweden
    \label{STOCKHOLM}}
\titlefoot{Dipartimento di Fisica Sperimentale, Universit\`a di
     Torino and INFN, Via P. Giuria 1, IT-10125 Turin, Italy
    \label{TORINO}}
\titlefoot{INFN,Sezione di Torino, and Dipartimento di Fisica Teorica,
     Universit\`a di Torino, Via P. Giuria 1,\\
     \indent~~IT-10125 Turin, Italy
    \label{TORINOTH}}
\titlefoot{Dipartimento di Fisica, Universit\`a di Trieste and
     INFN, Via A. Valerio 2, IT-34127 Trieste, Italy \\
     \indent~~and Istituto di Fisica, Universit\`a di Udine,
     IT-33100 Udine, Italy
    \label{TU}}
\titlefoot{Univ. Federal do Rio de Janeiro, C.P. 68528
     Cidade Univ., Ilha do Fund\~ao
     BR-21945-970 Rio de Janeiro, Brazil
    \label{UFRJ}}
\titlefoot{Department of Radiation Sciences, University of
     Uppsala, P.O. Box 535, SE-751 21 Uppsala, Sweden
    \label{UPPSALA}}
\titlefoot{IFIC, Valencia-CSIC, and D.F.A.M.N., U. de Valencia,
     Avda. Dr. Moliner 50, ES-46100 Burjassot (Valencia), Spain
    \label{VALENCIA}}
\titlefoot{Institut f\"ur Hochenergiephysik, \"Osterr. Akad.
     d. Wissensch., Nikolsdorfergasse 18, AT-1050 Vienna, Austria
    \label{VIENNA}}
\titlefoot{Inst. Nuclear Studies and University of Warsaw, Ul.
     Hoza 69, PL-00681 Warsaw, Poland
    \label{WARSZAWA}}
\titlefoot{Fachbereich Physik, University of Wuppertal, Postfach
     100 127, DE-42097 Wuppertal, Germany
    \label{WUPPERTAL}}
\addtolength{\textheight}{-10mm}
\addtolength{\footskip}{5mm}
\clearpage
\headsep 30.0pt
\end{titlepage}
%%%%%%%%%%%%%%%%%%%%%%%%%
%
% Change for the document body
%\pagestyle{heading} % for page numbering
\pagenumbering{arabic} % page numbering in number
\setcounter{footnote}{0} %
\large
%\linenumbers %%%CD
%
\section{Introduction}
\label{sec:introduction}
  The $\tau$ lepton, discovered in 1975~\cite{perl}, is the only lepton which is
sufficiently heavy to decay to final states containing hadrons.  
Predictions for the properties of such a heavy lepton have been made well in
advance of its discovery~\cite{tsai}.  The taus produce
intermediate and final-state hadrons with lower backgrounds
than most other low energy processes. 
%This enables studies
%of these hadron systems with relatively low ambiguity in the
%quantum numbers of the produced particles.
%Such    studies     include    the
%measurement~\cite{delphitaupolarisation} of the $\tau$ polarisation in
%$Z$ decays, an important ingredient  in the estimation of the Weinberg
%angle.      and     measurements     of    the     strong     coupling
%constant~\cite{alphas}.  Underlying these is  the need to separate the
%different  hadronic  final  states   in  $\tau$  decays.   

This  paper
describes a measurement of the decay rates of the $\tau$ lepton to the
different hadronic final states as  a function of both the charged-hadron
and
neutral-pion   multiplicities,  with  no   particle  identification
performed on the charged hadrons.
%
%Historically, there has been some discrepancy in the 
%completeness of the measured
%$\tau$ exclusive branching ratios; the exclusive branching ratios
%to decay modes containing one charged particle
%did not sum up to the inclusive branching ratio 
%to decay modes containing one charged particle. A number of previous analyses
%have attempted and succeeded in resolving this 
%question~\cite{pdg96,cellobr,alephbr}. 
%
Samples  of different  $\tau$-decay  final states  have  been selected
using  both ``sequential cuts''  methods  and  neural networks.   These
analyses  were  complementary, allowing  cross-checks  of the  results
and their uncertainties.  

The DELPHI  detector 
and data sample are described  in Section~\ref{sec:detector}.
The method used to determine the branching ratios is described in 
section~\ref{sec:method}.
The  techniques used  to separate charged leptons from hadrons
%and  hence  remove $\tau$-decays  containing
%electrons  and   muons  
are outlined   in~Section~\ref{sec:chargedparticles}.   
Section~\ref{sec:photons} describes the
reconstruction   of   photons and 
%Section~\ref{sec:neutralpions}  the
%reconstruction   of  
neutral   pions.
The selection of  $\eett$ events is outlined in  
Section~\ref{sec:tautauselection} and
the isolated $\tau$-decays  are classified according to their charged-particle  
multiplicity  in  Section~\ref{sec:topology}. 
The selection of $\tau$-decays as a function of the neutral
pion multiplicity is described in Section~\ref{sec:selection} and
the associated systematic uncertainties on the measured 
branching ratios are discussed in
Section~\ref{sec:systematics}. Section~\ref{sec:combination} presents the
results and conclusions are drawn in Section~\ref{sec:conclusions}.

%Some  of  the  exclusive   hadronic  decay  modes  have  theoretically
%predicted branching ratios  obtained from $\pi$ and $K$  decay or, for
%vector  final  states,  from   the  isovector  part  of  the  hadronic
%cross-section in $\ee$ annihilation  under the Conserved Vector Current
%(CVC) hypothesis~\cite{cvc}.  These predictions are  compared with the
%obtained measurements in Section~\ref{sec:discussion}.

DELPHI has previously published results on some of the decay
modes measured here using the 1990 data sample~\cite{tau90}.
This paper replaces those low-statistics results.
Similar analyses performed by other LEP experiments can be found in
~\cite{otherlep}.

\section{The DELPHI Detector and data sample}
\label{sec:detector} 
The  DELPHI detector and its performance
are described in  detail in~\cite{detect,performance}.
The  components relevant  to this  analysis are
summarised below. Unless specified otherwise, they 
covered the full solid angle of the barrel region used 
in this analysis ($43^\circ<\theta<137^\circ$) and lay in
a  1.2  Tesla solenoidal magnetic field  parallel  
to  the  beam\footnote{In the DELPHI reference
frame the origin was at the centre of the detector, coincident
with the ideal interaction region. The z-axis
was parallel to the e$^-$  beam, the  x-axis  pointed horizontally
towards  the centre  of  the LEP  ring  and the  y-axis was  vertically
upwards.   The  co-ordinates   r,$\phi$,z   formed  a cylindrical
coordinate system, while $\theta$ was the polar angle with 
respect to the z-axis.}.

The charged-particle track reconstruction  was based on four different
detector components. The principal track reconstruction
device was the Time Projection Chamber (TPC), a large
drift chamber covering the radial region 35 cm $<$ r $<$ 111 cm.
To enhance the precision of the TPC measurement, track reconstruction 
was supplemented
by a three-layer silicon Vertex Detector (VD) at radii between~6~and 12~cm, 
an Inner Detector (ID) between 12~and 28~cm  and 
the Outer Detector (OD) at radii between  197 and 206 cm from the
z-axis.  The TPC also provided up to 192 ionisation
measurements per charged particle track, useful for electron/hadron separation. 
It had boundary  regions between  read-out  sectors every $60^\circ$ in $\phi$ 
which were about
$1^\circ$ wide and which were covered by the VD, ID, and OD.

The main device for $\gamma$ and $\pi^0$ reconstruction
and electron/hadron separation, the High
density  Projection  Chamber (HPC) lay
between radii of 208~cm and 260~cm.
It consisted of 40 layers of 3~mm thick lead interspersed 
with 8~mm thick layers of gas  sampling volume, amounting to a minimum of about 18 radiation
lengths.
In the gas layers the ionising particles in a shower produced
electrons
which drifted in an electric field
into wire chambers.
In these wire chambers the induced signal on cathode pads 
gave a measurement of the deposited charge
with sampling granularity of 
10~mrad~$\times$~2~mrad~$\times$~1.0~$X_0$  
in~$\phi$~$\times$~$\theta$~$\times$~r
in the inner 4 radiation lengths and provided up to nine longitudinal samplings
of the energy deposition in a shower.
The
spatial precision for the  starting point of  an electromagnetic
shower was 1~mrad in
$\theta$ and 2~mrad in $\phi$.
The energy resolution was
$\Delta E/E =  0.31/E^{0.44} \oplus 0.027$.

%The electromagnetic calorimetry in the polar angle region 
%$0.800<|\qcth|<0.966$ was  performed by a lead-glass array. For smaller polar angles,
%$|\qcth|>0.966$ the calorimetry was carried out by the luminosity monitors,
%the SAT in 1992 and 1993, and the STIC in 1994 and 1995.

The  Hadron  Calorimeter (HCAL) was the instrumented flux return of the magnet. 
It was longitudinally segmented into 20 layers of iron and limited
streamer tubes. The tubes were grouped to give four longitudinal 
segments in the readout, with a granularity of 
$3.75^{\circ}$~$\times$~$2.96^{\circ}$ in $\phi$$\times$$\theta$.
Between the 18th and 19th 
HCAL layers and also outside the whole calorimeter, there
were drift chambers for detecting the muons which were expected to penetrate
the whole HCAL. The  barrel muon  chambers (MUB) 
covered  the   range  $|\qcth|\!<\!0.602$ while most azimuthal
zones in the range $0.602\!<\!|\qcth|$ 
were covered by forward muon chambers (MUF).

The Ring-Imaging Cherenkov detector (RICH), although not used in this analysis,
had an important effect on the performance of the calorimetry as it contained
the majority of the material in the DELPHI barrel region. Lying between the TPC
and OD in radius, it covered the complete polar angle region of this analysis.
The amount of material for particles of perpendicular incidence was equivalent
to 0.6 radiation lengths and 0.15 nuclear interaction lengths.

The data were collected in the years 1992 to 1995, 
at centre-of-mass energies $\sqrt{s}$ between 89 and 93~GeV
on or near to the Z resonance. 
It was required that
the  VD, TPC, HPC, MUB and  HCAL subdetectors be fully operational.
The integrated luminosity of the data sample
was 135~pb$^{-1}$  of which
about 100~pb$^{-1}$ was taken  
at  $\sqrt{s}\approx$~91.3~GeV, near the maximum  of the Z
production  cross-section. 

Selection requirements were studied
on simulated event samples after  a  detailed  simulation  of  the  detector
response~\cite{performance} and reconstruction 
by the same program as the
real  data.  Samples were simulated for the different detector conditions and
centre-of-mass energies in
every year of data taking and amounted to about 16 times the recorded
luminosities. The  Monte  Carlo  event generators  used  were:  KORALZ
4.0~\cite{jadach} for  $\eett$ events; DYMU3~\cite{dymu3}  for $\eemm$
events;    BABAMC~\cite{babamc} and BHWIDE~\cite{bhwide}
for    $\bhab$    events; 
JETSET 7.3~\cite{sjostrand}  for  $\eeha$  events;  
%PYTHIA 5.7~\cite{sjostrand}  for  $\eeha$  events;  
BDK~\cite{berends} for four-lepton final states;
%TWOGAM~\cite{teddy} for $\ee\to\ee\qq$
TWOGAM~\cite{twogam} for $\ee\to\ee\qq$
events.  The KORALZ generator incorporated the TAUOLA2.5~\cite{tauola}
package for modelling $\tau$-decays.

\section{Method}
\label{sec:method}
In an initial step, $\tau$-decays were selected
according to their charged-particle multiplicity from a
high-purity Z$\to\tt$ event sample. In 
decays containing only one charged particle, this particle can
be either an electron, muon or hadron. In higher charged-particle
multiplicity decays the initial charged particles are
hadrons. 

After rejection of one-prong decays containing muons and
electrons the following exclusive and semi-exclusive $\tau$
decay modes have been isolated and their branching ratios measured:
%as a function of the charged hadron and neutral pion multiplicity:
\begin{itemize}
\item
Charged multiplicity one: \\ 
$h^- \nu_\tau$,~~~                        
$h^- \pi^0 \nu_\tau$,~~~                       
$h^- 2\pi^0 \nu_\tau$, ~~~                      
%$h^- \geq 2\pi^0 \nu_\tau$,  ~~~                     
$h^- \geq 3\pi^0 \nu_\tau$;                       
\item
Charged multiplicity three: \\ 
$2h^-h^+ \nu_\tau$,      ~~~                  
$2h^-h^+ \pi^0 \nu_\tau$, ~~~                       
$2h^-h^+ \geq 2 \pi^0 \nu_\tau$;                        
\item
Charged multiplicity five: \\ 
$3h^-2h^+ \nu_\tau$,          ~~~              
$3h^-2h^+ \geq 1 \pi^0 \nu_\tau$.
\end{itemize}
where $h$ is either a $\pi$ or $K$ meson.
The charge conjugate decays were also included.

The $\pi^0$ mesons were detected and reconstructed via the photons
produced in the decay $\pi^0\to\gamma\gamma$. This $\pi^0$ decay mode
has a branching ratio of (98.798$\pm$0.032)\%, the remainder
decaying through the Dalitz process $\pi^0\to\gamma\ee$. Most of these
were also correctly identified with the conversion rejection algorithm,
and the fraction lost was a contribution to the inefficiency.

In the definition of these channels the presence of neutral kaons (reconstructed or not) was not
considered. For example the decay $h^- \nu_\tau$, included channels with one charged hadron and
none, one or more neutral kaons.
%The analyses described later 
%did not count neutral kaons, but the details of their decays
%or interactions were modelled. 
The presence of neutral kaons did not significantly affect the selection
efficiency, but was accounted for in the analysis. 
%Events with neutral kaons were included together with those with none.

Two complementary analyses were performed on each of
the samples of charged multiplicity one and three $\tau$-decays.
One analysis was based on sequential cuts and the other on neural
networks. The $\tau$-decays were classified as
a function of the $\pi^0$ multiplicity and the
branching ratios were obtained taking into account
statistical and systematic correlations.
Only a  sequential cuts analysis was performed for
 $\tau$-decays with charged multiplicity of five.
%For $\tau$-decays of charged multiplicity five only a  sequential cuts
%analysis was performed.

%At LEP on the $Z$ resonance
%it is  possible to isolate with good efficiency a high purity sample 
%of $\tau$-decays in the $\eett$ reaction.
%The branching ratio  $B(\tau\to X)$ for  the decay of
%the $\tau$ to a final state $X$ can be measured using the expression
%\begin{equation}
%\label{eqn:eq3}
%  B(\tau\to X) = \frac{N_X} {N_{\tau}} \cdot \frac{1-b_X} {1-b_{\tau}}
%                                     \cdot       \frac{\epsilon_{\tau}}
%                                     {\epsilon_X } ,
%\end{equation}
%where $N_X$  is the number of  identified decays of type  $X$ found in
%the  sample of  $N_{\tau}$ $\tau$-decay  candidates,  preselected with
%efficiency $\epsilon_{\tau}$ with a background fraction of $b_{\tau}$.
%$\epsilon_X$  is  the  total (preselection  $\times$  identification)
%efficiency for selecting the decay mode $\tau\to X$, with a background
%fraction of $b_X$, including background from other $\tau$-decay modes.
%In a $\tt$ event selection  without specific requirements on either of
%the  two $\tau$  candidates in  the event,  $\epsilon_{\tau}$  will be
%identical to the $\tt$ event selection efficiency.

The branching ratios were measured simultaneously with the following procedure.
Candidate $\tau$-decays can be classified using an estimator such as the maximum output
neuron from a neural network or the set of cuts of the sequential analysis. 
On real data all decays are assigned to the different classes, providing the
total number of events in each class: $N_{i,obs}$.
On simulated data,  a selection-probability  matrix 
$M_{ij}$ can be obtained, representing the probability 
for decay mode $j$ to be classified as decay mode $i$. 
This matrix could be diagonal, but in fact most of the off-diagonal 
terms are non-zero.
%These  classes (detailed in Section~\ref{sec:topology})
%typically  have  different  signal to  noise
%ratios for different types of decays and thus contain different levels
%of  information. 
To obtain  the Branching  Ratios~$B_j$, a  maximum-likelihood fit can  then be
performed to constrain 
the predicted number,~$N_{i,pred}$, of decays in~class~$i$ to $N_{i,obs}$ .
$N_{i,pred}$ is given by:
\begin{equation}
\label{eqn:classpred}
N_{i,pred} = N_\tau \sum_{j=1}^{n_c} M_{ij}\epsilon_j B_j
           + N_{i,bkg}~,
\end{equation}
where $N_\tau$ is the total number of produced $\tau$ particles, which is left
as a free parameter in the fit,
$\epsilon_{j}$  is the efficiency  for decay mode $j$ 
of the $\tt$ selection,
$N_{i,bkg}$ is the estimated background in class
$i$ due to  non-$\tt$ events, and $n_c$ is the
number of classes, synonymous with the number of decay
modes if all decays are classified. In this analysis
not all candidate $\tau$-decays were classified as a minimum
level was required on the maximum output neuron of
the neural network. Taking into account the track multiplicity, this led 
to three additional classes,
corresponding to those decays which were unclassified. 
Having three classes instead of just one for all the unclassified modes,
does not improve the precision on the measurement, but gives
additional information on the comparison of topological and exclusive branching
ratios.

If we do not take into account these three extra classes, the 
problem is undetermined, since there are $n_c+1$ unknowns (the
$n_c$ branching ratios and $N_\tau$) and only $n_c$ measurements. 
The inclusion of these three classes, corresponding to the
events not assigned to any given class, does not help, because, 
despite having three additional measurements, the equations
are nearly degenerate (the matrix is almost singular) and 
the resulting fit is highly unstable. We avoid the problem by
setting an additional constraint that all the branching ratios add to 1. 
In many previous measurements an alternative
procedure is proposed, which is not correct in the case of 
multiple branching ratios. Here $N_\tau$ is obtained from the
selected $\tau$ events, together with the expected efficiency 
($\epsilon_{\tau\tau}$)
and background ($b$), with the expression
$N_\tau=2\cdot\frac{N_{\tau\tau}}{\epsilon_{\tau\tau}}\cdot(1-b)$. 
However, this expression needs to assume  a
priori the branching ratios to estimate the $\tau\tau$ 
selection efficiency and nevertheless also makes an implicit assumption
on the sum of branching ratios when computing that efficiency. 
With the method described here, unexpected decays could
affect the goodness of the fit through its $\chi^2$ and in particular, 
with an excess in the extra classes mentioned above.

\boldmath
\section{Particle identification and detector calibration}
\label{sec:particleid}
\unboldmath
The detector response was studied using simulation
together with test samples of real data where the identity 
and momentum of
the particles was unambiguously known. Examples of such samples
consisted of $\bhab$ and $\eemm$ events,
the radiative processes $\bhab\gamma$ and
$\eemm\gamma$ and Compton events selected using kinematic constraints. 
Tau-decay test samples, which were selected taking advantage of
the redundancy of the detector, were also used. An example
is $\tau \rightarrow h ({\mathrm n} \pi^0) \nu$, (n$>$0), selected by
tagging the $\pi^0$ decay in the HPC. This gave 
a pure sample of charged hadrons to test the response of the
calorimetry, muon chambers, and ionisation loss in the TPC. 
The decays $\tau \rightarrow \mu\nu\nu$ 
selected with the 
calorimeters checked the muon chambers response and the TPC ionisation loss.
Various test samples were used
to calibrate the response of the model of the detector in the simulation 
program and where necessary to correct observed discrepancies.
%This was particularly important in the analyses using neural networks
%where it was necessary to estimate systematic uncertainties
%on the result after correcting or modifying a given input variable
%and  propagating the effect through the neural network.

Further details of electron, muon and charged-hadron
separation
in $\tau$-decays can be found in the analysis of the $\tau$
leptonic branching ratios~\cite{delphileptonicbr99}.

\subsection{Charged particles}
\label{sec:chargedparticles}
\subsubsection{Tracking}
\label{sec:pscale}
The precision on the component of the momentum transverse to
the beam direction, $p_t$, obtained with the DELPHI tracking detectors
was $\Delta (1/p_t) = 0.0008 \rm{(GeV/ \mathrm{c})^{-1}}$ for particles
with momentum close to 45 GeV/c. Calibration of the
momentum measurement was performed with $\eemm$ events. For lower 
momenta the masses of the $K^0_s$ and $\Lambda$
were reconstructed. For intermediate momenta three body decays ($\eemm\gamma$ 
and $\bhab\gamma$) were used. In these cases, the true energy of the
particles can be calculated to a good precision from energy and momentum 
conservation, using the accurate measurement of the particle direction only. 
The combination of all these methods 
gives an absolute momentum scale to a precision of 0.2\% over 
the full momentum range.

 Some 3\% of hadrons reinteract inelastically
with the detector material
before the TPC. These were reconstructed with an algorithm
which was designed to  find secondary reinteraction vertices using the
tracks   from   outgoing  charged   particles   produced  in   nuclear
interactions.
This is described in detail in the 
DELPHI analysis of the $\tau$ topological Branching 
Ratios~\cite{delphitopologicalbr},
where the efficiency of the algorithm, as well as the amount
of material in the detector in terms of nuclear interaction lengths, 
were studied. The  efficiency in the data was found to agree
well with the simulation prediction while there was an overestimate
by about 10\% in the simulation of the number of nuclear interaction
lengths before the TPC gas volume. The
correction factors obtained  have been applied
via reweighting techniques.
%, and the associated systematic uncertainties
%ascribed.

\subsubsection{TPC ionisation measurement}
The energy loss per unit path length due to ionisation, $dE/dx$,
of a charged particle traveling through the TPC gave
good separation between electrons and charged pions, particularly in
the low momentum range.
%This reduced the sample by a small amount primarily due
%to particles being close to the boundary regions of the TPC sectors where
%a narrow non-instrumented strip was located.  
The $dE/dx$ pull
variable, $\prod^j_{dE/dx}$, for a particular particle hypothesis ($j$
= e,$\pi$,K,p) is defined as
\begin{equation}
{\textstyle \prod^{j}_{dE/dx}} = \frac { {dE/dx}_{meas} -
                                                  {dE/dx}_{exp}(j) }
                            { \sigma(dE/dx) }~~,
\end{equation}
where ${dE/dx}_{meas}$ is the measured value, ${dE/dx}_{exp}(j)$ is
the expected momentum dependent value for a hypothesis $j$ and
$\sigma(dE/dx)$ is the resolution of the measurement. It
was required that there be at least 38 anode sense wires used in the
measurement. The  $dE/dx$ was calibrated as a function of particle
velocity, polar and azimuthal angle. The distributions in simulation
were tuned to agree with test samples of real data. 
The relative precision obtained 
was ~6.2\%. Fig.~\ref{fig:elecid} shows the distribution of $\pull^{\mathrm e}$ 
and of $\pull^{\pi}$
in an electron test sample
selected using calorimetric  cuts. 
Fig.~\ref{fig:pionid1} shows the same
distributions 
for a hadron test sample selected from $\tau$-decays.

\subsubsection{Electromagnetic calorimetry}
The calibration of the HPC 
for  the energy range from 0.5~GeV to 46~GeV
used test samples of electrons in Compton
events, both radiative and non-radiative Bhabha events, and
electrons tagged by the TPC $dE/dx$ measurement.
Since no difference was found in the response for electrons or photons, $\gamma$
samples were also used for the calibration. This will be described in
section~\ref{sec:eemphot}.

For electrons,  the associated energy deposited in the HPC (in GeV), $E_{ass}$,
should be equal to the measured value of the momentum (in GeV/c), within
experimental errors.  For hadrons the energy should be lower than the
measured momentum as hadrons typically traverse the HPC leaving only 
a small fraction of their energy. Muons deposit only a small amount of energy in the HPC.

\begin{figure}[tbp]
  \begin{center}
%    \leavevmode
%\put(236,330){\makebox(0,0){\Large\bf DELPHI}}
%\epsfig{figure=fig_elecid.eps,width=0.96\linewidth,height=12cm,bbllx=10,bblly=0,bburx=520,bbury=530} 
\epsfig{figure=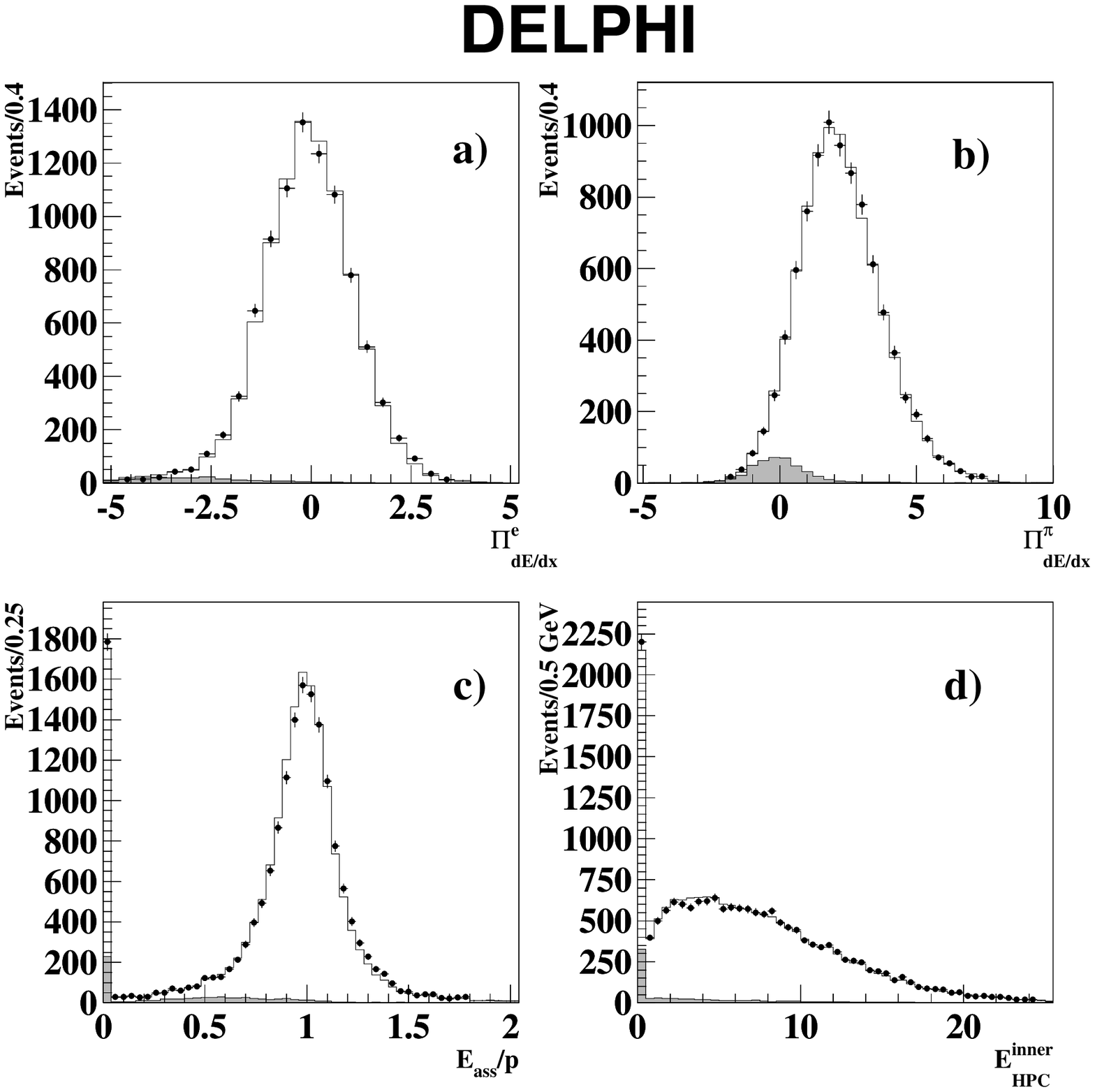,width=0.96\linewidth} 
    \caption{\it Distributions for electron 
    test samples in $\tau$-decays :
a) the variable $\pull^{\elec}$ ;
b) the variable $\pull^{\pi}$ ;
c) the variable $\frac{E_{ass}}{P}$ ;
d) the energy deposited in the innermost four layers of the HPC.
Data are shown as dots and simulation by a solid line. The shaded
region represents the simulated background including other $\tau$-decays.}
    \label{fig:elecid}
  \end{center}
\end{figure}

\begin{figure}[tbp]
  \begin{center}
%    \leavevmode
%\put(236,330){\makebox(0,0){\Large\bf DELPHI}}
%\epsfig{figure=fig_pionid1.eps,width=0.96\linewidth,height=12cm,bbllx=10,bblly=0,bburx=520,bbury=530} 
\epsfig{figure=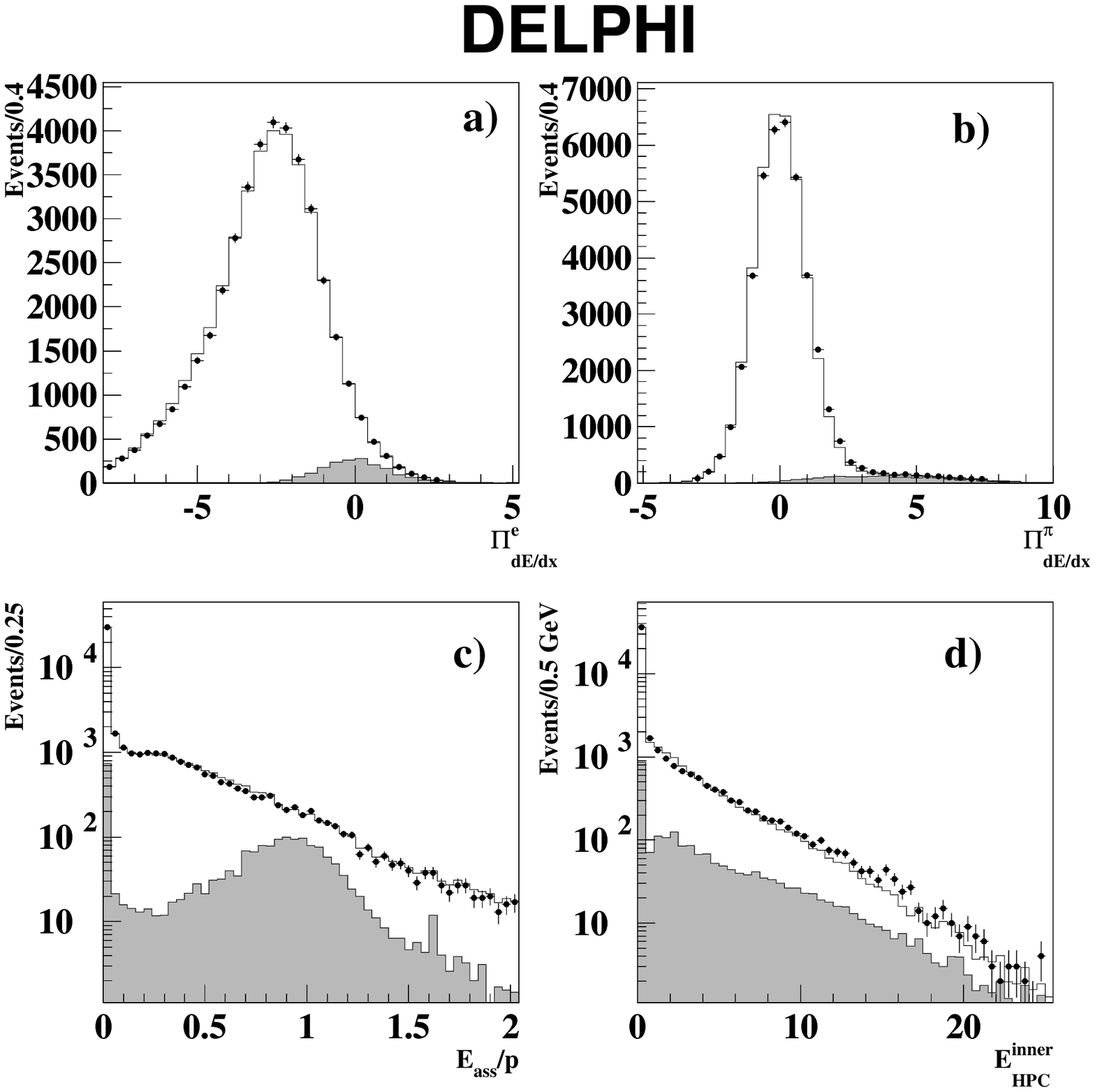,width=0.96\linewidth} 
    \caption{\it Distributions for hadron 
    test samples in $\tau$-decays of 
electron-hadron separation variables:
a) the variable $\pull^{\elec}$ ;
b) the variable $\pull^{\pi}$ ;
c) the variable $\frac{E_{ass}}{P}$ ;
d) the energy deposited in the innermost four layers of the HPC.
Data are shown as dots and simulation by a solid line. The shaded
region represents the simulated background including other $\tau$-decays.}
    \label{fig:pionid1}
  \end{center}
\end{figure}

The ratio of the energy deposition in the HPC to the reconstructed
momentum, $p$, has a peak at unity for electrons and a distribution rising
towards zero for hadrons. This is shown 
%the distributions, displayed
in  Fig.~\ref{fig:elecid}  for samples of electrons 
%selected using TPC $dE/dx$.
and Fig.~\ref{fig:pionid1} for samples of hadrons.
%The peak due to electrons is centred on zero with unit variance
%while the hadrons are confined to negative values.
It was also observed~\cite{delphileptonicbr99}  that the energy
deposition for hadronic showers starting before or inside the HPC
had to be downscaled by about 10\% in the simulation to get good
agreement with data. This is  due to an underestimate of the nuclear
interaction length of the material in some of the  subdetectors.

Electron rejection with high hadron selection efficiency
was performed using the associated energy deposition in 
only the first four layers of the HPC (corresponding to 6$X_0$ for 
perpendicular incidence) in which electrons deposited a significant 
amount of energy, while hadrons had a small interaction probability.
This is shown in Figs.~\ref{fig:elecid} and ~\ref{fig:pionid1} for 
electron and  hadron test samples from $\tau$-decays.

\subsubsection{Hadron calorimetry and muon identification}
\begin{figure}[tbp]
  \begin{center}
\epsfig{figure=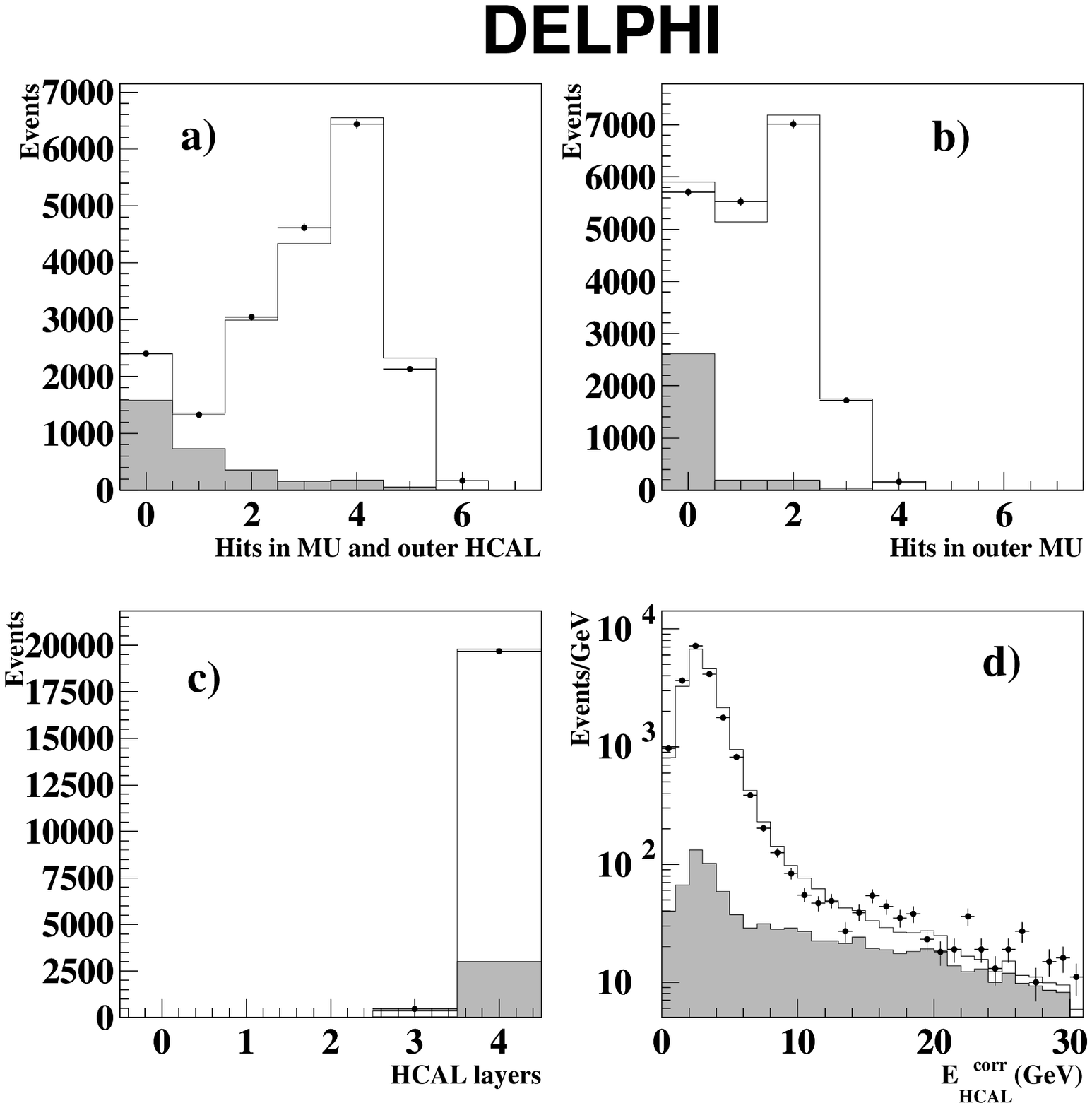,width=0.96\linewidth} 
    \caption{\it Muon-identification variables for muon test samples in $\tau$-decays:
a) number of muon chamber hits including the outer HCAL layer; 
b) number of muon chamber hits in the outer muon chambers;
c) number of layers in the HCAL;
d) corrected deposited energy in the HCAL.
Data are shown as dots and simulation after the corrections described in the text, 
by a solid line. The shaded area represents the non-muon background.}
    \label{fig:muonid}
  \end{center}
\end{figure}

\begin{figure}[tbp]
  \begin{center}
\epsfig{figure=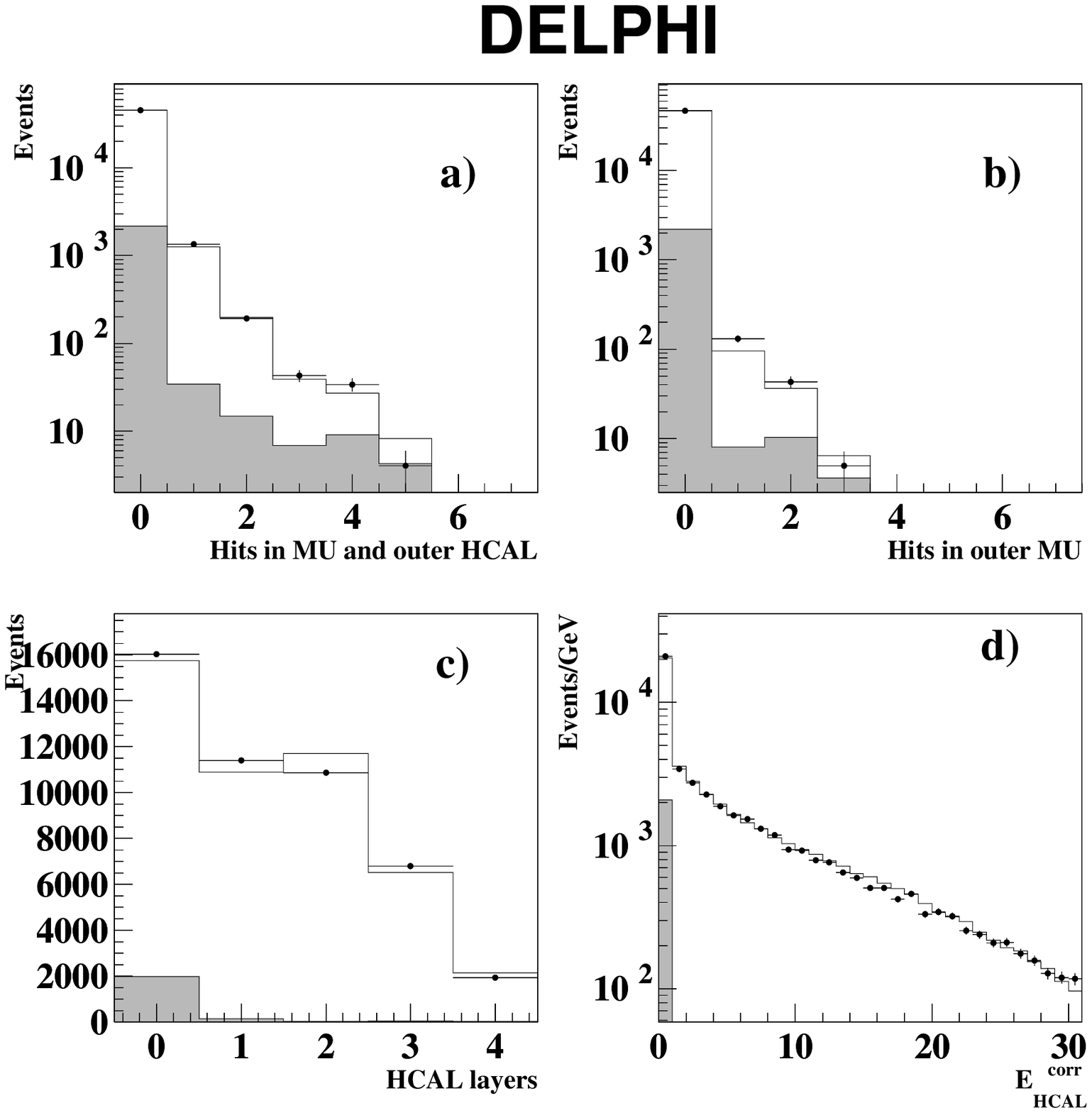,width=0.96\linewidth} 
    \caption{\it Muon-identification variables for hadron test samples in $\tau$-decays:
a) number of muon chamber hits including the outer HCAL layer; 
b) number of muon chamber hits in the outer muon chambers;
c) number of layers in the HCAL;
d) corrected deposited energy in the HCAL.
Data are shown as dots and simulation after the corrections described in the text, 
by a solid line. The shaded area represents the non-hadron background.}
    \label{fig:pionid2}
  \end{center}
\end{figure}

The signature of a muon
passing through the HCAL 
was that of a minimum-ionising particle,
leaving a roughly constant signal corresponding to an energy deposition of 
approximately 0.5~GeV 
in each of the four layers, and penetrating through into the muon chambers.
Hadrons, on the other hand, 
typically deposited most or all of their 
energy late in the HPC, the superconducting coil, or the
first layers of the HCAL, rarely penetrating through to the
muon chambers.
The  response of the  HCAL to hadrons 
depended  on the energy of the hadron  and where in the
detector it  interacted.   Studies  of the HCAL  response to  muons showed
good  agreement between data  and simulation.   For hadrons  the total
energy deposited  in the  HCAL was simulated well.  However  the depth
profile of the hadronic showers was not simulated well.  This is 
attributed to
cut-offs  in the modelling  of the  tails of  hadronic showers  in the
simulation  program.  These  had  a  negligible  effect  on  the  total
deposited energy but a significant  effect on the depth profile of the
shower.  This effect was  corrected for by artificially adding an
extra layer  hit in simulated hadronic
showers according to  the results obtained from a data sample of
charged  hadrons produced from a tightly-selected sample  of $\TRO$
decays.   An  additional HCAL layer  with  a  very  low   energy
deposition was  added in $(25.5\pm0.5)$\%  of hadronic $\tau$
decays.  This fraction and uncertainty were obtained from a fit of the
simulation  shower  depth  profile  to  the  data  test  sample. The distribution after this
correction is shown in  Figs.~\ref{fig:muonid}c) and ~\ref{fig:pionid2}c). 
%The
%branching  ratios obtained in  this analysis  were corrected  for this
%effect and the uncertainties were  obtained by varying the fraction by
%twice its fitted uncertainty

A number of different HCAL quantities gave hadron-muon separation,
such as the energy deposition in the outermost HCAL layer, or
the total energy in the HCAL,
$\AVHC$.  The total  
associated HCAL  energy, 
shown in Figs.~\ref{fig:muonid}d) and ~\ref{fig:pionid2}d), was corrected, as a function of
the number
of modules and the amount of material crossed by the particle, in such a way 
that the response for muons became independent
of the polar angle. 
%divided  by the
%number of  layers hit.
%Fig.~\ref{fig:muplot1} shows the HCAL
%response in a test sample of muons for: a) the energy deposit of
%the outermost layer of the HCAL; b) the maximum HCAL layer energy 
%associated to a charged particle; c) the average energy per hit layer
%in the HCAL.

The muon  chambers typically  had between  two  and five
layers  hit   by  a  penetrating   muon  (of  momentum   greater  than
2.5~GeV/$c$.) The  response to muons was  calibrated using dimuon
events. 
The simulation gave the same muon identification efficiency as the data.
Most
hadrons and  their resultant  shower did not  penetrate through  to the
muon  chambers,  especially  the  external  muon  chambers  which  lay
completely  outside  the magnet  yoke. However,  because  of the  poor
modelling of the  tails of hadronic showers in  the simulation program,
the probability that a hadron of  a given momentum would leave a signal
in the  muon chambers was higher in the data than in  the simulation. This was
studied using the same data sample of hadrons 
in tightly-tagged $\TRO$ events  and in three-prong
$\tau$-decays with  very  low muon  contamination.  Corrections  were
applied to the simulation for both  the inner and outer layers of muon
chambers.  These  were obtained by  adding extra muon  chamber hits
for  hadrons penetrating deeply into the  HCAL  so as  to obtain  good
agreement between data and simulation.  The fraction of extra hits was obtained
from a fit of the muon  chamber hit distribution in simulation to that
for  the data  test sample. Correlations with the corrections made to
the number of HCAL layers hit were taken into account.~Figs.~\ref{fig:muonid} and 
~\ref{fig:pionid2}, show the response of these
detectors for muon and hadron test samples.

\boldmath
\subsection{Photons and neutral pions}
\label{sec:photons}
\unboldmath 
%The $\pi^0$ has a branching ratio to two photons of (98.798$\pm$0.032)\%.
%The remainder decays via the Dalitz decay $\pi^0\to\gamma\ee$.
The reconstruction of photons and hence of $\pi^0$ mesons was based
principally on the HPC. 
Electromagnetic showers were
reconstructed using only the HPC information without any
prior knowledge of charged particles reconstructed 
in the tracking subdetectors and predicted to enter the HPC. 
Cuts based on the shower profile in the HPC were applied to photon candidates
to reduce the rate of fake photons from the interactions
of hadrons in the HPC. An algorithm~\cite{delphipizero,performance} was 
applied to individual HPC clusters to
see if they were compatible with having been 
produced by a single
$\pi^0$ decaying to two photons where the showers due to the 
two photons overlapped significantly. In addition,
photons which had converted to $\ee$ pairs in the 
detector material before the start of the HPC were reconstructed
using track segments from the tracking subdetectors.

\boldmath
\subsubsection{HPC shower reconstruction}
\label{sec:hpcpatrec}
\unboldmath 
The HPC gave up to nine longitudinal energy samples on a shower.  In
each  sample the  energy  deposition was  measured with  a granularity
of 2~cm  in r-$\phi$ and 3.5~mm in  z.  The shower pattern recognition
proceeded as follows.  All samplings in all nine layers were projected
on to a cylindrical grid of granularity 3.4~mm~$\times$~1.6~mrad in
z~$\times$~$\phi$.  Neighbouring bins were then added together into a
coarser grid of granularity $0.5^\circ$ by $0.5^\circ$ in
$\theta$ and $\phi$. A local maximum search was performed and contiguous
areas were separated if a significant minimum was found between two
local maxima.  All bins connected together after this were grouped
together into one cluster.  A fit was performed to the cluster
transverse profiles to estimate the position of the interacting
particle, together with the direction vector of the shower within the
HPC.  After the shower reconstruction, charged-particle tracks
reconstructed in the tracking system were extrapolated to the HPC and
associated to a cluster if it was compatible with having been produced
by that particle. To increase the efficiency for minimum-ionising
particles, additional low-energy clusters could be reconstructed along
the track extrapolation.

The substructure of each individual HPC cluster with energy greater
than 5~GeV was then studied to ascertain if it was compatible with
arising from a (typically high energy) neutral pion where the two
photons from the decay produced overlapping showers.

The high granularity of the HPC allowed  a measurement of the lateral
dimensions of a cluster.  For a cluster arising from two photons
entering  the  HPC  the  angular separation   of  the  two  photons is
about $m_{\pi^0}/E_{\pi^0}$ for symmetric  pair production (the most
difficult case).  This  is   about  7  mrad  for $E_{\pi^0}=20$~GeV,
similar  to  the granularity of  the detector.  To  search for cluster
substructure the energy  deposition inside a cluster was  plotted on
the $\phi-\theta$ plane with each depth layer of the cluster weighted,
giving the greatest weights to the more central layers,  which had the
most spatial-separation power.    This two-dimensional   distribution
of  weighted   charge deposition was then  fitted to a dipole function,
projected on to the main axis,  and two  Gaussian  distributions
fitted   to  the  projected distribution. The  invariant mass was
then calculated using the estimated energy deposition in each Gaussian
and the opening angle calculated from  the fit.  Some corrections
estimated from simulation were made  to account for detector  binning
effects and  biases in the fitting procedure.  The main background
came from photons converting just before the HPC and which were missed
by the photon conversion reconstruction algorithm.  This could give
rise to  a fake $\pi^0$ signal or a triple peak substructure  in the
cluster which  was  not properly  handled by  the algorithm.  Since
the magnetic  field deflected charged particles only in $\phi$, this problem was  mostly confined  to clusters with the
dipole axis lying  within 100~mrad of  the line with constant
$\theta$   passing  through  the cluster  barycentre.
To optimise the $\pi^0-\gamma$ separation with a single variable,
a neural network was used which had as inputs the estimated $\pi^0$ 
mass, the fraction of energy in the most energetic of the two 
photons and the angle of the dipole axis in the cluster.
The network had a single output neuron and was trained with a sample
of isolated photons in simulated $\mm\gamma$ final states to give 
a target output of zero and on tightly tagged $\pi^0$ candidates
in simulated $\TRO$ decays to give a target output of unity.

\begin{figure}[tbp]
  \begin{center}
    \leavevmode
\vspace{2 cm}
\epsfig{figure=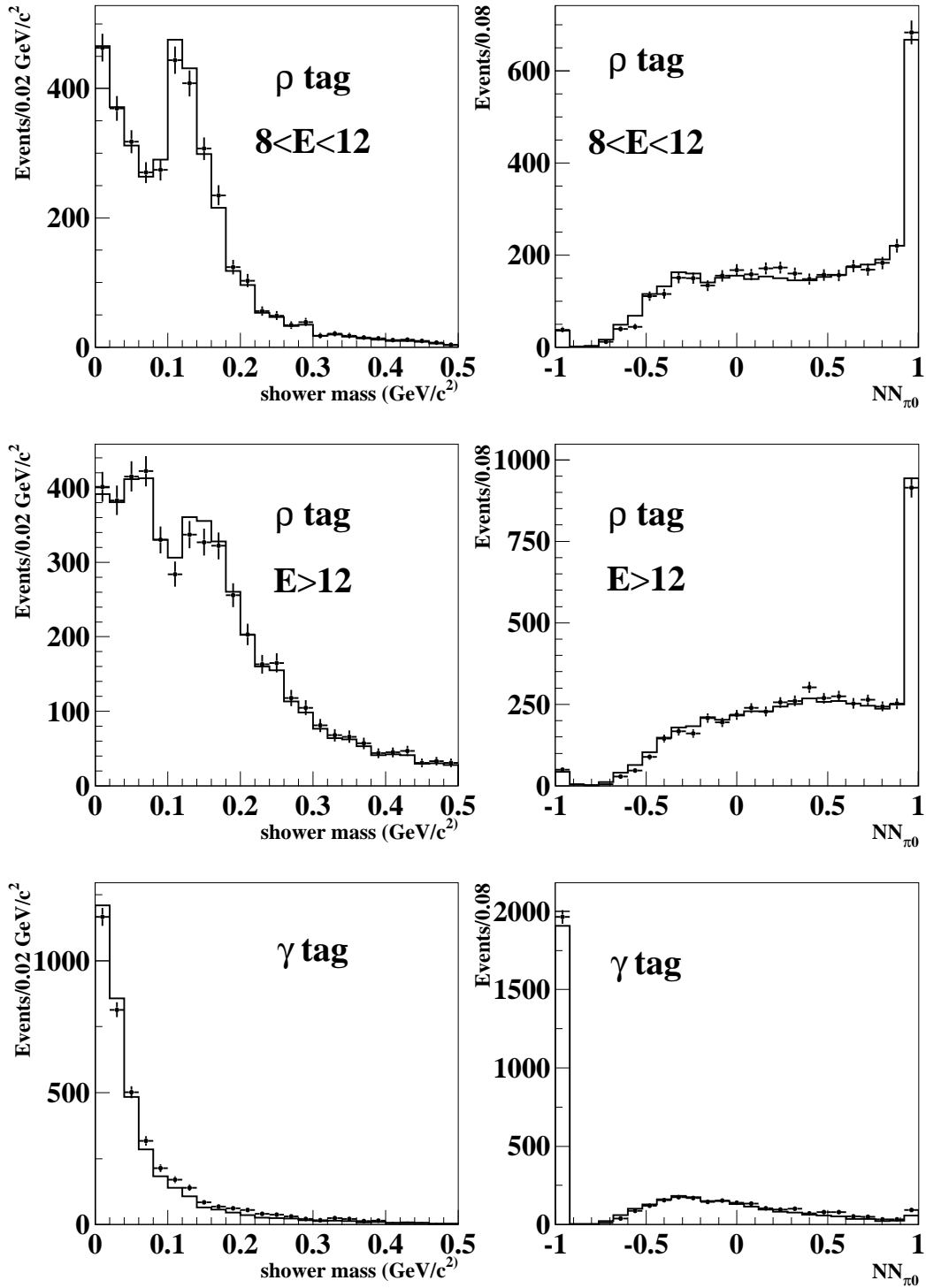,width=0.96\linewidth} 
\vspace{-2 cm}
    
    \caption{\it Distribution of reconstructed invariant
    mass in GeV/$c^2$ (left) and neural-network output (right) 
    reconstructed with the single-cluster
    algorithm for showers from the tight $\rho$ sample 
%for $8<E_{show}<12$ GeV  and     $E_{show}>12$ GeV 
at different energies and for showers from the isolated-$\gamma$ sample from $\mm\gamma$.
Data are shown as dots and simulation by a solid line.}
    \label{fig:singleclustpizero}
  \end{center}
\end{figure}

Fig.~\ref{fig:singleclustpizero} shows the invariant-mass distribution 
and neural network output
for single-cluster candidate $\pi^0$'s selected from a tightly-tagged
$\rho$ sample 
in two energy ranges ($8<E<12$ GeV
and $E>12$ GeV). This Figure also shows the same quantities 
for an isolated-$\gamma$ test sample from
$\mm\gamma$.

\begin{figure}[tbp]
  \begin{center}
    \leavevmode
\epsfig{figure=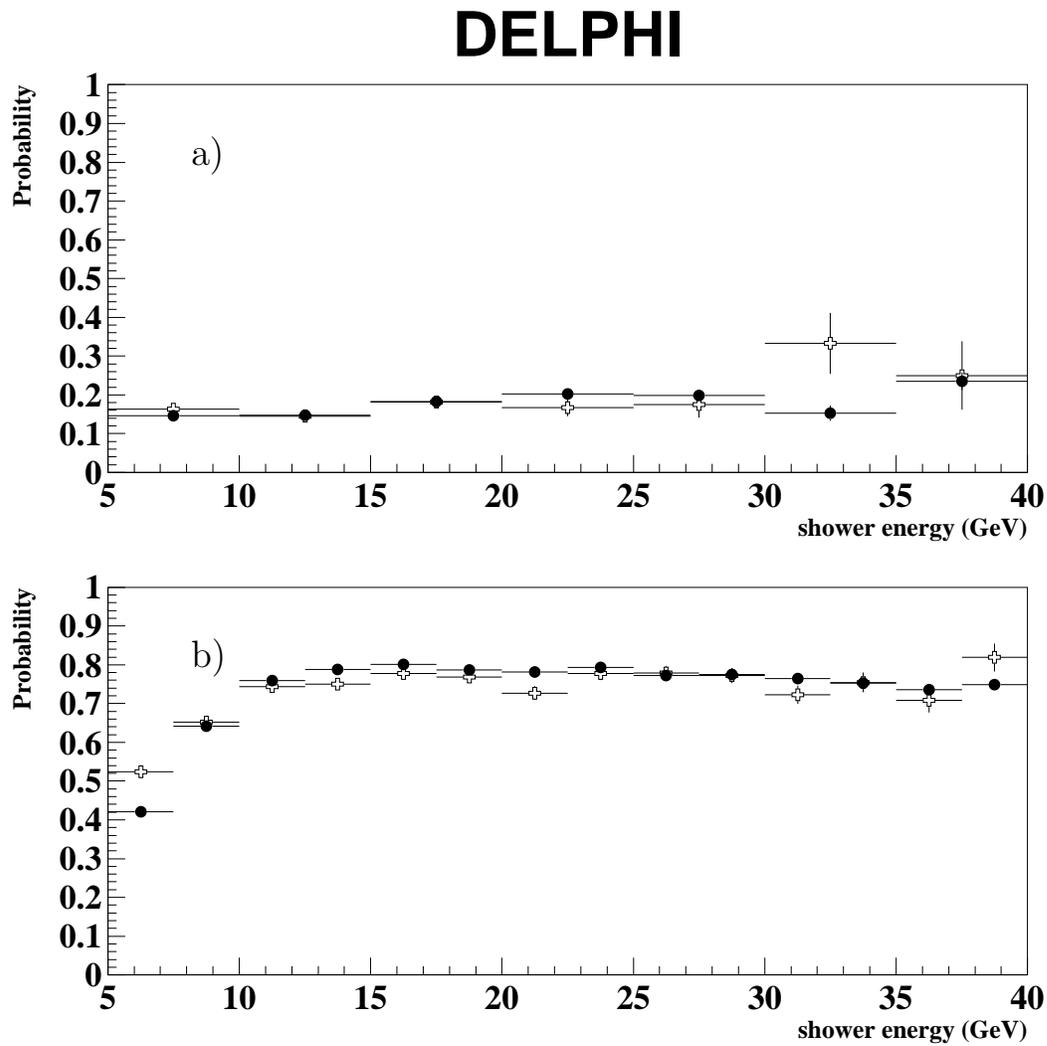,width=0.96\linewidth} 
\put(-360,350){\bf\Large a)}
\put(-360,160){\bf\Large b)}
    \caption{\it $\pi^0$-identification probability in a single shower
                 as a function of the shower energy, obtained
                 from isolated-$\gamma$ samples (a) and $\pi^0$ sample in tightly-tagged 
                 $\rho$ decays (b). 
                 Data are represented by crosses and simulation by filled
		 circles.}
    \label{fig:pizeroeff}
  \end{center}
\end{figure}

The HPC reconstruction was studied using isolated photons in
$\mm\gamma$ and $\ee\gamma$ final states. 
%The isolated photon sample was used to measure the
%probability for a photon to be  misidentified as a $\pi^0$.
%This occurred primarily via the single cluster algorithm.  
The 
probability to identify a single photon as $\pio$ is shown as a 
function of the reconstructed HPC cluster
energy in Fig.~\ref{fig:pizeroeff}a); on average it was $(16.8\pm0.6)$\% 
on data and
$(15.8\pm0.2)$\% in simulation.  The efficiency of the
algorithm was  studied in tightly-tagged $\tau$-decays containing one
charged hadron and a single energetic neutral HPC cluster with a 
combined mass compatible
with that of a $\rho$. Simulation
studies indicate that such a sample of HPC clusters  constituted a
$90.5$\% pure sample of $\pi^0\to\gamma\gamma$ decays. The probability
to identify a $\pi^0$ is also shown in
Fig.~\ref{fig:pizeroeff}b) as a  function of the reconstructed $\pi^0$
energy; on average it was $(69.7\pm0.5)$\% in data and
$(69.1\pm0.1)$\% in simulation.

The probability for a photon to be reconstructed as 
two HPC clusters was found to be a factor $1.15\pm0.02$ larger in the data, showing
an excess of unreconstructed conversions in the material in front
of the  HPC. The simulation was corrected according to this factor, following the
 reweighting technique described in~\cite{delphitopologicalbr}, and
a corresponding systematic uncertainty was assigned.
%From these distributions the
%uncertainties on the HPC efficiency and the probability to produce
%more than one  cluster were obtained.
% The angular precision  for high energy photons
%was 1~mrad in~$\theta$ and 2~mrad in~$\phi$.

\boldmath
\subsubsection{Converted photons}
\label{sec:conversions}
\unboldmath  Photons converting in the material before the HPC fell
into two classes, depending on whether the conversion took place before
or after the TPC sensitive volume.

About 7\%  of photons interacted in the  material before the TPC gas
volume giving an $\ee$ pair  detected in  some  of  the tracking
chambers. These were reconstructed using the tracks observed  in the
TPC.  A detailed study and description of the algorithm  and its
performance can be found in~\cite{delphitopologicalbr}. In simulation the
efficiency to reconstruct a converted photon was found to be
$(68.1\pm0.2)$\%
in one-prong $\tau$-decays and 
$(59.8\pm0.4)$\% in three-prong
$\tau$-decays. Good agreement between efficiencies in  data and
simulation was observed, while the simulation program underestimated
by about 10\% the material before the TPC in terms of radiation
lengths.  The photons obtained with this kinematic algorithm were in
general measured more precisely than those observed in the HPC.

A further  35\%/$\sin\theta$ of photons converted in the outer wall of
the TPC, the material of the RICH inner wall,  liquid radiator, drift
tube walls, mirrors, and outer walls, or in  the OD.  These
constituted a problem for the HPC pattern recognition as there was a
more limited possibility to reconstruct these conversions with the
tracking system as only the OD lay outside this region.  Such
conversion pairs were split in the  DELPHI magnetic field before
interacting  in the HPC to produce electromagnetic depositions. This
created a two-fold problem for  the neutral particle pattern
recognition: a  single  photon could produce two showers in the HPC,
one from each particle  of the $\ee$ pair.  These were reconstructed
as either one or two clusters by the HPC pattern recognition,
depending on the spatial separation of the showers.  Potentially,
both cases could be misidentified as a $\pi^0\rightarrow\gamma\gamma$
candidate.   Thus  the   number  of reconstructed photons was
incorrect. In  particular this splitting effect was important for
conversions in  the outer wall of  the TPC or the inner regions of the
RICH, far from the first sensitive plane of the HPC.

An algorithm reconstructed these converted photons from the
track segments  in the  OD. The OD consisted of five layers of
streamer tubes with a high efficiency for observing a charged
particle.  An  OD track element direction had a  resolution in
azimuthal  angle of  about  1~mrad and  thus  gave an unambiguous determination of the sign of  the charge
of a particle up to the beam momentum, if this particle originated
at radii smaller than 150 cm.  If there were two such track elements of
different sign of charge in the OD, unassociated to reconstructed charged
particles in  the TPC,  an algorithm which assumed  that both
track elements were produced by  an $\ee$ pair from a common conversion
point was run.   If  this common  conversion  point  was  compatible with  the
material structure in the TPC and the RICH and the OD track elements were
compatible in  polar angle, then this was regarded  as a photon.   If there
were HPC  clusters behind  the OD track  elements these clusters  had
to have energies which were compatible with the estimated e$^+$ and
e$^-$ energies derived from the  algorithm, in which case the
clusters were ignored  for further analysis.   This algorithm was
typically  about 25\% efficient.  
% More details are given in~\cite{delphitopologicalbr}.
%In Figure~\ref{fig:conv2}, data and
%simulation  are   compared  for  distributions  of  reconstructed
%conversion  radius  and  photon  energy  in  $\tau$  decays  for  this
%algorithm. The agreement is good.  
Studies of  efficiency 
using radiative  dimuon and dielectron 
events, showed the  ratio
of  post-TPC   conversion  reconstruction efficiency  in  data
compared  with  simulation  was  $0.95\pm0.07$, consistent with unity.
%In a further pass clusters were associated to any remaining OD hits
%and tagged as having arisen from a conversion.

\boldmath
\subsubsection{Hadronic shower rejection}
\label{sec:hpchadronrejection}
\begin{figure}[tbp]
  \begin{center}
    \leavevmode
\epsfig{figure=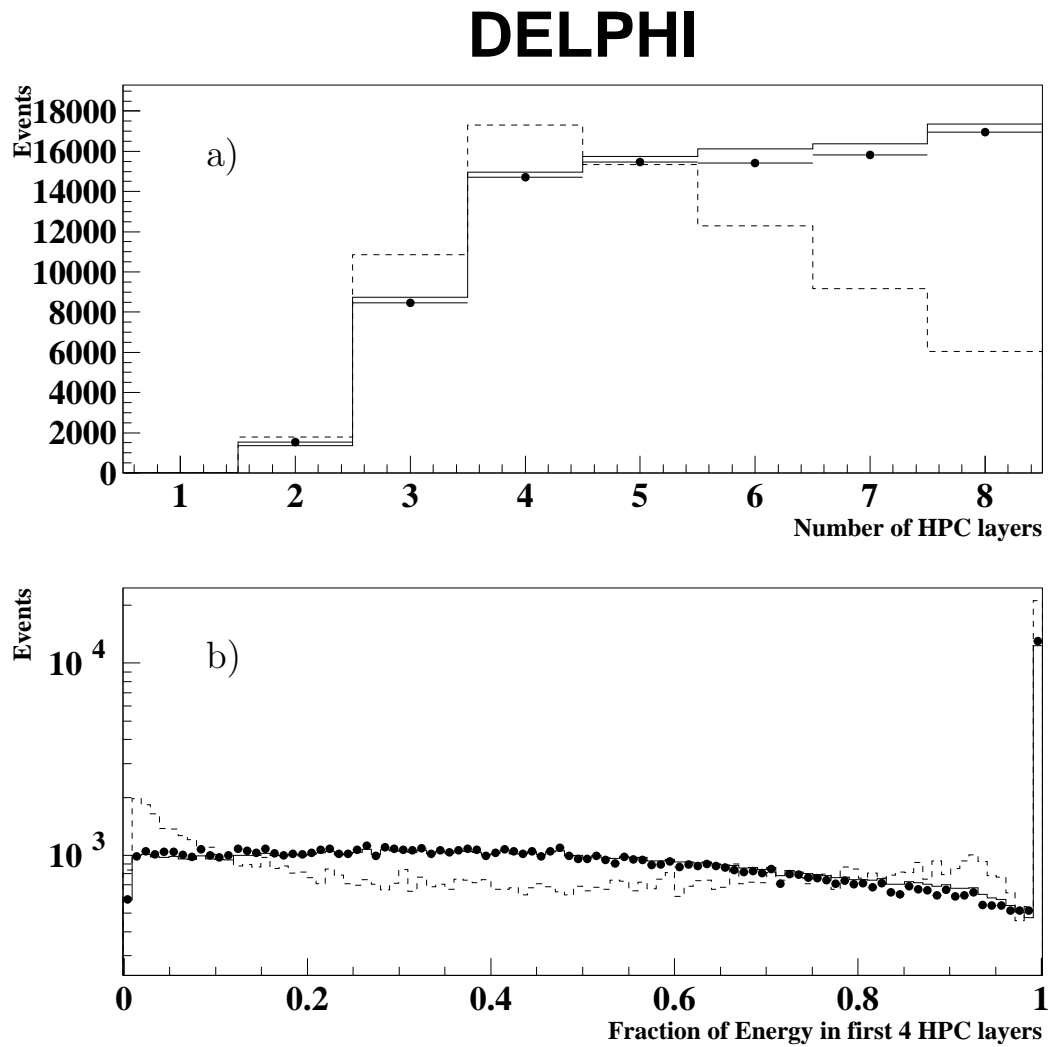,width=0.96\linewidth}        
\put(-360,350){\bf\Large a)}
\put(-360,160){\bf\Large b)}
    \caption{\it Distribution for candidate photon clusters for the inclusive sample of
	    $\tau$-decays:
           Number of HPC layers hit  (a);
            fraction of energy deposited in the four innermost layers (b).
     Data are shown as dots and simulation by a solid line. The dashed line shows 
     in arbitrary normalization the
     distributions for showers produced from charged hadrons.}
    \label{fig:hpclayers}
  \end{center}
\end{figure}

\begin{figure}[tbp]
  \begin{center}
    \leavevmode
\epsfig{figure=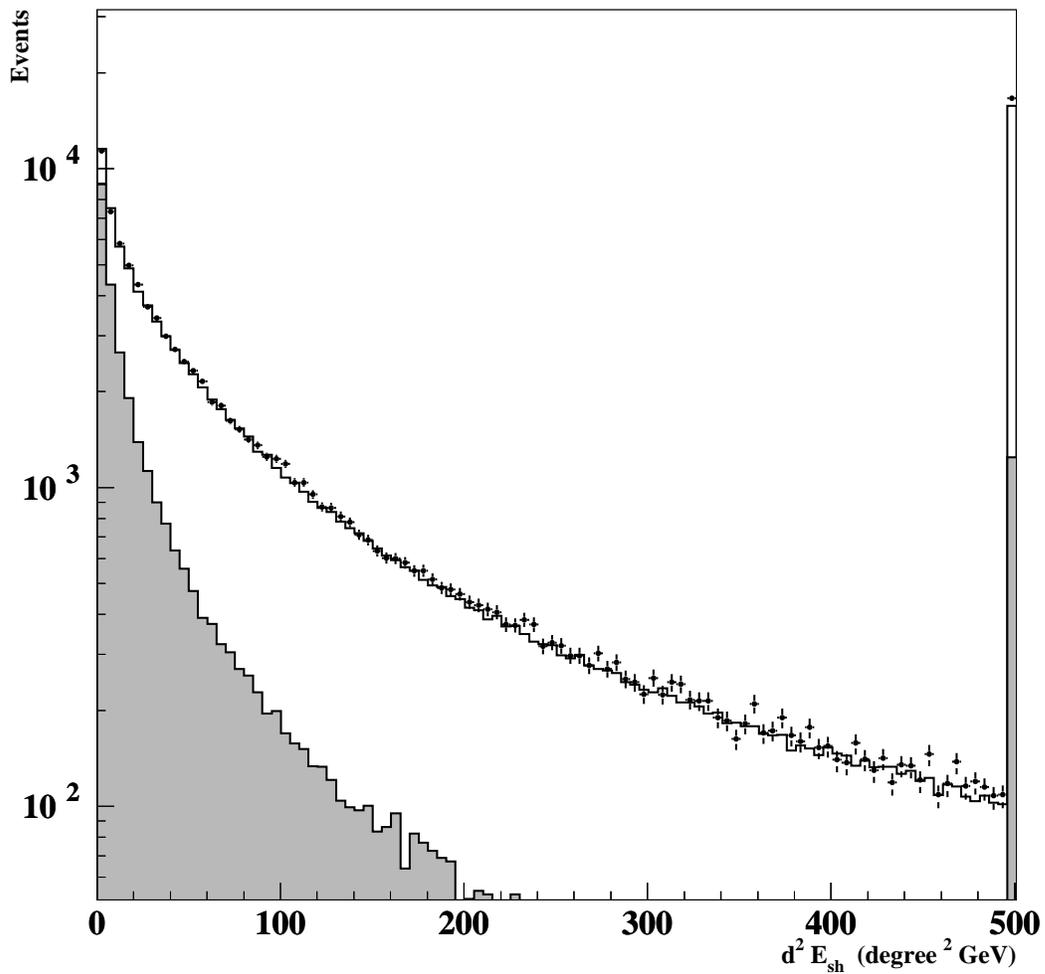,width=0.96\linewidth}        
    \caption{\it Distributions of the quantity $d^2_{sh-ch}E_{sh}$ used for rejection
             of hadronic showers in the HPC for the inclusive $\tau$ sample of photons with all identification
	     requirements applied except the cut on this variable. 
	     Data are shown as dots and simulation by a solid line.
             The shaded area shows the contribution from showers induced by charged hadrons. The
	     last bin shows the overflows.}
    \label{fig:hadronshowers}
  \end{center}
\end{figure}

\unboldmath  
The granularity of the HPC was used to remove many
clusters of a non-electromagnetic origin, such as hadronic showers
occurring in the HPC or before (in the RICH or OD).  These 
have different profiles in the detector due to the difference
between the nuclear interaction length and radiation length of lead,
and the difference in the sampling efficiency for the different processes through which
their energy is absorbed.  To be accepted as a photon shower a
cluster had  to have  both longitudinal and transverse profiles
consistent  with   those  expected  for  an electromagnetic
deposition~\cite{performance}. 
%In particular, it was required that there be at least three
%layers hit in a cluster, with at least two contiguous layers hit in
%the  first seven layers of the HPC, and that the longitudinal
%energy-weighted centre-of-gravity lie in the first seven layers of the
%HPC.  
This requirement rejected most showers from hadronic
interactions. 
In Fig.~\ref{fig:hpclayers}, the distributions of two quantities related to the
cluster profile in the HPC, namely the number of layers and the fraction of 
energy deposited in the first four layers,
are shown for candidate photons selected after this pattern recognition. A good agreement between
data and simulation is observed.
Because of the  high momentum of the charged  hadron and the proximity
to  the $\pi^0$'s, features  typical of  $\tau$-decays, additional
criteria  were applied to  reduce further  the contamination  from
hadronic showers.  Many hadronic showers were rejected by accepting
only those clusters for which the reconstructed  energy, $E_{sh}$,   was
greater than  500~MeV. The quantity $d^2_{sh-ch}E_{sh}$  had to be
greater than $10\deg^2$GeV, where $d_{sh-ch}$ was the opening angle
between  the cluster and  the track extrapolation at the HPC inner
surface. This variable tends to be strongly peaked at low values for fake showers originating 
from splits of
hadronic showers in the HPC, because they were typically of low energy and close to the
track entry point in the HPC, in contrast to those originating from a photon produced in a
$\tau$ decay. The distribution   of  this  quantity   is  shown  in
Fig.~\ref{fig:hadronshowers}, showing good  agreement between data and
simulation.  No hadronic rejection criteria were applied to HPC
clusters which were identified as candidate $\pi^0$ mesons with the
single-shower  $\pi^0$ algorithm, as such clusters benefited from a
low background.

\begin{figure}[tbp]
  \begin{center}
    \leavevmode
\epsfig{figure=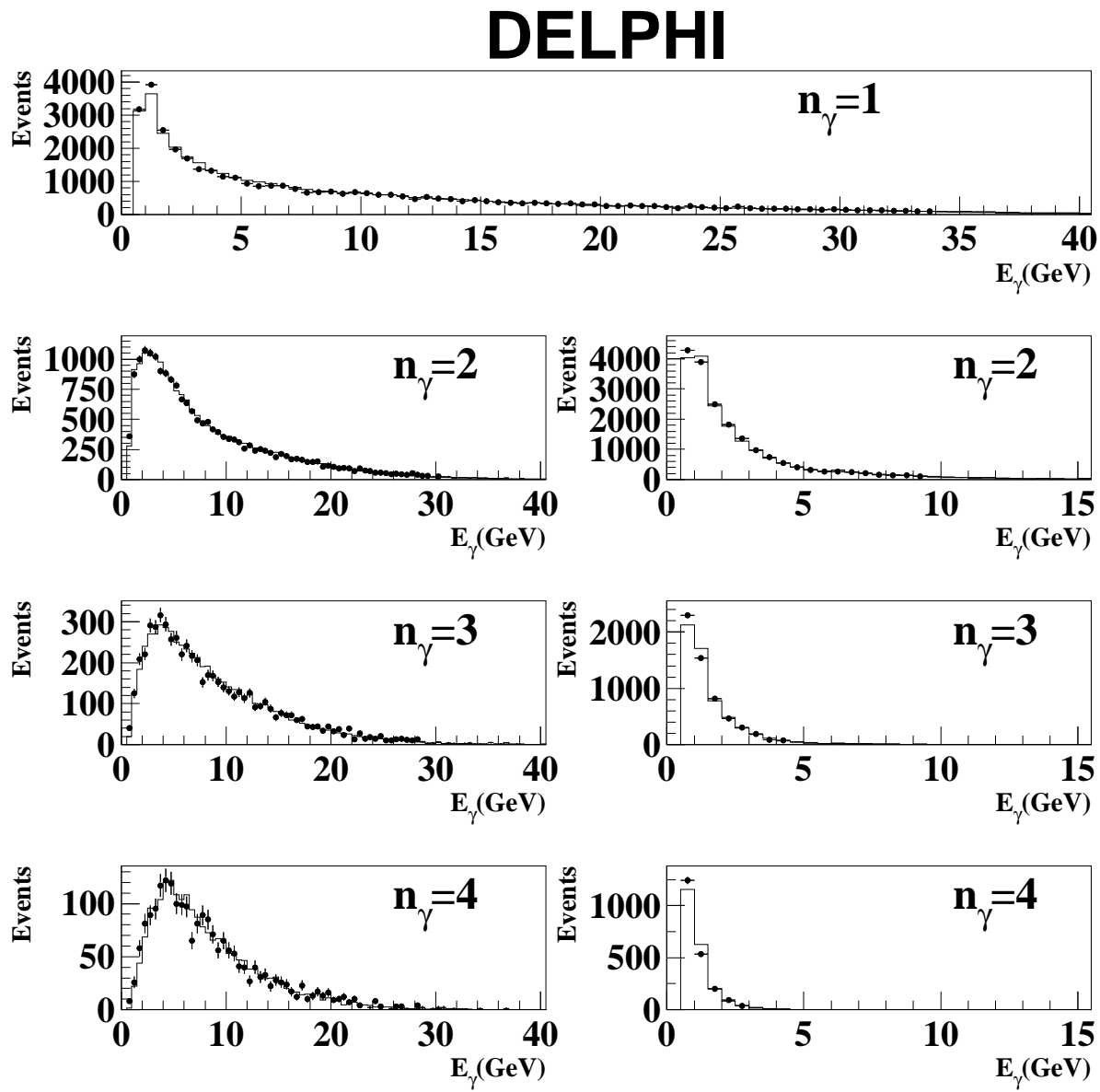,width=1.1\linewidth}        
    \caption{\it Energy distributions of identified photons in the HPC
             for $\tau$-decays containing 1, 2, 3 or 4 such 
             clusters. The figures on the left represent the most energetic
             cluster in the decay and those on the right the least energetic.
             Data are shown as dots and simulation by a solid line.}
    \label{fig:photonenergy}
  \end{center}
\end{figure}

In Fig.~\ref{fig:photonenergy}  the energy  spectra for selected HPC clusters
%passing the criteria for hadronic shower rejection  
are  shown  for the  maximum- and
minimum-energy photon  in  a $\tau$  decay  hemisphere, for  different
numbers  of reconstructed clusters in that hemisphere.  
The  agreement  between data  and
simulation is good in all cases  for both the low-energy region
and the high-energy region.

The full photon reconstruction efficiency was studied in two steps. First,
electron samples where the track had left a signal in the OD, with a small
probability of having interacted before reaching the HPC, were used to estimate
the shower reconstruction efficiency. Isolated $\gamma$ samples from radiative
$ee$ and $\mu\mu$ were used to check the shower profile cuts. 
The efficiency in the data was found to be
$(0.3\pm0.2)$\% less than in the simulation.

The production of fake photons from hadronic interactions was estimated from the
data and simulation agreement in the
distribution shown in Fig.~\ref{fig:hadronshowers}, for small values of the
variable, where the fake photons rate is comparable to that of 
the real photons. 
The simulation was found to
reproduce correctly the data to a relative $3$\%.

\boldmath
\subsubsection{Energy scale}
\label{sec:eemphot}
\unboldmath  

In addition to the previously measured electron samples, 
the HPC energy scale was studied 
using isolated photons in
$\mm\gamma$ and $\ee\gamma$ final states and $e\gamma$ compton scattering events. 
In these three cases
the direction  is well defined and the particle energy can 
be inferred with a very good
precision using kinematic constraints, independently from the energy
measurement in the calorimeter. This allowed the HPC energy
response to be calibrated as a function of energy.  A precision of 0.5\% or better was
obtained  on the energy scale throughout the entire energy spectrum.
The measured  energy resolution was $\sigma(E)/E = 0.31 \times
E^{-0.44} \oplus 0.027$.

\boldmath
\subsubsection{Spatial resolution}
\label{sec:hpcspatial}
\unboldmath

\begin{figure}[tbp]
  \begin{center}
    \leavevmode
\epsfig{figure=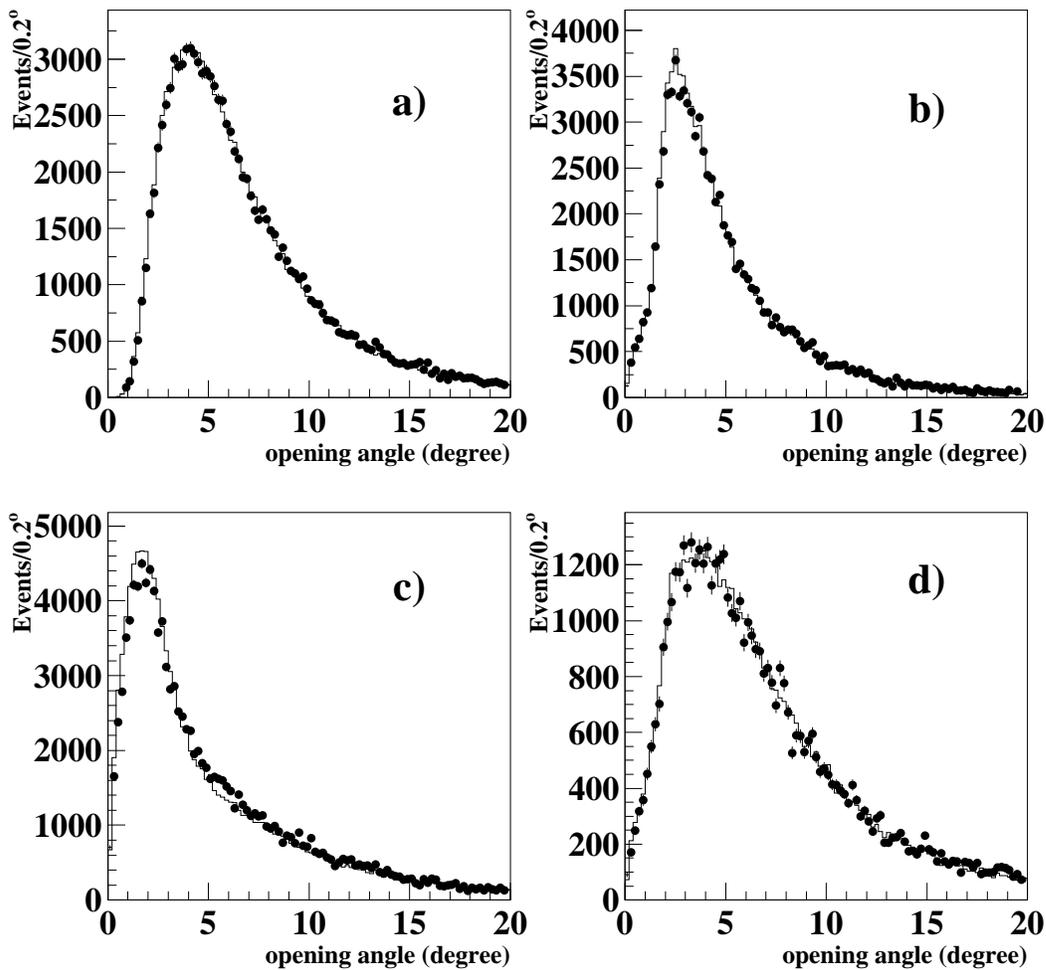,width=0.96\linewidth}        
    \caption{\it Distribution of opening angle between different types of
showers: a) a neutral shower fulfilling the $\gamma$ identification with a
shower associated to a track
b) two neutral showers fulfilling the $\gamma$ identification, 
c) a neutral shower failing the $\gamma$ identification with a
shower associated to a track
c) a neutral shower fulfilling the $\gamma$ identification with another one
failing it. 
Data are shown as dots and simulation by a solid line. 
}
    \label{fig:showersep}
  \end{center}
\end{figure}

The efficiency to reconstruct electromagnetic showers close to charged
hadron tracks and showers in the HPC is  important in $\tau$-decays
where the $\tau$-decay products are tightly collimated. 
To illustrate this, 
Fig.~\ref{fig:showersep} shows the minimum angular distance
between  different types of HPC clusters:
neutral clusters fulfilling the photon requirements, those failing them
and those associated to a charged particle.
The good agreement of data and simulation in the region of very small
opening angles demonstrates that all these effects are simulated correctly. 
%The
%distribution of the angular distance between a charged particle track
%extrapolation at
%the cylinder r~=~217~cm and the closest reconstructed neutral
%electromagnetic cluster is shown in Fig~\ref{fig:showersep} for showers 
%fulfilling the
%$\gamma$ requirements or failing them. 
%The
%shower reconstruction efficiency in simulation as a function of this
%distance is shown in Fig~\ref{fig:trackshowersep}b, showing a fall-off
%for distances below 10~cm.  
%Below this distance electromagnetic
%showers tended to be associated to the track. 

\boldmath
\subsubsection{Neutral pions} 
\begin{figure}[tbp]
  \begin{center}
    \leavevmode
\epsfig{figure=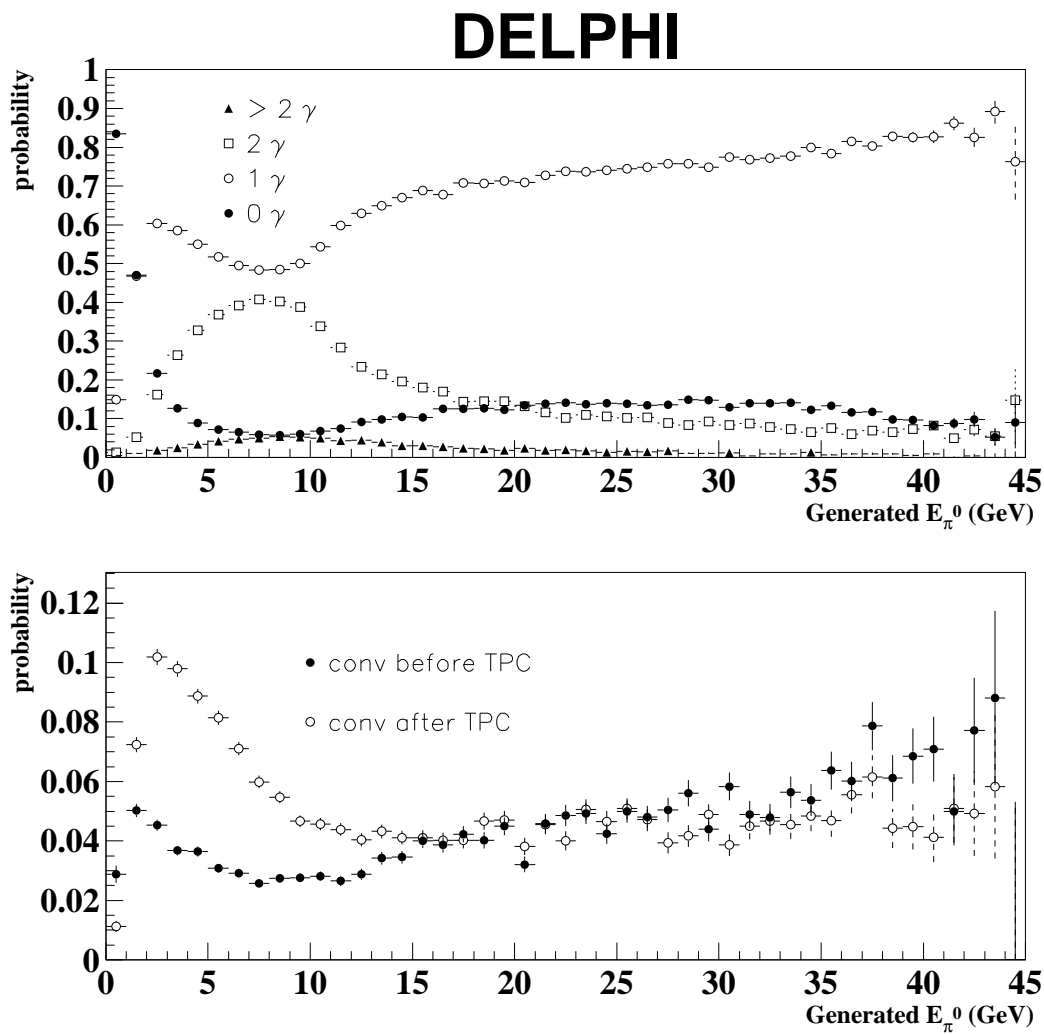,width=0.96\linewidth}        
    \caption{\it The fraction events,
    as a function
of $\pi^0$ energy with:
(top)  0,1,2, or more than 2,  photons reconstructed;
(bottom) a reconstructed conversion before or after the TPC, for
a sample of simulated non-radiative $\rho$-decays. }
    \label{fig:fracpizeros}
  \end{center}
\end{figure}

\unboldmath
\label{sec:neutralpions}
Fig~\ref{fig:fracpizeros} shows,
as a function of $\pi^0$ energy, the probability, in a simulated $\rho$
sample from $\tau$-decays, for
a $\pi^0$ to produce a given number of HPC  or
converted photons. The efficiency to observe one or more photons from one $\pi^0$ 
in the angular acceptance of the HPC is high,
dropping below 85\% only in the region below 3~GeV.

\begin{figure}[tbp]
  \begin{center}
    \leavevmode
\epsfig{figure=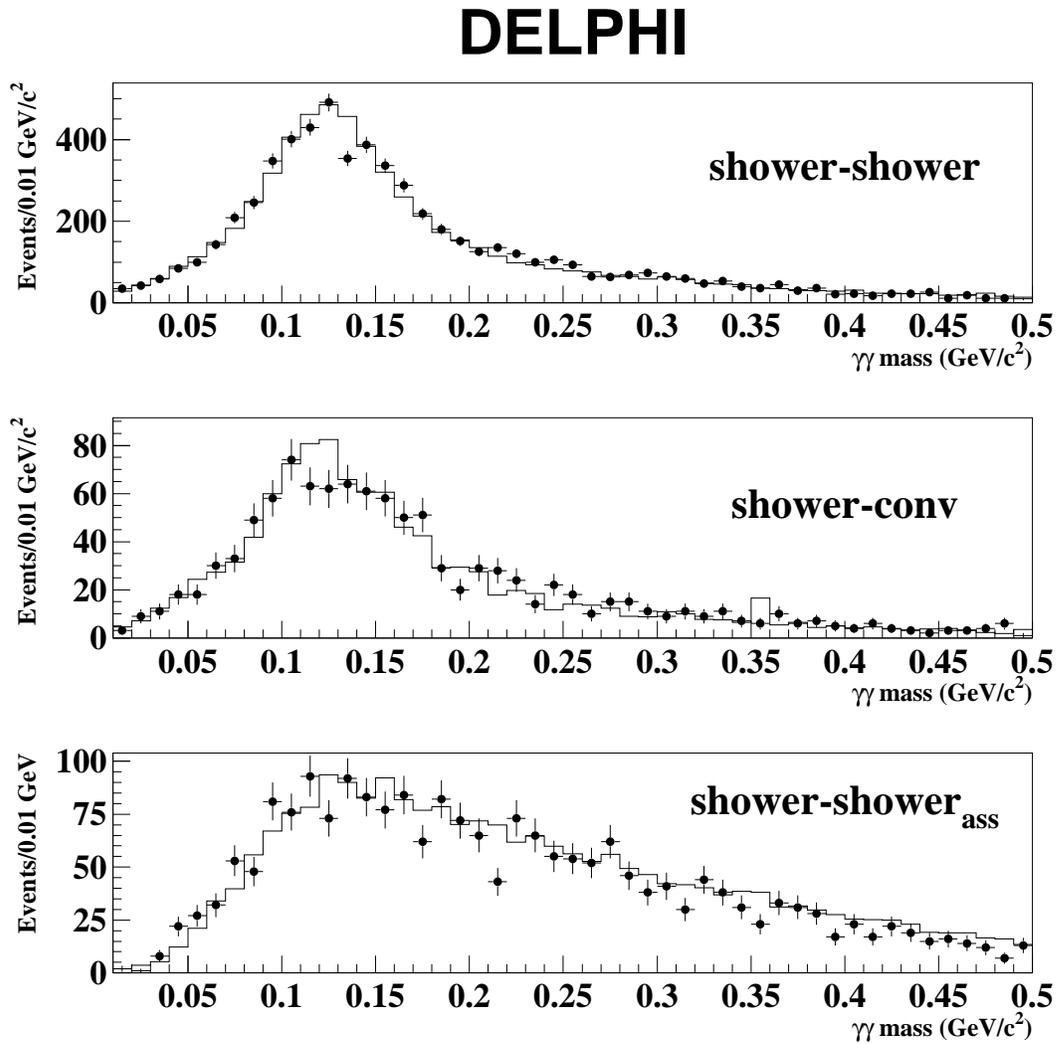,width=0.96\linewidth}        
\caption{\it Distribution of $\gamma$$\gamma$ invariant mass for $\pi^0$ 
            candidates for two neutral showers (top), neutral shower-converted
	    photon (middle) and charged-neutral shower (bottom) . 
	    Data are shown as dots and simulation by a solid line.
            }
    \label{fig:mpizero}
  \end{center}
\end{figure}

Reconstructed neutral pions fell into four
different categories. 
The first class (I) consisted of $\pi^0$ candidates
identified with the single cluster algorithm  
described in Section~\ref{sec:hpcpatrec}.
The second class (II) contained $\pi^0$ candidates 
reconstructed from
pairs of photons identified as separate HPC clusters,
while the third class (III)  contained  $\pi^0$ candidates 
reconstructed from
pairs of photons, of which at least  one
was a reconstructed converted photon.
The $\gg$ invariant-mass distributions for 
classes II and III of candidate $\pi^0$ are shown in Fig.~\ref{fig:mpizero}.
%No cut on the invariant mass was applied  for any of these three
%classes. 
Class I dominated for the high-energy region, 
the class II contributed significantly in the region 
below 10~GeV, while the class III had a rather flat energy dependence.

The fourth class (IV) recuperated photons
in single-prong $\tau$-decays where a photon
was accidentally associated to a charged-hadron track.
For $\tau$-decay hemispheres where the HPC cluster 
associated to the track satisfied the photon-candidate
requirements in all other respects, 
and where there was an additional photon candidate,
the HPC cluster was disassociated from the track,
provided that the invariant mass $m_{\gg}$
of the $\gg$ system
was greater than 70~MeV/$c^2$.
Simulation studies indicated that such decays were 
predominantly due to the $\pi^\pm \pi^0 \nut$ decay mode. 
The $m_{\gg}$ distribution for this
class of $\pi^0$ is also shown in Fig.~\ref{fig:mpizero},
before the mass cut.

\begin{figure}[tbp]
  \begin{center}
    \leavevmode
\epsfig{figure= 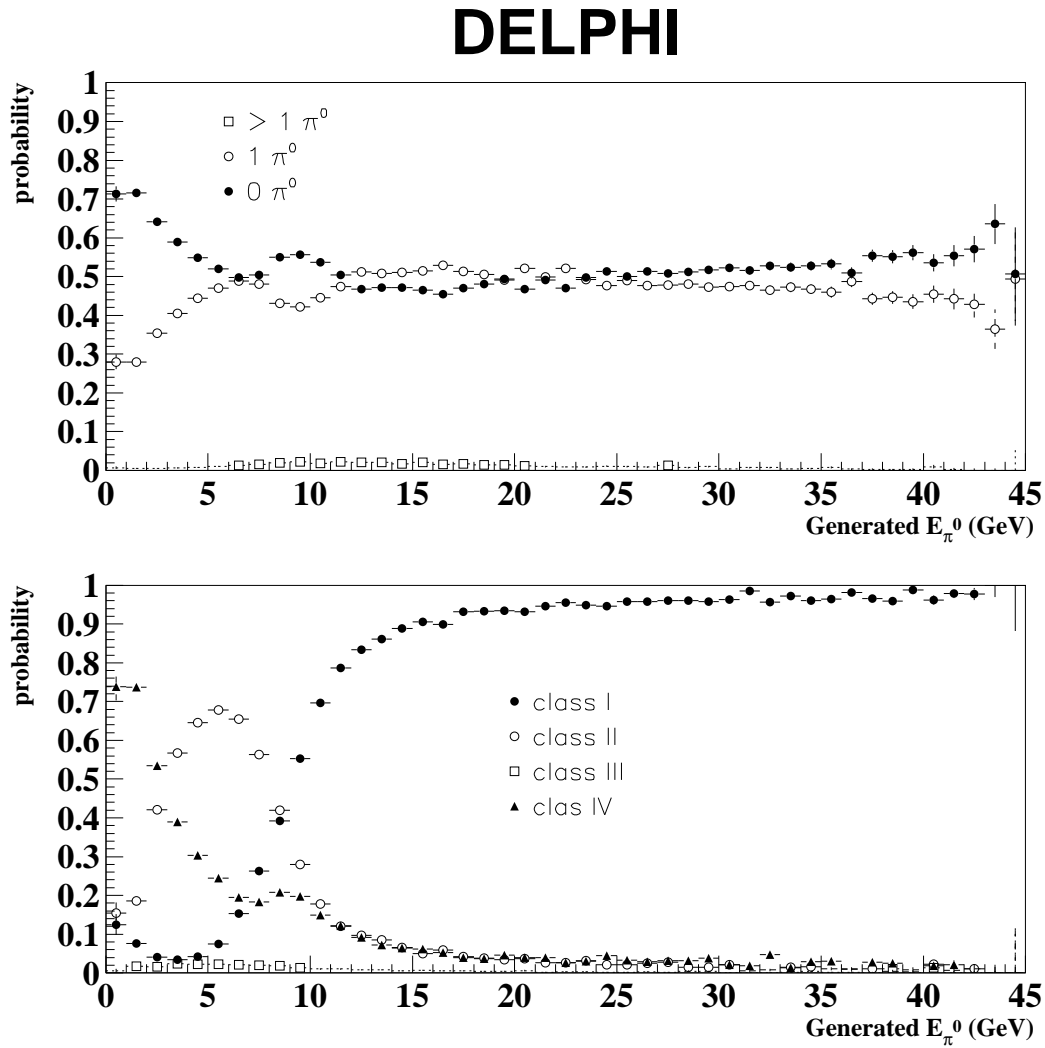,width=0.96\linewidth}        
    \caption{\it Total $\pi^0$ reconstruction
    efficiency (top) and probability to reconstruct a $\pi^0$ 
    in any of the classes described in the text
(bottom) as a function
of the $\pi^0$ energy for simulated $\rho$ decays.}
    \label{fig:fracpizeros2}
  \end{center}
\end{figure}

Fig~\ref{fig:fracpizeros2} shows the total identification efficiency as well as
the probability to classify
a $\pi^0$ in each of the four categories discussed above as a function of the
$\pi^0$ energy for simulated $\rho$ decays. 

It is important to note that many of the high energy showers, despite not being
resolved as $\pio$, are nevertheless most likely to come from a merged $\pio$.
%This fact is taken into account in the analyses in such a way that the
%``channel" selection efficiency may be higher than the $\pio$ efficiency. 
This accounted for in the analyses in such a way that, depending on other variables, a
single shower not identified as $\pio$ by any of the above criteria could be considered
as a $\pio$.

\boldmath
\section{Selection of $\eett$ events }
\label{sec:tautauselection}
\unboldmath
The selection of the $\eeztt$ event sample 
is identical to that used
in~\cite{delphitopologicalbr}. Only a summary is given here.

In the $\eeztt$ reaction at $\sqrt{s} =  M_Z$ , neglecting radiative
effects, the  $\tau^+$ and
$\tau^-$  are  produced back-to-back. 
The $\tau$'s each decay to one, three or five charged 
and one or more neutral particles in a tightly collimated jet.
Thus a $\tt$ event is characterised by two low-multiplicity
jets which appear 
back-to-back in the laboratory frame. Because each $\tau$ emits
at least one undetectable neutrino or anti-neutrino,
the full event energy is not observed in the detector.

Background events have various  signatures which enable them to be
separated  from the  signal.   For the  $\eeha$  channel, the  typical
charged-particle multiplicity is about 20, and quark fragmentation
produces less-collimated jets.   The $\bhab$ and $\eemm$ processes give
a  1 versus 1  charged-particle  topology, no  neutral electromagnetic
showers, and  contain the full  event energy measured in  the detector
due to the absence of final-state neutrinos. Two-photon events tend to
have low energy visible in the detector due
to the  loss of the $\ee$ pair  in the beam-pipe.  Cosmic  rays can be
removed using cuts on the distance  of closest approach to the interaction
region.

The data  were passed through the photon conversion algorithm
outlined in Section~\ref{sec:photons} to give
an improved estimate of the numbers of charged and neutral particles
in an event. To ensure that the $\tau$  products
lay in the acceptance of the relevant subdetectors it was demanded  that the
thrust axis of the event lie within the polar-angle region 
defined by $|\cos\theta| < 0.732$ and that there be at least one 
charged particle in the polar-angle 
region defined by  $|\cos\theta| >  0.035$. The event
was split into two hemispheres, each associated to a candidate $\tau$ decay, 
by a plane perpendicular to the thrust axis and passing
through the centre of the interaction region. It was required that
there be at least one charged particle in each hemisphere.

Hadronic decays of the Z were suppressed by requiring that there
be a maximum of eight charged particles in an event.
Background from four-fermion events was reduced, together with
a further suppression of Z hadronic decays, by requiring  that  the  event
isolation angle be  greater than $160^\circ$. The isolation angle
was  defined as the
minimum  angle  between  any  pair  of charged  particles  which  were
associated to opposite $\tau$-decay hemispheres.
Backgrounds from  $\mm$ and $\ee$ final states and cosmic rays were reduced 
by requiring that the isolation angle be less than $179.5^\circ$ for events 
with only two charged particles.

The  $\mm$ and $\ee$  contamination was  reduced further  by requiring
that both                         
%\mbox{$p_{rad}={(\frac{|\vec{p}_1|^2}{{p_{1}'}^2}+ 
%\frac{|\vec{p}_2|^2}{{p_{2}'}^2})}^{^{1/2}}$}
\mbox{$p_{rad}=\sqrt{\frac{|\vec{p}_1|^2}{{p_{1}'}^2}+ 
\frac{|\vec{p}_2|^2}{{p_{2}'}^2}}$}
 and
\mbox{$E_{rad}=\sqrt{\frac{E_1^2}{{E_{1}'}^2}+ \frac{E_2^2}{{E_{2}'}^2}}$} be
less than  unity.  The variables  $\vec{p}_1$ and $\vec{p}_2$  are the
momenta of the highest-momentum charged particles in  hemispheres
1~and~2 respectively.   
The   quantity  $p_{1}'$  was  obtained   from  the  formula
$p_{1}'=\sqrt{s}\sin\theta_2/(\sin\theta_1+
\sin\theta_2+|\sin(\theta_1+\theta_2)|)$,
and $p_{2}'$  by analogy  with the indices~1~and~2  interchanged.  The
angles $\theta_1$ and  $\theta_2$ are the polar angles  of the highest-momentum  charged particle in  
hemispheres 1~and~2  respectively.  The
variables  $E_1$  and $E_2$  are  the  total electromagnetic  energies
deposited in cones of half-angle $30^\circ$ about the momentum vectors
$\vec{p}_1$ and  $\vec{p}_2$ respectively, while $E_{j}'=cp_{j}'$,~for
$j=1,2$.
Much  of the remaining  background from  the dileptonic  channels came
from events containing hard radiation  lying far from the beam.  These
events  should lie  in a  plane.  Where  two charged  particles  and a
photon  were visible  in the  detector,  such events  were removed  
when the  sum of the angles between  the three particles was
greater than $359.8^\circ$.

Further reduction of the four-fermion contamination
was achieved by requiring that there be a minimum visible energy of 
$0.09\times\sqrt{s}$ in the event. Energy deposits recorded by the 
luminometers  (the SAT or STIC) at  angles of less than
$12^\circ$ from  the beam axis were excluded from this quantity.
For  events  with only  two
charged particles, the additional  condition that the
vectorial sum of the components of the
charged-particle  momenta transverse to the beam be greater  than 0.4
GeV/$c$ was applied. Two-photon events  typically  have very  low  values of  total
transverse  momentum  compared   with  $\tt$  events.  

Most cosmic rays were removed by  the  cut  on  isolation
angle. Further
rejection  was carried  out by  requiring  that at  least one  charged
particle in the  event have a perigee with  respect to the interaction
region of less  than~0.3~cm in the r-$\phi$ plane  and that both event
hemispheres  have a charged  particle whose  perigee point  lay within
4.5~cm of  the interaction region in~z and 1.5~cm in~r-$\phi$. 

In a final step,
a neural network was used to reduce the background from 
hadronic Z decays~\cite{delphitopologicalbr}.

The efficiency of the selection was estimated from simulation to be 
$(51.74\pm0.04)$\%. Within the angular acceptance it was 
about 85\%. A total of 80337 candidate $\eett$ events was selected.

The background levels were estimated from the data themselves by
fitting a normalisation factor to the background
contribution in variables sensitive to a particular background,
assuming that the shape of the background was that given by simulation, and where possible using
particle identification to isolate particular backgrounds.
The  total background was
estimated  to  be  $(1.51\pm0.10)$\%. The different contributions are shown in
Table.~\ref{tab:mode_bkg}.  
%with the  following  breakdown:
%$(0.11\pm0.01)\%$  from  $\mm$  final states;  $(0.40\pm0.07)\%$  from
%Bhabha events; $(0.29\pm0.01)\%$  from $\qq$ events; $(0.27\pm0.03)\%$
%from  $\ee\ee$ final states;  $(0.10\pm0.01)\%$ from  $\ee\mm$ events;
%$(0.27\pm0.03)\%$ from  $\ee\tt$ final states;  $(0.02\pm0.01)\%$ from
%$\ee\qq$   final   states;   $(0.05\pm0.01)\%$   from   cosmic   rays.
The backgrounds from $\mm\mm$, $\mm\tt$ and $\tt\tt$ final states were
negligible. 

\begin{table}[tbp]
\begin{center}
\begin{tabular}{l|r@{$\pm$}l}
\hline
 Source of
       & \multicolumn{2}{c}{$\tt$}      \\ 
 Background
       &\multicolumn{2}{c}{selection}\\
\hline                          
$\mm$               &  0.11& 0.01  \\
$\ee$               &  0.40& 0.07  \\
$\qq$               &  0.29& 0.01  \\
$\ee\ee$            &  0.27& 0.03  \\
$\ee\mm$            &  0.10& 0.01  \\
$\ee\tt$            &  0.27& 0.03  \\
$\ee\qq$            &  0.02& 0.01  \\
cosmic rays         &  0.05& 0.01  \\
\hline
\end{tabular}
\end{center}
\caption{\it Selected non-$\tt$ backgrounds, in percent, in the total
sample.}
\label{tab:mode_bkg}
\end{table}

\boldmath
\section{Charged-particle multiplicity selection}
\label{sec:topology}
\unboldmath  
The selection of  $\tau$-decays according to the charged-particle 
multiplicity was identical to that carried out for the
categories 1,~3 and~5 in the 
DELPHI measurement~\cite{delphitopologicalbr} 
of the $\tau$ topological branching ratios and only 
a brief description is given here. 
In the following a ``good'' track is defined as a track with
associated hits in either the TPC or OD.
The VD-ID tracks include not only tracks reconstructed
in the VD and ID without TPC or OD but also particles reconstructed from
the decay products of nuclear interactions in the detector
material.

A  one-prong $\tau$-decay  was  defined as  a  $\tau$-decay  hemisphere
satisfying any of the following criteria:
\begin{itemize}
\item 
only one good track  with at least one associated VD
hit, and no other tracks with associated VD hits;
\item
only one good track, without VD or ID hits, and one VD-ID track;
\item
no good tracks, and only one VD-ID track.
\end{itemize}
three-prong $\tau$-decays  were  isolated by  demanding
$\tau$-decay hemispheres satisfying at least one of the 
following sets of criteria:
\begin{itemize}
\item  
three, four or five good tracks,  of which
either two or three had associated VD hits;
\item 
two good tracks with associated VD  hits, plus one
VD-ID track;
\item
one good track with associated VD  hits, plus one or two
VD-ID tracks pointing within $3^\circ$ in azimuth to a TPC sector
boundary.
\end{itemize}
Candidate  five-prong $\tau$-decays   were selected  if they
satisfied at least one of the following topological criteria:
\begin{itemize}
\item 
five good tracks of which at least 
four had two or more associated VD hits;
\item 
four good tracks with associated VD hits, 
and one other VD-ID track.
\end{itemize}
Additional criteria were applied in the   selection  of five-prong $\tau$-decays due to the large potential background from  hadronic Z decays
and misreconstructed three-prong $\tau$-decays. The background
originating from $ 3h^\pm \ge 1 \pi^0 \nut$ final states with a Dalitz
decay was expected  to occur  at  a similar  level to  the signal.
Electron-rejection criteria based on $E_{ass}/P$
and $dE/dx$ described   in
Section~\ref{sec:chargedparticles} 
reduced this background by about 70\%,
and it was further suppressed by requiring that all  good tracks had a
reconstructed 
momentum greater  than~1~GeV/$c$.  To  reject Z$\to\qq$ events  it was
required that the total momentum  of the the five-prong system be greater
than 20~GeV/$c$. Only good tracks  were included in the calculation of
this quantity.

%These three classes accounted for 97.6\% of candidate $\tau$-decays in
%the $\tt$ sample.  The remaining 2.4\% of candidate $\tau$-decays were
%mostly one-prong and three-prong $\tau$-decays with some pattern recognition
%failure  or detector  inefficiency. 
%In~\cite{delphitopologicalbr}
%these were further split into classes with less separation power 
%than the main classes. In this analysis such decays are 
%added to the corresponding unclassified
%categories.

Table~\ref{tab:mode_eff} contains the  efficiencies of  these
selection requirements for the different exclusive $\tau$-decay modes 
and the charged particle multiplicity selections, as obtained
from simulation and after corrections for observed 
discrepancies between data and simulation in the rate 
and reconstruction efficiency of material reinteractions.

%The estimated non-$\tt$ backgrounds in the different charge multiplicity samples
%are given in Table~\ref{tab:mode_bkg}.
\begin{table}[tbp]
\begin{center}
\begin{tabular}{l|r@{$\pm$}l|r@{$\pm$}l|r@{$\pm$}l|r@{$\pm$}l}
\hline
 true $\tau$ 
       & \multicolumn{2}{c}{$\tt$}      
       & \multicolumn{6}{|c}{Charged Multiplicity Classification}   \\ 
\cline{4-9}
 decay mode
       &\multicolumn{2}{c}{selection}
       &\multicolumn{2}{|c}{1}  
       &\multicolumn{2}{|c}{3} 
       &\multicolumn{2}{|c}{5} \\
\hline                          
$\EL$               & 50.60& 0.07 & 99.95& 0.00 &  0.00& 0.00 &  0.00& 0.00 \\%&  0.02& 0.00 &  0.02& 0.00 \\
\hline                          
$\MU$               & 53.31& 0.07 & 99.96& 0.00 &  0.00& 0.00 &  0.00& 0.00 \\%&  0.01& 0.00 &  0.03& 0.00 \\
\hline                          
$\pi^- \nut$        & 49.69& 0.09 & 99.88& 0.01 &  0.04& 0.01 &  0.00& 0.00 \\%&  0.00& 0.00 &  0.08& 0.01 \\
$K^- \nut $         & 49.43& 0.36 & 99.90& 0.03 &  0.02& 0.02 &  0.00& 0.00 \\%&  0.01& 0.01 &  0.06& 0.03 \\
$\pi^-\kol\nut$     & 53.10& 0.48 & 99.79& 0.06 &  0.07& 0.03 &  0.00& 0.00 \\%&  0.00& 0.00 &  0.14& 0.05 \\
$K^-\kol\nut$       & 54.60& 0.87 & 99.78& 0.11 &  0.11& 0.08 &  0.00& 0.00 \\%&  0.00& 0.00 &  0.11& 0.08 \\
$\pi^-\kos\nut$     & 52.17& 0.48 & 94.48& 0.30 &  4.30& 0.27 &  0.00& 0.00 \\%&  0.16& 0.05 &  1.06& 0.14 \\
$K^-\kos\nut$       & 52.38& 0.86 & 94.50& 0.54 &  4.42& 0.49 &  0.00& 0.00 \\%&  0.06& 0.06 &  1.02& 0.24 \\
$\pi^-\kol\ko\nut$  & 52.82& 1.04 & 95.12& 0.62 &  3.72& 0.54 &  0.00& 0.00 \\%&  0.08& 0.08 &  1.07& 0.30 \\
$\pi^-2\kos\nut$    & 46.34& 1.80 & 86.72& 1.80 & 10.45& 1.63 &  0.00& 0.00 \\%&  0.00& 0.00 &  2.82& 0.88 \\
\hline                          
$\pi^-\pio\nut$     & 51.77& 0.06 & 97.87& 0.03 &  0.60& 0.01 &  0.00& 0.00 \\%&  0.84& 0.02 &  0.69& 0.01 \\
$K^-\pio\nut$       & 51.40& 0.47 & 97.66& 0.20 &  0.85& 0.12 &  0.00& 0.00 \\%&  0.80& 0.12 &  0.69& 0.11 \\
$\pi^-\pio\kol\nut$ & 51.85& 0.73 & 97.32& 0.33 &  0.78& 0.18 &  0.00& 0.00 \\%&  0.78& 0.18 &  1.11& 0.21 \\
$K^-\pio\kol\nut$   & 52.66& 1.24 & 96.71& 0.61 &  0.94& 0.33 &  0.00& 0.00 \\%&  0.94& 0.33 &  1.41& 0.40 \\
$\pi^-\pio\kos\nut$ & 50.78& 0.73 & 92.64& 0.54 &  4.65& 0.43 &  0.00& 0.00 \\%&  0.55& 0.15 &  2.16& 0.30 \\
$K^-\pio\kos\nut$   & 51.32& 1.32 & 92.56& 0.97 &  5.01& 0.80 &  0.00& 0.00 \\%&  0.41& 0.23 &  2.03& 0.52 \\
\hline                          
$\pi^-2\pio\nut$    & 51.07& 0.11 & 95.88& 0.06 &  1.25& 0.03 &  0.00& 0.00 \\%&  1.42& 0.04 &  1.45& 0.04 \\
$K^-2\pio\nut$      & 50.42& 1.12 & 94.65& 0.71 &  2.28& 0.47 &  0.00& 0.00 \\%&  1.39& 0.37 &  1.68& 0.40 \\
\hline                          
$\pi^-3\pio\nut$    & 48.89& 0.25 & 94.36& 0.16 &  1.68& 0.09 &  0.00& 0.00 \\%&  1.83& 0.10 &  2.13& 0.10 \\
%\hline                                                                      
%one-prong             & 51.42& 0.03 & 98.71& 0.01 &  0.42& 0.01 &  0.00& 0.00 \\%&  0.44& 0.01 &  0.43& 0.01 \\
\hline                          
\hline                                                                    
$\PPP$              & 54.71& 0.11 &  0.90& 0.03 & 90.26& 0.09 &  0.01& 0.00 \\%&  2.10& 0.04 &  6.74& 0.07 \\
$K^-\pi^-\pi^+\nut$ & 54.64& 0.56 &  1.03& 0.15 & 90.35& 0.45 &  0.00& 0.00 \\%&  1.70& 0.20 &  6.92& 0.38 \\
$K^-K^+\pi^+\nut$   & 53.87& 0.90 &  2.08& 0.35 & 87.23& 0.82 &  0.00& 0.00 \\%&  1.10& 0.26 &  9.59& 0.73 \\
\hline                                                                    
$\PPPZ$             & 53.88& 0.13 &  1.26& 0.04 & 86.39& 0.12 &  0.10& 0.01 \\%&  3.15& 0.06 &  9.10& 0.10 \\
\hline                                                                    
$\PPPZZ$            & 53.14& 0.46 &  1.37& 0.15 & 83.64& 0.46 &  0.22& 0.06 \\%&  3.68& 0.24 & 11.09& 0.39 \\
\hline                                                                    
$\PPPZZZ$           & 52.13& 1.06 &  1.46& 0.35 & 78.73& 1.20 &  0.17& 0.12 \\%&  4.74& 0.62 & 14.90& 1.05 \\
%\hline                                                                    
%3-prong             & 54.32& 0.08 &  1.06& 0.02 & 88.51& 0.07 &  0.05& 0.00 \\%&  2.54& 0.03 &  7.84& 0.06 \\
\hline                                                                    
\hline                                                                    
$\PPPPP$            & 49.63& 1.19 &  0.11& 0.11 & 12.63& 1.13 & 57.52& 1.67 \\%&  0.23& 0.16 & 29.51& 1.55 \\
\hline                                                                    
$\PPPPPZ$           & 48.91& 2.23 &  0.00& 0.00 & 15.04& 2.28 & 52.85& 3.18 \\%&  0.81& 0.57 & 31.30& 2.96 \\
%\hline                                                                    
%5-prong             & 49.47& 1.05 &  0.09& 0.09 & 13.16& 1.01 & 56.49& 1.48 \\%&  0.36& 0.18 & 29.90& 1.37 \\
\hline
\end{tabular}
\end{center}
\caption{\it Estimates of the $\tt$ selection- and topology-classification 
efficiencies, in percent,
for different exclusive decay modes, as obtained from simulation. The efficiencies are 
corrected for observed 
discrepancies between data and simulation in the rate 
and reconstruction efficiency of material reinteractions. The quoted uncertainties
are from the simulation~statistics~only. When no events are classified in a given class the Poissonian
upper bound is taken as error. Numbers smaller than 0.005\% are represented in the table as 0.00.}
\label{tab:mode_eff}
\end{table}

%\begin{table}[tbp]
%\begin{center}
%\begin{tabular}{l|r@{$\pm$}l|r|r|r}
%\hline
% Source of
%       & \multicolumn{2}{c}{$\tt$}      
%       & \multicolumn{3}{|c}{Charged Multiplicity}   \\ 
%\cline{4-6}
% Background
%       &\multicolumn{2}{c}{selection}
%       &\multicolumn{1}{|c}{1}  
%       &\multicolumn{1}{|c}{3} 
%       &\multicolumn{1}{|c}{5} \\
%\hline                          
%$\mm$               &  0.11& 0.01 & ~~~0.00& ~~~0.00 &  ~~~0.00 \\
%$\ee$               &  0.40& 0.07 & 99.95& 0.00 &  0.00 \\
%$\qq$               &  0.29& 0.01 & 99.96& 0.00 &  0.00 \\
%$\ee\ee$            &  0.27& 0.03 & 99.88& 0.01 &  0.04 \\
%$\ee\mm$            &  0.10& 0.01 & 97.87& 0.03 &  0.60 \\
%$\ee\tt$            &  0.27& 0.03 & 95.88& 0.06 &  1.25 \\
%$\ee\qq$            &  0.02& 0.01 & 94.36& 0.16 &  1.68 \\
%cosmic rays         &  0.05& 0.01 & 99.90& 0.03 &  0.02 \\
%\hline
%\end{tabular}
%\end{center}
%\caption{\it Selected non-$\tt$ backgrounds, in percent, in the total
%sample and in the samples of different
%charged multiplicity.}
%\label{tab:mode_bkg}
%\end{table}

%The branching ratios obtained were:
%\[\begin{array}{llcr}
%B_1 & \equiv B(\tau^- \to {\mathrm (particle)}^-  
%      \!\geq\!0\pi^0 \!\geq\!0K^0\nu_\tau(\bar{\nu})) & =&(85.316\pm0.093\pm0.048)\%;\\
%B_3 & \equiv B(\tau^- \to 2h^-h^+  \!\geq\!0\pi^0 \!\geq\!0K^0\nu_\tau) & =&(14.569\pm0.093\pm0.046 )\%;\\
%B_5 & \equiv B(\tau^- \to 3h^-2h^+ \!\geq\!0\pi^0 \!\geq\!0K^0\nu_\tau) & =&(0.115\pm0.013\pm0.006)\%,
%\end{array}\]

In this analysis the quality of reconstruction of the charged particle
tracks, especially  their momentum and precision  of the extrapolation
to  the calorimeters, was important for identification pourposes.  
Thus an additional requirement was made that candidate
one-prong $\tau$-decays should contain a ``good'' track.
This rejected candidate $\tau$-decays
reconstructed with only a VD-ID track or 
with the inelastic nuclear interaction
reconstruction algorithm.
These
have been extensively studied in~\cite{delphitopologicalbr}
and the necessary corrections for any data/simulation discrepancies
were applied, and the related uncertainties estimated.

The data sample of $\tau$-decays contained 134421 candidate one-prong decays,
23847 candidate three-prong decays and 112 candidate five-prong decays.

\boldmath
\section{Selection of (semi-)exclusive $\tau$-decay modes}
\label{sec:selection}
\unboldmath
Analyses using
sequential cuts and neural networks
identified the different decay modes.
%These analyses were carried out simultaneously on the 
%data sample in order to facilitate the estimation of
%the correlations
%between the statistical and systematic uncertainties.
In both cases, the different channel selections were applied simultaneously to
take into account  statistical and systematic correlations.

The following decay modes were selected using
sequential cuts (where $h=\pi$ or $K$):
$h^- \nut$,
$h^- \pi^0 \nut$,
$h^- \geq\! 2 \pi^0 \nut$,
$2h^- h^+  \nut$,
$2h^- h^+ \geq 1 \pi^0 \nut$,
$3h^- 2h^+ \nut$ and
$3h^- 2h^+ \pi^0 \nut$.
The neural-network analysis was only performed
for the one- and three-prong decays
and included the following additional modes:
$h^- 2 \pi^0 \nut$,
$h^- \geq\! 3 \pi^0 \nut$,
$2h^- h^+ \pi^0 \nut$ and
$2h^- h^+ \geq 2 \pi^0 \nut$.
It also included a measurement of the
electronic and muonic branching ratios.
Although no dedicated selection is present, we also quote the branching ratio
for the inclusive channel $h^-  \geq\! 1 \pi^0 \nut$, obtained by adding all the
modes with at least one $\pio$.

%No $\pi^-/K^-$  separation  was performed,  results included  both
%$\pi^-$ and $K^-$ contributions.   The branching ratios were varied by
%the  uncertainties  for  the measurements  in~\cite{delphikaon1};  the
%effect  on  the branching  ratios  caused  by  the difference  in  the
%selection efficiency between the  two components was negligible in all
%cases.   

In this analysis there is no explicit $K^0$ rejection or identification and the
selection efficiencies were, to first order, independent of the presence of
neutral kaons. These decays were therefore included in the
equivalent class without $K^0$. 
This was done regardless of the $K^0$ decay (even for
 the decay  mode $\Tauto h^-K^0\nut
\to      h^-\pi^0\pi^0\nut$) or of their interaction in the detector. 
For other mesons, the decays were classified according to the number of charged
pions, charged kaons and neutral pions except for the decay modes containing 
$\eta$ with subsequent decay to $\gamma\gamma$ or $\pi^+\pi^-\gamma$ and
$\omega$   with subsequent decay to $\pi^0\gamma$.  
These decay modes
are difficult to isolate from the decay modes measured in this analysis,
but are treated as background. Their total branching ratio was~\cite{pdg2002}
$(0.289\pm0.027)$\%, $(0.266\pm0.027)$\% in one-prong decays and 
$(0.023\pm0.003)$\% for three-prongs. The branching ratios have been
corrected for these backgrounds.

\subsection{Sequential-Cuts Analysis}
\label{sec:seq_cuts}
The various hadronic decay modes were selected with the cuts described below.
The selection efficiencies and cross-talk between channels are given
in Table~\ref{tab:seqeffa} for the one- and three-prong
modes, together with the backgrounds from non-$\tt$ sources.
Table~\ref{tab:seqeffb} contains the analogous information for the
five-prong decay modes. The analysis for leptonic decays is described 
in~\cite{delphileptonicbr99}.
\begin{table}[p]
\begin{center}
\begin{tabular}{l|r@{$\pm$}l|r@{$\pm$}l|r@{$\pm$}l|r@{$\pm$}l|r@{$\pm$}l}
\hline
 true $\tau$ 
       & \multicolumn{10}{|c}{Sequential cuts decay classification}   \\ 
\cline{2-11}
 decay mode
       &\multicolumn{2}{|c}{$h^- \nut$}  
       &\multicolumn{2}{|c}{$h^-\pio\nut$} 
       &\multicolumn{2}{|c}{$h^- \geq 2\pio\nut$} 
       &\multicolumn{2}{|c}{$3h^\pm \nut$}  
       &\multicolumn{2}{|c}{$3h^\pm\!\! \geq\!\! 1\pio\nut$} \\
\hline                          
$\EL$              & 0.11& 0.01 &  0.89& 0.02 &  0.03& 0.00  & 0.00& 0.00 &  0.00& 0.00   \\
\hline                                                                                      
$\MU$              &  1.62& 0.03 &  0.22& 0.01 &  0.00& 0.00 &  0.00& 0.00&  0.00& 0.00 \\
\hline                                                                                      
$\pi^- \nut$       & 49.69& 0.13 &  1.44& 0.03 &  0.20& 0.01 & 0.03& 0.01 &  0.01& 0.00  \\
$K^- \nut $        & 50.82& 0.53 &  1.18& 0.12 &  0.21& 0.05 & 0.05& 0.02 &  0.01& 0.01  \\
$\pi^-\kol\nut$    & 28.45& 0.61 &  7.86& 0.37 &  0.70& 0.11 & 0.05& 0.03 &  0.02& 0.02  \\
$K^-\kol\nut$      & 29.73& 1.40 &  7.20& 0.79 &  0.48& 0.21 & 0.00& 0.09 &  0.13& 0.11  \\
$\pi^-\kos\nut$    &  5.30& 0.31 & 13.92& 0.48 &  2.23& 0.20 & 3.37& 0.25 &  0.51& 0.10   \\
$K^-\kos\nut$      &  6.88& 0.78 & 11.77& 0.99 &  3.06& 0.53 & 3.48& 0.56 &  0.55& 0.23  \\
$\pi^-\kol\ko\nut$ &  7.64& 0.79 & 13.23& 1.00 &  3.96& 0.58 & 0.09& 0.06 &  1.05& 0.21  \\
$\pi^-2\kos\nut$   &  0.43& 0.34 & 14.33& 1.81 &  9.40& 1.51 & 5.89& 1.22 &  5.22& 1.15  \\
\hline                                                                                      
$\pi^-\pio\nut$    &  1.37& 0.02 & 44.08& 0.09 &  3.03& 0.03 & 0.21& 0.01 &  0.36& 0.01  \\
$K^-\pio\nut$      &  1.22& 0.13 & 30.79& 0.56 &  2.33& 0.18 & 0.25& 0.06 &  0.40& 0.08  \\
$\pi^-\pio\kol\nut$&  0.89& 0.19 & 39.13& 0.98 &  8.23& 0.55 & 0.09& 0.06 &  1.05& 0.21  \\
$K^-\pio\kol\nut$  &  0.43& 0.22 & 13.45& 1.13 &  4.70& 0.70 & 0.34& 0.19 &  1.19& 0.36  \\
$\pi^-\pio\kos\nut$&  0.08& 0.06 & 26.10& 0.90 & 16.08& 0.75 & 0.45& 0.14 &  3.96& 0.40  \\
$K^-\pio\kos\nut$  &  0.22& 0.17 & 15.07& 1.29 &  8.45& 1.00 & 1.26& 0.40 &  3.43& 0.65 \\
\hline                                                                                      
$\pi^-2\pio\nut$   & 0.05& 0.01 & 19.30& 0.10 & 25.50& 0.12  & 0.12& 0.01 &  1.81& 0.04  \\
$K^-2\pio\nut$     & 0.00& 0.10 & 17.26& 1.18 & 23.08& 1.31  & 0.00& 0.10 &  2.20& 0.46  \\
\hline                                                                                      
$\pi^-3\pio\nut$   & 0.02& 0.01 & 10.65& 0.25 & 41.23& 0.40  & 0.05& 0.02 &  2.16& 0.12  \\
\hline                                                                                      
\hline                                                                                      
$\PPP$             & 0.02& 0.00 &  1.82& 0.03 &  0.13& 0.01  & 71.82& 0.10 &  6.72& 0.05  \\
$K^-\pi^-\pi^+\nut$& 0.00& 0.02 &  1.55& 0.19 &  0.09& 0.05  & 73.05& 0.68 &  7.31& 0.40  \\
$K^-K^+\pi^+\nut$  & 0.00& 0.02 &  1.58& 0.19 &  0.05& 0.05  & 73.58& 0.81 &  7.65& 0.49  \\
\hline                                                                                      
$\PPPZ$            & 0.00& 0.00 &  1.14& 0.04 &  0.90& 0.04  & 18.71& 0.16 & 45.79& 0.21  \\
\hline                                                                                      
$\PPPZZ$           & 0.00& 0.01 &  0.38& 0.07 &  1.98& 0.16  & 6.26& 0.28 & 61.84& 0.56  \\
\hline                                                                                      
$\PPPZZZ$          & 0.00& 0.08 &  0.08& 0.08 &  2.94& 0.48  & 2.33& 0.43 & 64.63& 1.37  \\
\hline                                                                                      
\hline                                                                                      
$\PPPPP$           & 0.00& 0.21 &  0.16& 0.21 &  0.20& 0.21  & 13.67& 1.56 & 14.40& 1.59  \\
\hline                                                                                      
$\PPPPPZ$          & 0.00& 0.83 &  1.00& 0.91 &  0.00& 0.83  & 1.89& 1.24 & 21.66& 3.76  \\
\hline                                                                                      
\hline                                                                  
Source
       & \multicolumn{10}{|c}{Non-$\tt$ backgrounds}   \\                  
\cline{2-11}                                                               
\hline                                                                                      
$\mm$              & 0.02& 0.01 & 0.06& 0.01 & 0.02& 0.01 & 0.00& 0.00 & 0.00& 0.00  \\
$\ee$              & 0.05& 0.02 & 0.10& 0.02 & 0.03& 0.02 & 0.00& 0.00 & 0.00& 0.00  \\
$\qq$              & 0.15& 0.03 & 0.09& 0.02 & 0.17& 0.04 & 0.29& 0.03 & 1.20& 0.12  \\
$4f$               & 0.39& 0.07 & 0.31& 0.04 & 0.23& 0.06 & 0.14& 0.03 & 0.11& 0.05  \\
\hline
\end{tabular}
\end{center}
\caption{\it For the sequential-cuts analysis, classification 
efficiencies, in percent,
for different exclusive one- and three-prong 
decay modes, as obtained from simulation
after correction for the data/simulation discrepancies discussed in the text.
The bottom part shows the backgrounds in percent in each class from non-$\tt$
sources. The quoted uncertainties
are from the simulation~statistics~only. When no events are classified in a given class the Poissonian
upper bound is taken as error. Numbers smaller than 0.005\% are represented in the table as 0.00.
}
\label{tab:seqeffa}
\end{table}
\begin{table}[tb]
\begin{center}
\begin{tabular}{l|r@{$\pm$}l|r@{$\pm$}l}
\hline
 true $\tau$ 
       & \multicolumn{4}{|c}{Decay classification}   \\ 
\cline{2-5}
 decay mode
       &\multicolumn{2}{|c}{$5h^\pm \nut$}  
       &\multicolumn{2}{|c}{$5h^\pm \!\!\geq\!\!1 \pio\nut$} \\
\hline                                                                                                    
$\PPP$             & 0.00& 0.00 &  0.00& 0.00   \\
$K^-\pi^-\pi^+\nut$& 0.00& 0.02 &  0.00& 0.02   \\
$K^-K^+\pi^+\nut$  & 0.00& 0.02 &  0.00& 0.02   \\
\hline                                             
$\PPPZ$            & 0.11& 0.01 &  0.01& 0.00   \\
\hline                                             
$\PPPZZ$           & 0.10& 0.04 &  0.05& 0.03   \\
\hline                                             
$\PPPZZZ$          & 0.00& 0.08 &  0.17& 0.12   \\
\hline                                             
\hline                                             
$\PPPPP$           & 55.26& 2.25 &  3.63& 0.85   \\
\hline                                             
$\PPPPPZ$          & 35.60& 4.37 & 17.68& 3.48   \\
\hline
\hline
Source
       & \multicolumn{4}{|c}{Non-$\tt$ backgrounds}   \\ 
\cline{2-5}
\hline                                                                                                    
$\mm$              & 0.00& 0.00 & 0.00& 0.00  \\
$\ee$              & 0.00& 0.00 & 0.00& 0.00  \\
$\qq$              & 4.55& 2.63 & 0.00& 0.00  \\
$4f$           & 0.00& 0.00 & 0.00& 0.00  \\
\hline
\end{tabular}
\end{center}
\caption{\it For the sequential cuts analysis, the top part
contains
estimates of classification 
efficiencies, in percent,
for different exclusive five-prong 
decay modes, as obtained from simulation
after correction for the data/simulation discrepancies  discussed in the text.
The bottom part shows
backgrounds from non-$\tt$
sources. 
The quoted uncertainties 
are from the simulation~statistics~only.}
\label{tab:seqeffb}
\end{table}

\boldmath
\subsubsection{One-prong decays}
%\subsubsection{$\TH$}
\label{sec:hnuexclusive}
\unboldmath
In the selection of $\TH$ decays,
the separation  of a single  hadron from electrons and  muons requires
the  use  of most  of  the components  of  the  DELPHI detector.   The
detector     quantities     used     have    been     discussed     in
Section~\ref{sec:chargedparticles}.   The  main  background  
arises  from  $\THZ$
decays where the $\pi^0$  remains undetected, due to threshold effects
or dead regions in the calorimeter.

It was  required that the  charged particle have a  momentum exceeding
$0.05\times\pbeam$. The  mean energy per layer deposited  in the HCAL,
$\AVHC$,  was  used  to  classify  the charged-particle  tracks  into
candidate and non-candidate minimum-ionising particles (MIP).  For particles
consistent with a MIP,  
$\AVHC <  8$~GeV,  a  strong  muon veto  was  applied,
excluding all  particles which were  observed in the muon  chambers or
the  outer layer of  the HCAL.   For the  non-MIP region,  $\AVHC \geq
8$~GeV, with  less muon  contamination, a muon  veto was  applied by
excluding particles only if they  were observed in the outer layers of
the muon chambers.

For electron rejection it was required that the electromagnetic energy
deposited by the charged particle in the first four HPC layers did not
exceed  350~MeV, and  that the  $dE/dx$  did not  exceed the  expected
signal of  a pion by more  than two standard  deviations: $\pull^\pi <
2$.  (This $dE/dx$ requirement  was tightened for charged particles near
to  the  azimuthal  boundaries  between  HPC modules,  where  the  HPC
criterion gave poor rejection.)  It  was also required that the charged
particle was either observed in the HCAL or deposited at least 500~MeV
in the last five layers of the HPC.
 
Hadronic $\tau$-decays containing $\pi^0$'s were rejected by insisting
that  there be  no  candidate photon,  reconstructed  as described  in
Section~\ref{sec:photons}, in a cone  of half angle $18^{\circ}$ about
the charged particle.

\begin{figure}[tbp]
  \begin{center}
    \leavevmode
\epsfig{figure=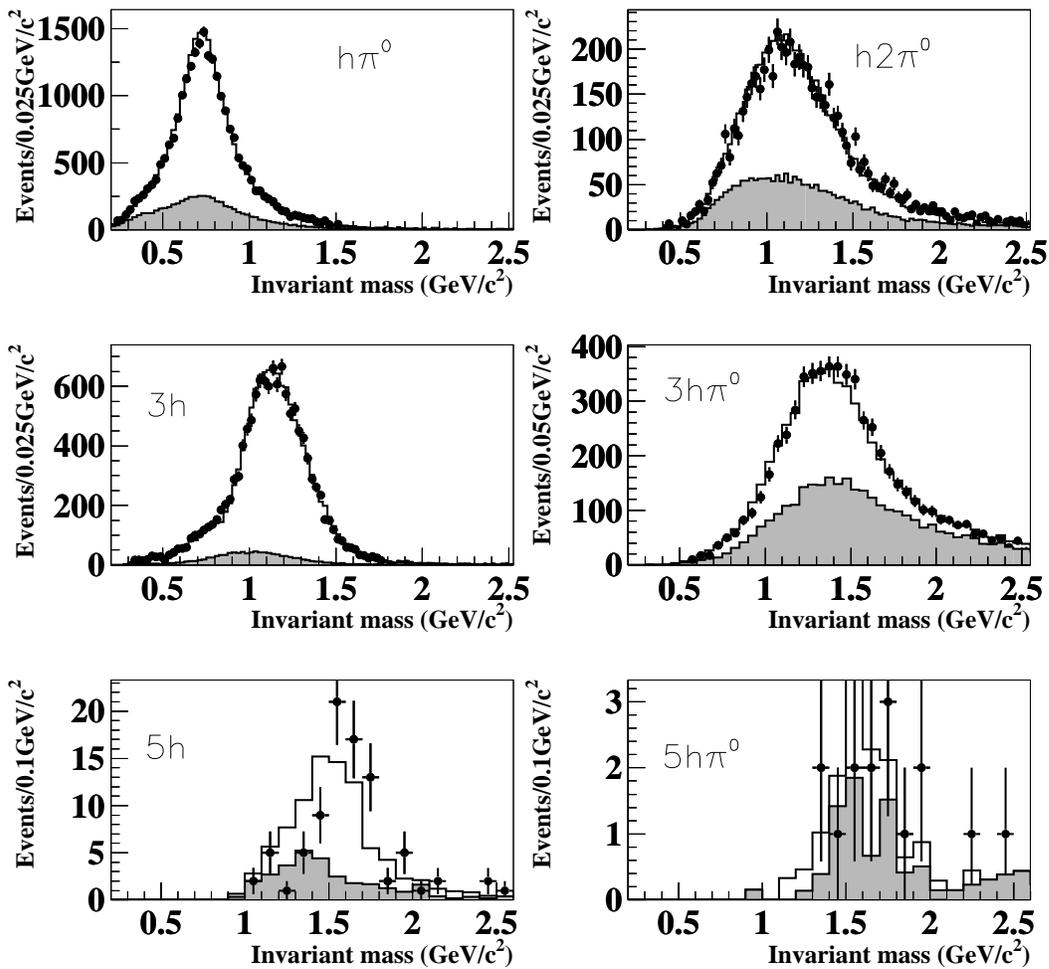,width=0.96\linewidth}
\caption{\it Invariant-mass distributions for the decays  
         selected with sequential cuts, excluding the cuts directly related to this variable. 
	 Data are shown as dots, simulation by a solid line. The
         shaded area shows the background prediction from simulation.}
    \label{fig:invmasscuts}
  \end{center}
\end{figure}

The  $\tau$-decay  to $\HZ$  was  selected by  requesting an  isolated
charged particle with an  accompanying $\pi^0$ candidate.  The charged
particle had to have a reconstructed momentum greater than 0.5~GeV/$c$
and to be incompatible  with the electron hypothesis using  the loose cut of
$\pull^\pi<3.5$. 
%discussed in Section~\ref{sec:chargedparticles}.  
Candidate
$\pi^0$'s were subdivided into three different classes, described below:
\begin{enumerate}
\item
two photons, where each photon was measured as a separate
electromagnetic cluster in the HPC  or was a reconstructed conversion.
The photons had to be separated by less than 10$^\circ$
and the reconstructed $\pi^0$ candidate had to have a reconstructed 
mass in the range 0.04~GeV/$c^2$ to 0.3~GeV/$c^2$;
\item
one shower with energy greater than 6 GeV and passing
the  criteria described in Section~\ref{sec:photons}.
This may happen either when a very energetic $\pi^0$ is not
recognised as such by the $\pi^0$ reconstruction algorithms or when one
of the photons enters a dead region of the calorimeter or is of too low energy
to be observed in the calorimeter. The energy of the shower was taken as the
energy of the $\pi^0$;
\item
An identified $\pi^0$ as described in Section~\ref{sec:neutralpions}.
\end{enumerate}
The $h^-\pi^0$ invariant-mass distribution, calculated 
assuming the pion mass for the
charged particle, is shown in Fig.~\ref{fig:invmasscuts}. To reduce background
it was required that the reconstructed $h^-\pi^0$ invariant mass
lie in the range $0.48~ $GeV$/c^2$ to $1.20~ $GeV$/c^2$ and that the angle
between the charged-particle direction
and the $\pi^0$ direction be less than 20$^\circ$.

The $\tau$-decay to  $\HGZZ$ was selected  by requiring  an isolated
charged  particle with  two or  more accompanying  $\pi^0$ candidates.
The charged particle had to have a reconstructed momentum greater than
0.5~GeV/$c$.
The   candidate   $\pi^0$'s  were   reconstructed   as  described   in
Section~\ref{sec:neutralpions}.    Furthermore,  decays  with
only one reconstructed $\pi^0$ candidate were accepted if there was at
least  one  well-reconstructed   photon  candidate  (as  described  in
Section~\ref{sec:photons})  which was not  used in  the reconstruction
of~a~$\pi^0$.

This semi-exclusive mode had little background from non-$\tau$ sources
or  from  $\tau$-decay  modes  containing  electrons  and muons.   The
background was  dominated by the $\HZ$ decay  mode.  Further rejection
of the background  was performed by requiring that  the invariant mass
of the  $h^-\pi^0\pi^0$ system be greater than  0.8~GeV/$c^2$ and that
the total reconstructed  energy be greater than 10~GeV.  The pion mass
was  assumed for the  charged particle  and the  $\pi^0$ mass  for the
$\pi^0$ candidate(s).

\subsubsection{Three-prong decays}
%\subsubsection{\boldmath $\tau^-\to 2h^-h^+\nu_\tau$ \unboldmath}
\label{sec:schhh}
The signature of the decay  $\THHH$ is of three charged particles with
no accompanying electromagnetic showers. 
A candidate $\HHH$ decay had
to  have three  charged-particle  tracks in  a hemisphere. 
The vector sum of
the three charged-particle momenta 
had to have  a magnitude  greater  than $0.166\times\sqrt{s}$.
It   was  required  that  there  be  no
reconstructed photon of energy  greater than 1.5~GeV within $10^\circ$
of  the three-charged-particle  system momentum  vector and  that the
total  neutral  electromagnetic energy  in  a  cone  of half-angle  
10$^\circ$
around the  three-charged-particle system  be less  than 0.3
times the  momentum of the  three-charged-particle system. To reject
cases  where a photon  or $\pi^0$  was associated  to a  charged-particle track
extrapolation  in  the HPC  it  was  required  that the  total  energy
associated to the three tracks in  the first five layers of the HPC be
less than 0.3 times the momentum of the three-charged-particle system.

The $\tau$-decay  to $2h^-h^+ \geq 1 \pi^0 \nu_\tau$ 
was selected by  requesting three charged-particle tracks together with a $\pi^0$ candidate.  
The $\pi^0$  candidate had to lie in    the    barrel    region,
$|\cos\theta|<0.732$,
within  a  cone  of  half-angle
$30^\circ$  about the  highest-momentum  charged
particle. 

\boldmath
\subsubsection{Five-prong decays}
%\subsubsection{$\tppppp$ and $\tpppppno$}
\unboldmath
\label{sec:sc5h}
The exclusive decays $\tppppp$ 
and $\tpppppno$ were selected from the inclusive five-prong sample.

Decays with a total momentum greater than 40~GeV/$c$, 
an invariant mass of 
the five-charged-particle system greater than 1.5~GeV/$c^2$ or in which 
all photons had an energy less than 1.5~GeV  were considered as 
$\tppppp$ decays. 
Otherwise the decay was classified as $\tpppppno$.

\subsubsection{Results of the sequential-cuts analysis}
\label{sec:seq_res}

The branching ratios were extracted from the data with a maximum-likelihood fit as described in Section~\ref{sec:method}.

The numbers of candidate $\tau$-decays in each class are given in 
Table~\ref{tab:seq_res}, together with 
the branching ratio obtained. The uncertainties
quoted are statistical and take into account
correlations between different channels.
\begin{table}[tbp]
\begin{center}
\begin{tabular}{l|r|r@{$\pm$}l}
\hline
decay mode & \multicolumn{1}{|c}{Number}  & \multicolumn{2}{|c}{branching ratio (\%)}   \\ 
\hline
$h^- \nut$                      &  9727  & 12.765 & 0.129 \\
$h^-\pio\nut$                   & 21098  & 26.243 & 0.227 \\
$h^- \!\!\geq\!\! 2\pio\nut$    &  6187  & 10.928 & 0.193 \\
$3h^\pm \nut$                   & 12761  &  9.352 & 0.097 \\
$3h^\pm\!\! \geq\!\! 1\pio\nut$ &  5363  &  5.162 & 0.091\\
$5h^\pm \nut$                   &    96  & 0.097  & 0.015 \\
$5h^\pm\!\! \geq\!\! 1\pio\nut$ &    13  & 0.016  & 0.012 \\
\hline
\end{tabular}
\end{center}
\caption{\it For the sequential-cuts analysis, the numbers of selected events
in each class and branching ratios obtained. The quoted uncertainties
are statistical only.}
\label{tab:seq_res}
\end{table}

The invariant-mass distributions of the different classes of selected 
decays are shown in~Fig.~\ref{fig:invmasscuts} for all cuts applied 
except those directly related to the mass. 
%The agreement between the data
%and the simulation is good for the $\HZ$ and $\HZZ$ classes, while
%the data in $\HGZZZ$ is peaked at a slightly higher mass.

\subsection{Neural-Net Analysis}
\label{sec:nn}
The decay modes were also selected with the help of neural
networks. Two different neural networks were designed, one for
one-prong decays and another for three-prongs. They are described in this
section. The events were initially separated according to their track
multiplicity and then the selection with the corresponding neural net
was applied. For five-prongs the sequential-cut analysis described in Sec.\ref{sec:sc5h} was applied. 

\subsubsection{One-prong decays}
\label{sec:nn_oneprong}

For the one-prong decay modes,
a total of 43 input variables that could help the identification were
studied: general variables  (such as neutral multiplicities, invariant masses,
and number of  identified $\pi^0$), charged-particle  variables 
(such as momentum,
$dE/dx$, and calorimetric energies)  or  neutral-particle  quantities
(such as energy, and shower-profile variables). This number  was reduced first
using  a  principal-component  analysis,  removing linearly-redundant
variables   after    testing   that   they   did    not   affect   the
performance. Then, the network was trained and tested with and without
variables which appeared to be  less significant; they were removed if
the  results were  not degraded.  Finally, 15  variables were  used as
input.  These variables were:
 \begin{itemize}
 \item
the total invariant mass including charged and neutral particles;
 \item 
the number of reconstructed photons;
 \item 
the number of reconstructed $\pi^0$;
 \item 
the number of reconstructed photons not linked to any $\pi^0$;
 \item 
the magnitude of the momentum of the charged particle;
 \item
the polar angle of the charged particle;
 \item
the azimuthal angle, modulo 15$^\circ$, of the extrapolation of the 
charged-particle
track to the HPC;
 \item
the pion hypothesis $dE/dx$ pull variable, $\pull^\pi$;
 \item
the number of muon chamber layers with hits associated to the charged particle;
 \item
the number of muon chamber outer layers with hits associated to 
the charged particle;
 \item
the total electromagnetic energy deposited in a cone of half-angle 30$^\circ$ 
around the charged-particle track;
 \item
the energy in the HPC associated to the charged particle;
 \item 
the energy in the inner four layers of the HPC associated to 
the charged particle;
 \item 
the total hadron calorimetric energy associated to the charged particle;
 \item 
the number of layers in the HCAL associated to the charged particle.
 \end{itemize}

A feed-forward  neural network~\cite{odorico}  with one
input  layer, one hidden  layer and  one output  layer was  used.  The
input  layer had  15 neurons,  each one  corresponding to  one  of the
variables listed  above.  All the  input variables were  normalised to
the  range $[-1,1]$.  Several  structures were  tested. Finally  a net
with one hidden  layer of 40 neurons was used as  the optimum in terms
of  efficiency and  simplicity.   The output  layer  consisted of  six
neurons. The  assigned target  value of these  neurons was +1  for the
corresponding class  and $-1$ for the rest.   Each neuron corresponded
to   one  of  the   following  decay   modes:  $\EL$;   $\MU$;  $\HH$;
$\HZ$;~$\HZZ$;~$\HGZZZ$.  
%Decay  modes containing $\eta$  and $\omega$
%particles were excluded from these categories.
  
A training procedure was performed  on about 3000 simulated events for
each  of the  classes, optimising  the network  parameters to  give an
answer in the output layer as  close as possible to $+1$ in the neuron
corresponding  to the  generated class  and $-1$  in all  others. 
%More
%details of the procedure are given in~\cite{nimarticle}.

The total sample of simulated events, excluding those used for
the  training, was  used to  evaluate the  probabilities that  a given
decay be identified  in a given class.  The selection efficiencies
of the different classes  and the misidentification probabilities are
shown in Table~\ref{tab:nn_eff1}.
\begin{table}[p]
\begin{center}
\begin{tabular}{l|r@{$\pm$}l|r@{$\pm$}l|r@{$\pm$}l|r@{$\pm$}l|r@{$\pm$}l|r@{$\pm$}l}
\hline
 true $\tau$ 
       & \multicolumn{12}{|c}{Neural network decay classification}   \\ 
\cline{2-13}
 decay mode
       &\multicolumn{2}{|c}{$\EL$}  
       &\multicolumn{2}{|c}{$\MU$} 
       &\multicolumn{2}{|c}{$h^- \nut$}  
       &\multicolumn{2}{|c}{$h^-\pio\nut$} 
       &\multicolumn{2}{|c}{$h^- 2\pio\nut$} 
       &\multicolumn{2}{|c}{$h^- \!\!\geq\!\! 3\pio\nut$} \\
\hline                          
$\EL$              & 89.86& 0.06 &  0.02& 0.00 &  1.32& 0.02 &  0.51& 0.01 &  0.16& 0.01 &  0.01& 0.00  \\
\hline                                                                                                    
$\MU$              &  0.10& 0.01 & 88.02& 0.07 &  2.50& 0.03 &  0.41& 0.01 &  0.01& 0.00 &  0.00& 0.00  \\
\hline                                                                                                    
$\pi^- \nut$       &  2.07& 0.04 &  1.80& 0.04 & 78.59& 0.11 &  5.15& 0.06 &  0.22& 0.01 &  0.02& 0.00  \\
$K^- \nut $        &  0.46& 0.07 &  3.33& 0.19 & 82.95& 0.40 &  5.84& 0.25 &  0.27& 0.06 &  0.04& 0.02  \\
$\pi^-\kol\nut$    &  1.32& 0.16 &  1.80& 0.18 & 68.45& 0.67 & 14.60& 0.59 &  1.46& 0.16 &  0.14& 0.05   \\
$K^-\kol\nut$      &  0.57& 0.23 &  1.85& 0.41 & 74.51& 1.46 & 11.83& 1.26 &  1.44& 0.36 &  0.00& 0.09  \\
$\pi^-\kos\nut$    &  5.49& 0.31 &  1.43& 0.16 & 38.08& 0.62 & 20.92& 0.64 &  6.57& 0.34 &  0.40& 0.09   \\
$K^-\kos\nut$      &  4.68& 0.65 &  3.77& 0.58 & 36.16& 1.35 & 20.84& 1.42 &  6.41& 0.75 &  0.59& 0.23   \\
$\pi^-\kol\ko\nut$ &  3.29& 0.53 &  1.10& 0.31 & 38.84& 1.34 & 25.56& 1.41 &  6.40& 0.72 &  1.86& 0.40  \\
$\pi^-2\kos\nut$   &  6.36& 1.26 &  0.72& 0.44 & 17.37& 1.35 & 22.23& 2.42 & 10.35& 1.57 &  2.28& 0.77  \\
\hline                                                                                                    
$\pi^-\pio\nut$    &  1.18& 0.02 &  0.43& 0.01 &  7.40& 0.05 & 68.51& 0.08 &  7.04& 0.05 &  0.20& 0.01  \\
$K^-\pio\nut$      &  0.94& 0.12 &  1.09& 0.13 & 11.18& 0.38 & 66.57& 0.57 &  5.63& 0.28 &  0.25& 0.06  \\
$\pi^-\pio\kol\nut$&  0.61& 0.16 &  0.21& 0.09 &  5.19& 0.45 & 64.99& 0.96 & 13.57& 0.69 &  1.19& 0.22   \\
$K^-\pio\kol\nut$  &  0.55& 0.25 &  0.45& 0.22 & 13.48& 1.13 & 57.61& 1.64 &  9.49& 0.97 &  0.73& 0.28   \\
$\pi^-\pio\kos\nut$&  2.12& 0.29 &  0.72& 0.17 &  4.38& 0.42 & 40.91& 1.00 & 21.23& 0.84 &  3.41& 0.37  \\
$K^-\pio\kos\nut$  &  3.56& 0.67 &  2.57& 0.57 &  7.06& 0.92 & 41.17& 1.77 & 13.52& 1.23 &  3.32& 0.64   \\
\hline                                                                                                    
$\pi^-2\pio\nut$   &  0.84& 0.02 &  0.15& 0.01 &  1.39& 0.03 & 33.92& 0.13 & 38.33& 0.13 &  4.22& 0.05   \\
$K^-2\pio\nut$     &  0.84& 0.29 &  0.42& 0.20 &  1.35& 0.36 & 35.18& 1.49 & 35.45& 1.49 &  3.92& 0.60  \\
\hline                                                                                                    
$\pi^-3\pio\nut$   &  0.62& 0.06 &  0.07& 0.02 &  0.69& 0.07 & 18.76& 0.32 & 42.33& 0.41 & 15.97& 0.30   \\
\hline                                                                                                    
$\PPP$             &  0.08& 0.01 &  0.03& 0.00 &  0.29& 0.01 &  2.03& 0.03 &  0.26& 0.01 &  0.02& 0.00  \\
$K^-\pi^-\pi^+\nut$&  0.13& 0.05 &  0.03& 0.03 &  0.33& 0.09 &  1.79& 0.20 &  0.18& 0.07 &  0.02& 0.02   \\
$K^-K^+\pi^+\nut$  &  0.17& 0.08 &  0.05& 0.04 &  0.31& 0.10 &  1.64& 0.23 &  0.17& 0.08 &  0.00& 0.03   \\
\hline                                                                                                    
$\PPPZ$            &  0.09& 0.01 &  0.01& 0.00 &  0.10& 0.01 &  1.70& 0.05 &  1.60& 0.05 &  0.20& 0.02  \\
\hline                                                                                                    
$\PPPZZ$           &  0.06& 0.03 &  0.02& 0.02 &  0.03& 0.02 &  1.08& 0.12 &  2.13& 0.17 &  1.09& 0.12  \\
\hline                                                                                                    
$\PPPZZZ$          &  0.00& 0.08 &  0.00& 0.08 &  0.08& 0.08 &  0.23& 0.14 &  2.30& 0.43 &  2.75& 0.47  \\
\hline                                                                                                    
$\PPPPP$           &  0.00& 0.21 &  0.00& 0.21 &  0.00& 0.21 &  0.32& 0.26 &  0.12& 0.21 &  0.00& 0.21  \\
\hline                                                                                                    
$\PPPPPZ$          &  0.00& 0.83 &  0.00& 0.83 &  0.00& 0.83 &  1.00& 0.91 &  0.00& 0.83 &  0.00& 0.83  \\
\hline                                                                                                    
\hline
Source
       & \multicolumn{12}{|c}{Non-$\tt$ backgrounds}   \\ 
\cline{2-13}
\hline                                                                                                    
$\mm$              & 0.03& 0.01 & 0.39& 0.02 & 0.02& 0.00 & 0.06& 0.01 & 0.05& 0.01 & 0.00& 0.00   \\
$\ee$              & 1.27& 0.19 & 0.01& 0.01 & 0.16& 0.03 & 0.06& 0.01 & 0.05& 0.02 & 0.08& 0.08  \\
$\qq$              & 0.02& 0.01 & 0.07& 0.01 & 0.09& 0.02 & 0.15& 0.02 & 0.11& 0.03 & 0.26& 0.13   \\
4f                 & 1.91& 0.19 & 0.84& 0.08 & 0.37& 0.05 & 0.44& 0.04 & 0.25& 0.05 & 0.18& 0.13   \\
\hline
\end{tabular}
\end{center}
\caption{\it For the neural-networks analysis, the top part contains
estimates of classification 
efficiencies, in percent,
for different exclusive one-prong 
decay modes, as obtained from simulation
after correction for the data/simulation discrepancies discussed in the text.
The bottom part shows backgrounds from non-$\tt$
sources.
The quoted uncertainties
are from the simulation~statistics~only. When no events are classified in a given class the Poissonian
upper bound is taken as error. Numbers smaller than 0.005\% are represented in the table as 0.00.
}
\label{tab:nn_eff1}
\end{table}

Each of  the preselected one-prong  decays was processed  through the
neural network and the decay  was identified as belonging to the class
whose corresponding output  was largest.  
%In
%0.5\% of  decays more than  one output neuron  had a value  of greater
%than zero. This  was well reproduced by simulation  where the fraction
%was 0.4\%.  In  such cases an event was put  into both classes. 
Events
with no  output neuron  above zero were  not classified. The number of events
with two or more output neurons above zero was negligible.  
%These events
%corresponded  to $7.5\pm0.2\%$ {\bf  to be  corrected} of  all decays,
%consistent    with    the    fraction   expected    from    simulation
%of~$7.0\pm0.1\%$.

\begin{figure}[tbp]
  \begin{center}
    \leavevmode    
\epsfig{figure=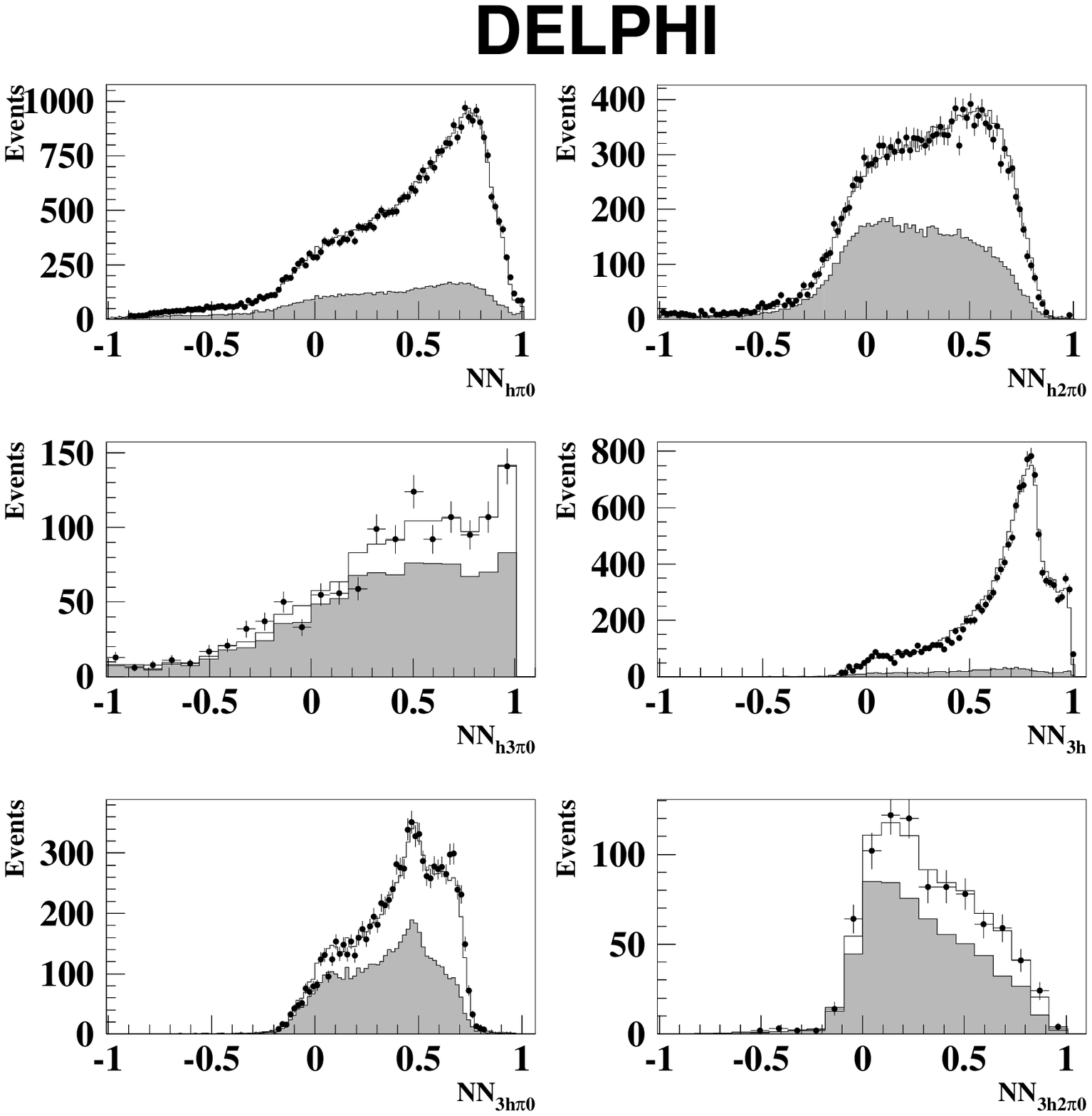,width=1.1\linewidth}        
    \caption{\it Maximum-output neuron value in one-prong and three-prong neural-net analyses.
             For each event the output of the class whose output neuron is maximum is represented. 
              Data are shown as dots and simulation by a solid line. The shaded area represents the background events.}
    \label{fig:neuronop2}
  \end{center}
\end{figure}

The distributions of the maximum value of the output  
neuron for each decay mode for all
decays are shown in Figs.~\ref{fig:neuronop1} and ~\ref{fig:neuronop2}. In most cases, 
this shows satisfactory agreement between data  and simulation. 

\subsubsection{Three-prong decays}
\label{sec:nn_threeprong}
The three-prong $\tau$-decay
candidates selected were divided into three classes:
$2h^-h^+ \nu_\tau$,
$2h^-h^+ \pi^0 \nu_\tau$ and
$2h^-h^+ \geq 2 \pi^0 \nu_\tau$.

A simpler  network   was   used   in  this   case,   with  all   the
electron/muon/hadron-identification variables  removed and  the
remaining variables kept, giving a total
of seven variables:
\begin{itemize}
\item 
the total momentum of the charged-particle system;
\item
the total electromagnetic energy associated to the charged-particle tracks;
\item
the total electromagnetic energy deposited in a cone of half-angle 15$^\circ$
around the momentum vector of the charged-particle system including that associated
to the charged-particle tracks;
\item
the number of reconstructed photons;
\item
the number of reconstructed $\pi^0$;
\item
the number of reconstructed photons not used in a reconstructed $\pi^0$;
\item 
the total invariant mass.
\end{itemize}

The photons and $\pi^0$ had to lie in a cone of half-angle 30$^\circ$ about
the highest-momentum charged particle.
The hidden layer had 15 neurons  and three output neurons were used. 
The
network  was trained  with  3000 events  of each  of  the signal
classes optimising the network as  for the one-prongs, to give outputs
close to $+1$  in the neuron corresponding to  the generated class and
$-1$ in the others. Here,  to reduce the background from other decays,
the network was also trained with 3000 one-prong events that fulfilled
the three-prong selection  requirements, to give answers as  close to $-1$
in  all the output  neurons.  

The  event classification  from  the output
neuron  values was  performed in  an equivalent  way to  the one-prong
case.
% with the  threshold value set to 0.2  to optimise the efficiency
%and backgrounds for the two  classes. 
The efficiencies and background
levels for the different decay classes are given in 
Table~\ref{tab:nn_eff3}.
\begin{table}[p]
\begin{center}
\begin{tabular}{l|r@{$\pm$}l|r@{$\pm$}l|r@{$\pm$}l|r@{$\pm$}l}
\hline
 true $\tau$ 
       & \multicolumn{6}{|c}{Neural network decay classification}   \\ 
\cline{2-9}
 decay mode
       &\multicolumn{2}{|c}{$3h^\pm \nut$}  
       &\multicolumn{2}{|c}{$3h^\pm \pio\nut$} 
       &\multicolumn{2}{|c}{$3h^\pm \!\!\geq\!\! 2\pio\nut$}
       &\multicolumn{2}{|c}{Unclassified} \\
\hline                          
$\EL$              & 0.01& 0.00 &  0.01& 0.00 &  0.00& 0.00  &  8.11& 0.06 \\
\hline                                                                                                    
$\MU$              &  0.03& 0.00 &  0.00& 0.00 &  0.00& 0.00 &  8.92& 0.06\\
\hline                                                                                                    
$\pi^- \nut$       & 0.08& 0.01 &  0.03& 0.00 &  0.00& 0.00  & 12.04& 0.09\\
$K^- \nut $        & 0.07& 0.03 &  0.01& 0.01 &  0.00& 0.01  &  7.03& 0.27\\
$\pi^-\kol\nut$    & 0.12& 0.05 &  0.06& 0.03 &  0.00& 0.02  & 12.05& 0.44 \\
$K^-\kol\nut$      & 0.05& 0.09 &  0.13& 0.11 &  0.04& 0.09  &  9.56& 0.90\\
$\pi^-\kos\nut$    & 4.03& 0.27 &  1.07& 0.14 &  0.00& 0.02  & 21.99& 0.57\\
$K^-\kos\nut$      & 3.90& 0.59 &  1.01& 0.31 &  0.14& 0.12  & 22.50& 1.28\\
$\pi^-\kol\ko\nut$ & 2.24& 0.44 &  1.30& 0.33 &  0.49& 0.21  & 18.92& 1.16\\
$\pi^-2\kos\nut$   & 6.41& 1.27 &  6.47& 1.27 &  0.77& 0.45  & 27.05& 2.30\\
\hline                                                                        
$\pi^-\pio\nut$    & 0.41& 0.01 &  0.73& 0.02 &  0.07& 0.00   & 14.01& 0.06\\
$K^-\pio\nut$      & 0.44& 0.08 &  0.92& 0.12 &  0.12& 0.04   & 12.84& 0.41\\
$\pi^-\pio\kol\nut$& 0.22& 0.10 &  1.15& 0.21 &  0.15& 0.08   & 12.71& 0.67\\
$K^-\pio\kol\nut$  & 0.79& 0.29 &  1.46& 0.40 &  0.11& 0.11   & 15.34& 1.20\\
$\pi^-\pio\kos\nut$& 0.39& 0.13 &  5.45& 0.46 &  0.41& 0.13   & 20.98& 0.83\\
$K^-\pio\kos\nut$  & 1.29& 0.41 &  5.00& 0.78 &  0.15& 0.14   & 22.37& 1.50\\
\hline                                                                        
$\pi^-2\pio\nut$   & 0.27& 0.01 &  2.20& 0.04 &  0.75& 0.02   & 17.93& 0.10\\
$K^-2\pio\nut$     & 0.22& 0.15 &  2.38& 0.48 &  0.62& 0.24   & 19.62& 1.24 \\
\hline                                                                        
$\pi^-3\pio\nut$   & 0.10& 0.03 &  2.05& 0.12 &  1.73& 0.11   & 17.69& 0.31\\
\hline                                                                        
$\PPP$             & 78.11& 0.09 & 14.10& 0.08 &  0.24& 0.01  &  4.84& 0.05 \\
$K^-\pi^-\pi^+\nut$& 77.79& 0.64 & 14.21& 0.54 &  0.18& 0.07  &  5.34& 0.35\\
$K^-K^+\pi^+\nut$  & 74.53& 0.80 & 15.72& 0.67 &  0.26& 0.09  &  7.17& 0.48  \\
\hline                                                                        
$\PPPZ$            & 16.51& 0.16 & 69.06& 0.19 &  3.62& 0.08  &  6.99& 0.11  \\
\hline                                                                        
$\PPPZZ$           &  4.31& 0.24 & 59.12& 0.57 & 24.80& 0.50  &  7.22& 0.30 \\
\hline                                                                        
$\PPPZZZ$          &  1.63& 0.36 & 40.66& 1.41 & 46.68& 1.43  &  5.51& 0.65 \\
\hline                                                                        
$\PPPPP$           & 18.47& 1.76 & 19.64& 1.80 &  0.51& 0.32  &  2.05& 0.64  \\
\hline                                                                        
$\PPPPPZ$          & 4.68& 1.93 & 30.12& 4.19 &  3.21& 1.61   &  7.72& 2.44 \\
\hline                                                                                                    
\hline
Source
       & \multicolumn{6}{|c}{Non-$\tt$ backgrounds}   \\ 
\cline{2-7}
\hline                                                                                                    
$\mm$              & 0.00& 0.00 & 0.00& 0.00 & 0.00& 0.00 & 0.24& 0.02  \\
$\ee$              & 0.00& 0.00 & 0.00& 0.00 & 0.00& 0.00 & 0.60& 0.09  \\
$\qq$              & 0.75& 0.05 & 1.89& 0.12 & 5.11& 0.67 & 1.07& 0.11 \\
4f                 & 0.26& 0.04 & 0.22& 0.05 & 0.82& 0.33 & 0.53& 0.04\\
\hline
\end{tabular}
\end{center}
\caption{\it For the neural-networks analysis,  classification 
efficiencies, in percent,
for different exclusive three-prong 
decay modes, as obtained from simulation
after correction for the data/simulation discrepancies discussed in the text.
The last column represents the percentage of events not classified in any
of the classes by the neural network, including the sequential-cuts
selection of five-prong modes.
The bottom part shows backgrounds in percent in each class from non-$\tt$
sources. The quoted uncertainties
are from the simulation~statistics~only. When no events are classified in a given class the Poissonian
upper bound is taken as error. Numbers smaller than 0.005\% are represented in the table as 0.00.
}
\label{tab:nn_eff3}
\end{table}

The distributions  of  the largest  value  of the  output
neurons  in each  decay    are   shown
in~Fig.~\ref{fig:neuronop2},  showing in most cases good agreement  in shape between data
and simulation.

\subsubsection{Results of the neural-network analysis}
\label{sec:nn_res}
As explained in Section~\ref{sec:method}, a simultaneous fit for the branching ratios
was performed by fitting the predicted
number of candidate $\tau$-decays in each class to the observed number.
In this case, the information of the neural-net output was also used 
in the fit, where the sum
over classes was extended to run over classes and bins in the neural-net output.
Only positive values of this output were taken into account for the quoted
results. The  minimum value used in the fit  
was varied  through  the full  range from  $-1$  to $0$
without  any variation  on the branching  ratio obtained,  beyond that
expected from statistical fluctuations.
For the five-prong case the sequential-cuts analysis was used.
The numbers of selected candidate $\tau$-decays in each class are given in 
Table~\ref{tab:nn_res}, together with the branching ratio obtained. 
The uncertainties
quoted are statistical and take into account
correlations between different channels.
Despite the fact that the fit was not minimizing a $\chi^2$, it was a maximum-likelihood fit, a $\chi^2$ is calculated to estimate of the
goodness of the fit.
Accounting only for statistical errors, a 
$\chi^2=808$ for 490 $d.o.f.$ was obtained, with the contribution from each channel presented
in Table~\ref{tab:nn_res}. The effect of systematic errors on the $\chi^2$ will be
discussed in Section~\ref{sec:syssum}.
\begin{table}[tbp]
\begin{center}
\begin{tabular}{l|r|r@{$\pm$}l|r@{ (}c@{)}}
\hline
decay mode & \multicolumn{1}{|c}{Number}  & \multicolumn{2}{|c}{branching ratio} &\multicolumn{2}{|c}{$\chi^2$ (bins)}  \\ 
\hline
$e^- \nut\nue$                  & 25529  & 17.803 & 0.108 & 54.9  & 55\\
$\mu^- \nut\num$                & 25860  & 17.350 & 0.104 & 160.1 & 55\\
$h^- \nut$                      & 19212  & 12.780 & 0.120 & 68.6  & 55\\
$h^-\pio\nut$                   & 34675  & 26.291 & 0.201 & 85.1  & 55\\
$h^- 2\pio\nut$                 &  9504  &  9.524 & 0.320 & 59.0  & 55\\
$h^- \!\!\geq\!\! 3\pio\nut$    &  1083  &  1.403 & 0.214 & 92.1  & 55\\
\hline
$3h^\pm \nut$                   & 12176  &  9.340 & 0.090 & 152.5 & 55\\
$3h^\pm \pio\nut$               &  8909  &  4.545 & 0.106 & 77.8  & 55\\
$3h^\pm\!\! \geq\!\! 2\pio\nut$ &  1272  &  0.561 & 0.068 & 51.1  & 55\\
\hline
$5h^\pm \nut$                   &    96  &  0.097 & 0.015 & 0.0   & 1\\
$5h^\pm\!\! \geq\!\! 1\pio\nut$ &    13  &  0.016 & 0.012 & 1.7   & 1\\
\hline
\end{tabular}
\end{center}
\begin{center}
\begin{tabular}{l|r|r@{}l|r@{ (}c@{)}}
\hline
unclassified & \multicolumn{1}{|c}{Number}  & \multicolumn{2}{|c}{expected} &\multicolumn{2}{|c}{$\chi^2$ (bins)}  \\ 
\hline
 1-prong &   18558 & 18857.7 & & 2.2& 1\\
 3-prong &    1517 & 1455.1  & & 1.6& 1\\
 5-prong &       3 &    5.2  & & 1.6& 1\\
\hline
\end{tabular}
\end{center}
\caption{\it For the neural-networks analysis, numbers of selected events
in each class and branching ratios obtained. The quoted uncertainties
are statistical only. The last column shows the contribution 
of each to the total $\chi^2$, computed with
statistical errors only. In parenthesis is shown the 
number of data points used in each case. The last
three lines compare the measured number of 
unclassified events with the expectation after the fit.}
\label{tab:nn_res}
\end{table}

\begin{figure}[tbp]
  \begin{center}
    \leavevmode
\epsfig{figure=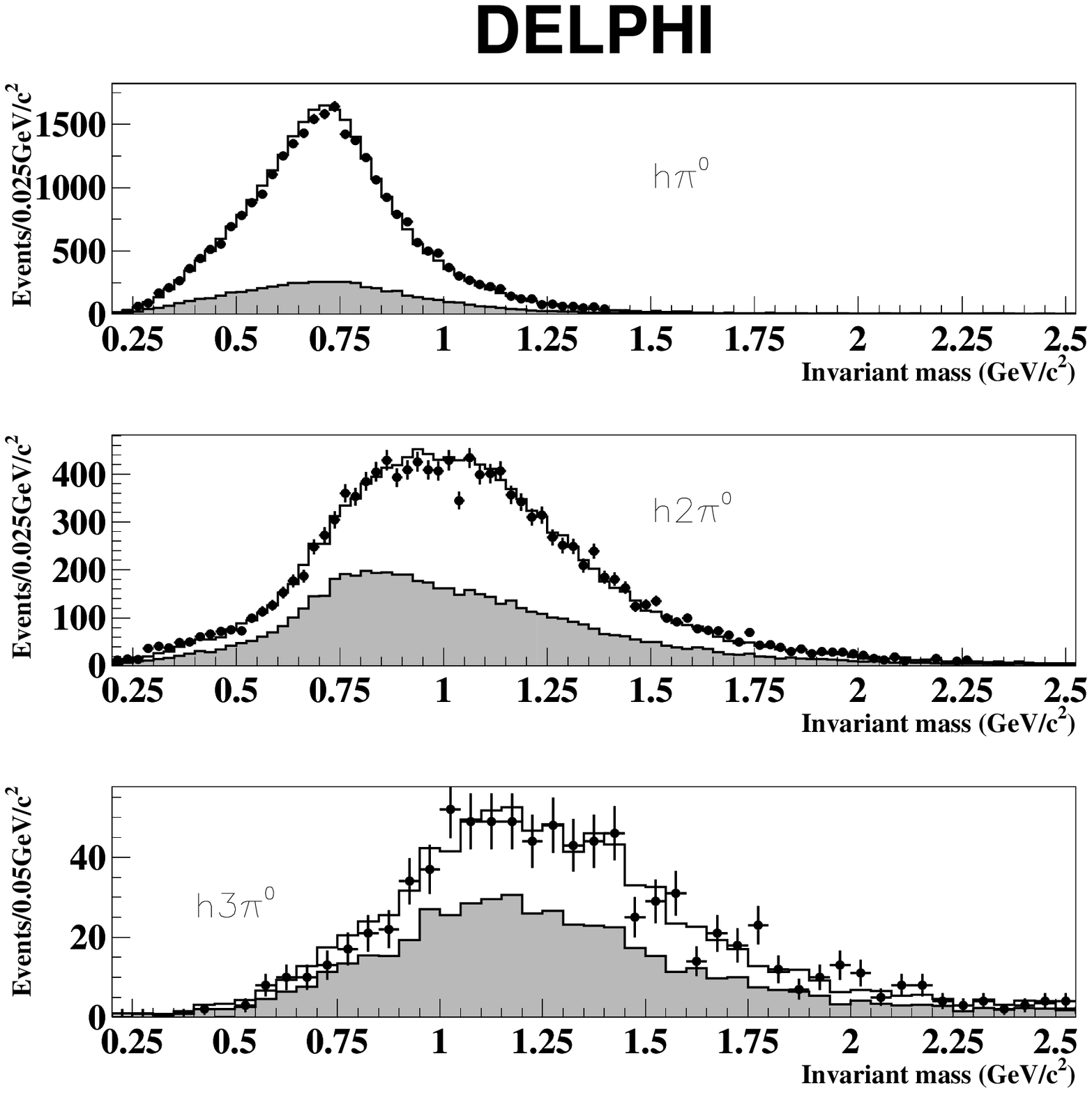,width=0.96\linewidth}
\caption{\it Invariant-mass distributions for the one-prong decays  
         selected with the neural network. Data are shown as dots, simulation by a solid line. The
         shaded area shows the background prediction from simulation.}
    \label{fig:nninvmass}
  \end{center}
\end{figure}

\begin{figure}[tbp]
  \begin{center}
    \leavevmode
\epsfig{figure=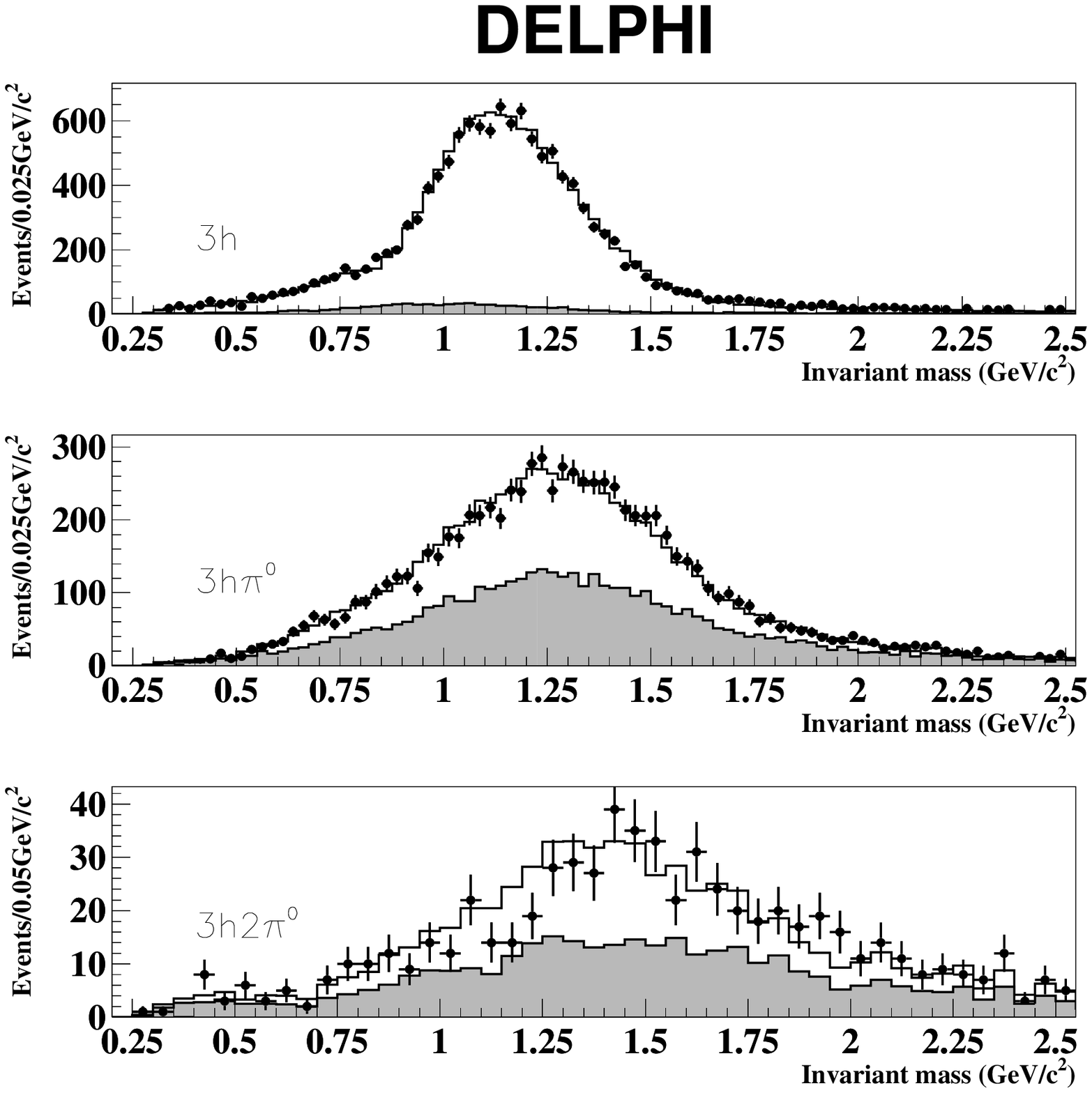,width=0.96\linewidth}
\caption{\it Invariant-mass distributions for the three-prong decays  
         selected with the neural network. Data are shown as dots, simulation by a solid line. The
         shaded area shows the background prediction from simulation.}
    \label{fig:nninvmass2}
  \end{center}
\end{figure}

The invariant-mass distributions of the different classes of selected 
decays are shown in~Figs.~\ref{fig:nninvmass} and ~\ref{fig:nninvmass2}.

\section{Systematics}
\label{sec:systematics}
The systematic uncertainties due to any specific source
were estimated simultaneously for all measured decay modes in
the neural-network and sequential-cuts analyses.
This was also the case for inclusive branching ratios, where the calculated
systematic errors accounted for the existing cancellations between the
errors of the different channels involved, leading in many cases to 
smaller errors.
%They were also estimated separately for each of the analyses
%to ensure that there was no major difference in the sensitivity of the
%analyses to any particular effect.

The systematic errors  were evaluated using test samples  of events as
discussed  in  Sections~\ref{sec:chargedparticles}~and~\ref{sec:photons}.   
Where
appropriate the  relevant
input variables were varied by the observed uncertainty and the selection and 
fit were
repeated.  The variation  in the  results  was taken  as an
estimate of the systematic
effect on the branching ratios. 
The  effects of the  external background and  the preselection
efficiency  were also  checked.  
The potential  sources of  systematic
uncertainties are discussed below and 
summarised in  Table~\ref{tab:sys_oneprong}. 

\boldmath
\subsection{$\tt$ selection and non-$\tt$ backgrounds}
\unboldmath
\label{sys_ttsel}
 The background
levels from channels other than  $\tt$ were varied by the uncertainties
given   in  Section~\ref{sec:tautauselection} and the fit was repeated.
The observed changes on the results for the variation in 
each of the background types were added in quadrature to obtain
the estimate of the systematic error. 

The probability of identifying a hemisphere from a background event in a given
class was checked with the electron and muon test samples described in 
Section~\ref{sec:particleid}. The $\qq$ background was checked with 
low-multiplicity $\qq$ test samples selected 
applying the $\tt$ selection, except the isolation cut, which was
changed to $120^\circ<\theta_{iso}<150^\circ$. The classification rates were compared
between real and simulated data and the systematic error was estimated 
conservatively as the largest of the statistical error and the difference
between both.

\subsection{Charged-particle reconstruction}
\label{sys_multsel}
The sources of systematic uncertainty associated with the charged-particle multiplicity selection 
have been studied in~\cite{delphitopologicalbr}.
For track reconstruction,
the sources investigated include: the efficiencies of the different
tracking subdetectors to be included on a reconstructed track, both
for isolated tracks and for tracks in higher track density
topologies; effects of the TPC inter-sector boundary regions;
the two-track resolution of the tracking system and the 
efficiency to reconstruct a multi-prong $\tau$-decay as a function
of the minimum opening angle between any two particles; the
candidate $\tau$ charge reconstruction. 

\subsection{Material reinteractions}
\label{sys_conversions}
Uncertainties from the photon conversion reconstruction 
were particularly important for those decay modes 
containing $\pi^0$'s. The effect on the branching ratios was estimated
by  varying by their uncertainties the  correction  factors for  the
reconstructed    and    unreconstructed    conversions 
given    in
Table~1 of~\cite{delphitopologicalbr}, which  were obtained from data
test  samples of radiative dilepton events.
%   A
%contribution  for the  uncertainty in  the incident  photon  rates was
%included.    
The  resultant   uncertainties  are   dominated   by  the
contribution from the unreconstructed conversions.  A similar approach
was taken for the  nuclear interactions, with
the correction factors given by Table~2 in~\cite{delphitopologicalbr}.

\boldmath
\subsection{Relative efficiency of exclusive modes}
\unboldmath
\label{sys_releff}
Due to  mass effects and decay dynamics the momentum  
distributions of $\pi^\pm$
and  $K^\pm$ are different even for otherwise identical final states.

To estimate the size of
these effects the  proportions of charged  pions and kaons  in a
given decay  mode  were  varied  by  an  amount  consistent  with  the
uncertainties quoted in the Particle Data Listings~\cite{pdg2002}, the
selection  efficiency for  that  class recalculated  and the  fit
repeated. The change in the  measured branching ratio was taken as the
systematic uncertainty.

Within many classes there were a number of exclusive decay modes which
differ  in  $K^0$  multiplicity,  and  which may  not  have  the  same
selection  efficiency. To  estimate  the uncertainty  on the  measured
branching ratios, the exclusive branching ratios in a given class were
varied  within   the  uncertainties   quoted  in  the   Particle  Data
Listings~\cite{pdg2002}.   The
uncertainty on the  three-prong modes also included a  contribution due to
the  decay  modes  $K^-\pi^-\pi^+\pio\nut$  and  $K^-K^+\pi^-\pio\nut$
which were not included in the simulation.

Similarly, the decays containing  $\eta$ and  $\omega$
mesons were  varied by the  uncertainties on the world  average to   obtain
systematic uncertainties on the measured branching ratios.

\boldmath
\subsection{Decay modelling }
\unboldmath
\label{sys_eta}
The uncertainties associated with the  modelling of the 
decays  involving 
%$\aone$, $\eta$ and $\omega$ 
several pions or kaons
were  estimated by  
correcting  the  efficiencies taking  into
account  differences   between  data  and   simulated  invariant-mass
distributions. In addition, the hadronic structure of the $3\pi$ final
state was  varied between  the default TAUOLA~\cite{tauola}  model and
that obtained in the DELPHI analysis of the $3\pi$ structure in $\tau$-decays~\cite{delphithreepi}.    For  the   $3\pi\pio$   structure  the
parameterisation of  Model~1 of~\cite{cleostruct}  was used and,  as a
cross-check,     the     parameterisation     of    $3\pi\pio$     used
in~\cite{delphithreepi} was used to  reweight the distributions of the
minimum opening angle.

The  charged particle(s) produced  in the  various $\tau$-decay modes
have  different momentum  spectra for  the different  helicity states.
This leads to differences in acceptance as a function of the 
$\tau$ polarisation due to cuts in the $\tt$ selection.
This is  especially the  case for $\tau\to\pi\nut,K\nut$ where
the momentum spectra differ most between the two helicity states.
The analysis used the result and uncertainty from the
DELPHI analysis on $\tau$ polarisation~\cite{delphitaupolarisation}.

\subsection{Trigger}
\label{sys_simstat}
The trigger  efficiency for $\tt$ final  states was $(99.98\pm0.01)$\%
for events within the  polar-angle acceptance.  Studies indicated that
this inefficiency  was due to  events where both $\tau$'s  decayed via
the   $\tau\to\mu\nu\nu$  mode.    This can be extrapolated to an inefficiency
of $(6\pm3)\times 10^{-4}$ for the channel $\TMU$.   
The associated systematic uncertainty was
obtained by varying the inefficiency by its error.

\subsection{Energy and momentum scale and resolution}
\label{sys_hpcscale}
The HPC energy scale was altered in  the simulation 
by the uncertainty described in Section~\ref{sec:eemphot}
and the  complete analysis re-performed.  The changes in  the obtained
branching ratios  were taken as  the uncertainty. In a  similar manner
the   simulation  energy   was  smeared   and  the   branching  ratios
re-estimated. This  took into account, with the correct correlation,
different effects related to the electromagnetic calorimetry:
$\ee$ rejection, $\TEL$ identification and rejection through $E_{ass}$,
$\pio$ identification and total invariant masses.
The same procedure was followed with the momentum scale and resolution 
as given in Section~\ref{sec:pscale}.

\subsection{HCAL, muon chambers and dE/dx}
\label{sys_hcal}
%The systematics relating to the charged particle identification
%had the largest impact on the $\EL$, $\MU$ and $\HH$ decay modes
%where there were no $\pi^0$'s.
 
The correction in simulation to the tails of hadronic showers 
in the HCAL and muon chambers
was modified by the uncertainties derived in 
Section~\ref{sec:chargedparticles}. The analysis was repeated,
and the observed variations  in  the
branching ratios obtained were taken as  uncertainties.

The fraction  of extra layers added  in the simulation  to give better
data/simulation    agreement   was varied by the uncertainty 
obtained    in
Section~\ref{sec:chargedparticles}  and the analysis repeated.
The  uncertainties  were  taken  from  the variations  in  the branching ratios obtained.
For the tails of showers penetrating into the 
muon chambers, the efficiency was  varied by the uncertainty observed in
the test samples for both muons and hadrons. 

In a similar way, the response of the HCAL and muon chambers for muons was
varied within the uncertainties obtained in~\ref{sec:chargedparticles} with muon
test samples.

The $dE/dx$  was varied in simulation 
for each particle according to the errors in the tuning
described in~\ref{sec:chargedparticles}
and the  analysis re-performed.  The  uncertainties were
taken from the changes in the branching ratios obtained. This affected
mostly the  $\TEL$ and $\TH$ branching ratios  whose separation depended
most on $dE/dx$.

\subsection{Photon and neutral-pion reconstruction}
\label{sys_photon}
The photon efficiency, the probability to split one photon into two,
the probability to create fake photons from a hadron, as well as the $\pi^0$
reconstruction efficiency and fake probability were checked with different
test samples, as described in Section~\ref{sec:photons}. The different errors
were propagated to the efficiency tables and the fits were repeated. The
observed difference was taken as systematic error.

\subsection{Summary of systematic uncertainties}
\label{sec:syssum}
A  contribution to  the systematic  uncertainty was  included  for the
statistical uncertainty  on the components of  the selection-efficiency
matrices due to the finite simulation sample size.

The systematic uncertainty associated with each source and
for each measured decay mode is shown for the neural network analysis 
in Table~\ref{tab:sys_oneprong}. The total systematic error was 
calculated as the quadratic sum of these contributions, since they
were essentially independent.
The errors for the
sequential-cuts analysis were similar, but slightly larger in general.

\begin{figure}[tbp]
  \begin{center}
    \leavevmode    
\epsfig{figure=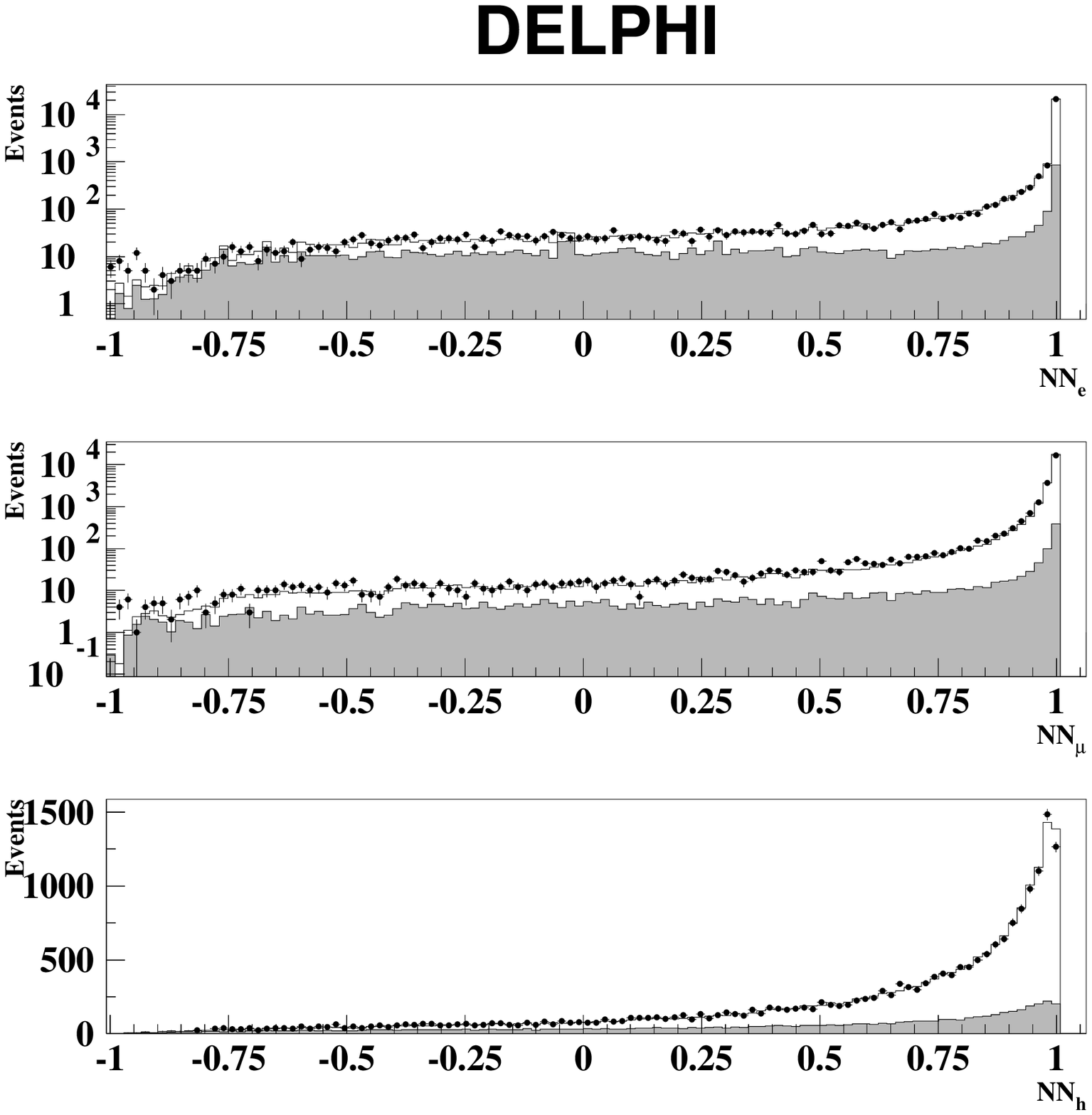,width=1.1\linewidth}        
    \caption{\it Maximum-output neuron value in one-prong analyses.
             For each event the output of the class whose output neuron is maximum is represented. 
              Data are shown as dots and simulation by a solid line. The shaded area represents the background events.}
    \label{fig:neuronop1}
  \end{center}
\end{figure}

An attempt to estimate the effect of systematic errors on the goodness of the 
fit was made under the following procedure. The systematic errors were 
estimated bin by bin as the
observed difference in the simulated distributions when the previously discussed 
systematic effects were
varied within their uncertainties. A $\chi^2=397$ for 490 $d.o.f.$ 
was thus evaluated, neglecting the bin to bin
correlations (which slightly underestimates the $\chi^2$).
The major contributions to the $\chi^2$ reported in Table~\ref{tab:nn_res} 
came from distortions of the neural network output in regions 
far from the cut, and
where the signal and
background separation was very clear and therefore did not affect the results
significantly compared to the quoted systematic errors. In particular, the largest contribution,
arising from the $\TMU$ channel is due to the slight widening of the sharp peak on the neural
network output distribution (Fig.~\ref{fig:neuronop1}) caused by the HCAL and muon 
chamber response systematics, with a migration of a small fraction of events from values close to 1
to the region from 0.5 to 1. This has a large impact on the $\chi^2$, but very small one on the
results, since they are nevertheless identified clearly. Similar arguments apply to the second largest
contribution, from the $\THHH$, but in this case the neural network peak (Fig.~\ref{fig:neuronop1})
is narrower in data.

\begin{table}[tbp]
\newcommand{\IR}[1]{\rule{1em}{0pt}\makebox[0cm][c]{\rotatebox{90}{\ #1}}}
\newcommand{\MK}[1]{\makebox[0.06\linewidth][r]{#1}}
  \begin{center}
    \leavevmode
    \begin{tabular}[center]{l|r|r|r|r|r|r|r|r}
\hline
           & \multicolumn{8}{|c}{1-prong decay mode}  \\
\cline{2-9}
Source of systematic  
       &\multicolumn{1}{|c}{\IR{$\EL$~~~}}  
       &\multicolumn{1}{c}{\IR{$\MU$~~~}} 
       &\multicolumn{1}{c}{\IR{$h^- \nut$~~~~~}  }
       &\multicolumn{1}{c}{\IR{$h^-\pio\nut$~~} }
       &\multicolumn{1}{c}{\IR{$h^- 2\pio\nut$~}} 
       &\multicolumn{1}{c}{\IR{$h^- \!\!\geq\!\! 1\pio\nut$}} 
       &\multicolumn{1}{c}{\IR{$h^- \!\!\geq\!\! 2\pio\nut$}}
       &\multicolumn{1}{c}{\IR{$h^- \!\!\geq\!\! 3\pio\nut$}} \\
\hline
Non-$\tau$ background   scale   &  \MK{26}     & \MK{8}   &    \MK{2} &  \MK{7}     & \MK{6}   &    \MK{11} & \MK{4}   &    \MK{2} \\
Non-$\tau$ background classification   &   9     &  3   &    3 &   8     &    2   &    8 &    2 &    5 \\
Tracking and VD efficiency       &  10     &  3   &   15 &  33     &  121   &   70 &   50 &   93  \\
Material reinteractions          &  16     & 12   &   13 &  38     &   28   &   25 &   48 &   28  \\
Exclusive BRs                    &  13     & 13   &   38 &  41     &   28   &   47 &   24 &    7  \\
Decay modelling                  &   1     &  2   &    1 &  17     &   22   &    8 &   13 &   10  \\
Trigger                          &   4     &  30  &    3 &   7     &    3   &    10&    3 &   $<$1  \\
Energy and momentum calibration  &  90     &  10  &   13 &  81     &  193   &    33&   63 &  155  \\
HCAL and muon chamber response   &   1     & 70   &   70 &   7     &    4   &    2 &    4 &    8  \\
$dE/dx$ calibration              &  54     & 14   &   42 &   2     &   12   &   30 &   13 &   37  \\
Photon and $\pi^0$ reconstruction&  23     &   7  &   32 &  49     &  116   &    37&   34 &  109  \\
Simulation statistics            &  28     &  27  &   31 &  57     &   88   &    39&   51 &   58  \\
\hline                                                                                      
Total systematic                 &  116    &  85  &  103 &  130    &  274   &   116&  116 &  224  \\  
\hline
    \end{tabular}
    \end{center}
\vspace{1.0cm}
\begin{center}
    \begin{tabular}[center]{l|r|r|r|r|r|r}
\hline
  & \multicolumn{6}{|c}{3- or 5-prong decay mode}  \\
\cline{2-7}
Source of systematic  
       &\multicolumn{1}{|c}{\IR{~$3h^\pm \nut$~~~~~~}}  
       &\multicolumn{1}{c}{\IR{~$3h^\pm \pio\nut$~~~} }
       &\multicolumn{1}{c}{\IR{~$3h^\pm \!\!\geq\!\! 2\pio\nut$}}
       &\multicolumn{1}{c}{\IR{~$3h^\pm \!\!\geq\!\! 1\pio\nut$}}
       &\multicolumn{1}{c}{\IR{~$5h^\pm \nut$~~~~~~}  }
       &\multicolumn{1}{c}{\IR{~$5h^\pm \!\!\geq\!\!1 \pio\nut$~}} \\
\hline
Non-$\tau$ background  scale  &  \MK{ 5}     & \MK{2}   &    \MK{3} &  \MK{5}     & \MK{0}   &    \MK{0}\\
Non-$\tau$ background  classification  &   4     & 18   &   40 &  40     &    0     &      0\\
Tracking and VD efficiency       &  15     & 30   &   29 &  70     &    2.3   &    5.1  \\
material reinteractions          &  27     &  8   &   19 &  22     &    1.5   &    1.1   \\
Exclusive BRs                    &  11     & 39   &   30 &  23     &    0.0   &    0.0   \\
Decay modelling                  &   3     &  5   &    1 &   6     &    1.0   &    1.0   \\
Trigger                          &   3     &  2   &    0 &   2     &    0.0   &    0.0  \\
Energy and momentum calibration  &  17     &  37  &   27 &  10     &    0.3   &    0.3  \\
HCAL and muon chamber response   &   1     &  3   &    2 &   1     &    0.0   &    0.0  \\
$dE/dx$ calibration              &  17     &  0   &   10 &  23     &    0.0   &    0.0  \\
Photon and $\pi^0$ reconstruction&  62     &  70  &   60 &  44     &    0.8   &    0.8  \\
Simulation statistics            &  27     &  38  &   28 &  24     &    4.4   &    3.5  \\
\hline                                                                                      
Total systematic                 &   79    & 103  &   95 &  103    &    5.4   &    6.4  \\  
\hline
    \end{tabular}
    \caption{\it Contributions in units of ${\mathit{10^{-5}}}$ 
to the systematic 
uncertainties on the branching ratios. }
    \label{tab:sys_oneprong}
  \end{center}
\end{table}

\section{Results}
\label{sec:combination}

The neural-network analysis gave for all hadronic channels 
better precision both in statistics and
systematics, and included more channels. 
Therefore the results
from this analysis were taken as the basic measurement, 
while the sequential analysis
(except for the five-prong channels) was kept only as a 
cross-check. However, the
performance for the leptonic decays is slightly worse than 
in~\cite{delphileptonicbr99}
and therefore those results 
are not updated. 
Taking into account the statistical and 
systematic correlation of the channels with one or
several $\pi^0$ some inclusive branching ratios were also derived.

The results are shown in Table~\ref{tab:res}.

\begin{table}[tbph]
\begin{center}
\begin{tabular}{l|r}
 \hline
~~~~~~~Decay mode & Branching Ratio(\%)\\
 \hline
$\tau^-\to h^- \geq 0 \ko \nut                      $ &  $ 12.780\pm0.120\pm0.103   $ \\
$\tau^-\to h^-\pio\geq 0 \ko \nut                   $ &  $ 26.291\pm0.201\pm0.130   $ \\
$\tau^-\to h^- 2\pio\geq 0 \ko \nut                 $ &  $  9.524\pm0.320\pm0.274   $ \\
$\tau^-\to h^-\geq 1\pio\geq 0 \ko \nt              $ &  $ 37.218\pm0.155\pm0.116   $ \\
$\tau^-\to h^-\geq 2\pio\geq 0 \ko \nt              $ &  $ 10.927\pm0.173\pm0.116   $ \\
$\tau^-\to h^-\geq 3\pio\geq 0 \ko \nt              $ &  $  1.403\pm0.214\pm0.224   $ \\
 \hline
$\tau^-\to 3h^\pm \geq 0 \ko \nut                   $ &  $  9.340\pm0.090\pm0.079   $ \\
$\tau^-\to 3h^\pm \pio\geq 0 \ko \nut               $ &  $  4.545\pm0.106\pm0.103   $ \\
$\tau^-\to 3h^\pm\!\! \geq\!\! 1\pio\geq 0 \ko \nut $ &  $  5.106\pm0.083\pm0.103   $ \\
$\tau^-\to 3h^\pm\!\! \geq\!\! 2\pio\geq 0 \ko \nut $ &  $  0.561\pm0.068\pm0.095   $ \\
 \hline
$\tau^-\to 5h^\pm \geq 0 \ko \nut                   $ &  $  0.097\pm0.015\pm0.005 $ \\
$\tau^-\to 5h^\pm\!\! \geq\!\! 1\pio\geq 0 \ko \nut $ &  $  0.016\pm0.012\pm0.006 $ \\
 \hline
\end{tabular}
\caption{\it Measured branching ratios in percent.
The uncertainties are statistical followed by systematic.}
\label{tab:res}
\end{center}
\end{table}

The correlation matrix for 
the combined statistical and systematic uncertainties is shown in Table~\ref{tab:corr_nnandseq}.
\begin{table}[tbp]
\begin{center}
\newcommand{\IR}[1]{\rotatebox{90}{\ #1}}
\begin{tabular}{l|rrrrrr|rrrr|rr}
\hline
%mode & \multicolumn{8}{|c}{correlation coefficient} \\
 & \multicolumn{1}{|c}{\IR{$h^-\nut$                       }}
 & \multicolumn{1}{|c}{\IR{$h^-\pio\nut$                   }}
 & \multicolumn{1}{|c}{\IR{$h^- \!\!\geq\!\! 1\pio\nut$    }}
 & \multicolumn{1}{|c}{\IR{$h^- 2\pio\nut$                 }}
 & \multicolumn{1}{|c}{\IR{$h^- \!\!\geq\!\! 2\pio\nut$    }}
 & \multicolumn{1}{|c}{\IR{$h^- \!\!\geq\!\! 3\pio\nut$    }}
 & \multicolumn{1}{|c}{\IR{$3h^\pm \nut$                   }}
 & \multicolumn{1}{|c}{\IR{$3h^\pm \pio\nut$               }}
 & \multicolumn{1}{|c}{\IR{$3h^\pm\!\! \geq\!\! 1\pio\nut$ }}
 & \multicolumn{1}{|c}{\IR{$3h^\pm\!\! \geq\!\! 2\pio\nut$ }}
 & \multicolumn{1}{|c}{\IR{$5h^\pm \nut$                   }} 
 & \multicolumn{1}{|c}{\IR{$5h^\pm \!\!\geq\!\! 1\pio\nut$ }} \\
\hline
$h^-\pio\nut$                  &$-$0.34&       &       &       &       &       &       &       &      &      &&\\
$h^- \!\!\geq\!\! 1\pio\nut$   &$-$0.47&   0.56&       &       &       &       &       &       &       &      &&\\
$h^- 2\pio\nut$                &   0.06&$-$0.66&   0.15&       &       &       &       &       &       &      &&\\
$h^- \!\!\geq\!\! 2\pio\nut$   &$-$0.03&$-$0.74&   0.15& 0.81  &       &       &       &       &       &      &&\\
$h^- \!\!\geq\!\! 3\pio\nut$   &$-$0.06&   0.38&   0.11&$-$0.86&$-$0.36&       &       &       &       &      &&\\
\hline
$3h^\pm \nut$                  &$-$0.07&$-$0.08&   0.15&   0.00&$-$0.03&$-$0.02&       &       &       &      &&\\
$3h^\pm \pio\nut$              &$-$0.02&$-$0.01&$-$0.05&$-$0.03&$-$0.02&   0.03&$-$0.53&       &       &      &&\\
$3h^\pm\!\! \geq\!\! 1\pio\nut$&$-$0.04&$-$0.04&$-$0.13&$-$0.04&$-$0.06&$-$0.02&$-$0.56&   0.75&       &      &&\\
$3h^\pm\!\! \geq\!\! 2\pio\nut$&$-$0.01&$-$0.01&$-$0.04&   0.03&$-$0.02&$-$0.06&   0.26&$-$0.78&$-$0.16&      &&\\
\hline
$5h^\pm \nut$                  &$-$0.01&$-$0.01&   0.01&   0.00&   0.00&   0.00&$-$0.02&$-$0.03&$-$0.01&   0.03&&\\
$5h^\pm \!\!\geq\!\! 1\pio\nut$&   0.00&   0.00 &   0.00&   0.00&   0.00&   0.00&    0.01&   0.00&$-$0.05&$-$0.05&$-$0.57&\\
\hline
$B_1$        &   0.09&    0.10&   0.26&   0.04&   0.11&   0.03&$-$0.50&$-$0.25&$-$0.39& $-$ 0.06& $-$ 0.03& 0.00\\
$B_3$        &$-$0.09& $-$0.10&$-$0.26&$-$0.04&$-$0.11&$-$0.03&   0.50&   0.25&   0.39&     0.06&     0.03& 0.00\\
$B_5$        &$-$0.02&   0.00 &   0.00&   0.00&   0.00&   0.00&$-$0.03&   0.03&   0.00&     0.00&     0.72&0.40\\

\hline
\end{tabular}
\end{center}
\caption{\it Correlation matrix of the combined statistical and systematic uncertainties. The last three rows show the
correlation with the topological branching ratios presented in~\cite{delphitopologicalbr}.}
\label{tab:corr_nnandseq}
\end{table}

Using the world averages~\cite{pdg2002} for the channels involving $\ko$ and neglecting this contribution for channels with more than
three charged pions or kaons, we can derive the branching ratios shown in Table~\ref{tab:resnok}.
In this subtraction, the total error on the world average was added in quadrature to the systematic 
error of these measurements.

\begin{table}[tbph]
\begin{center}
\begin{tabular}{l|r}
 \hline
~~~~~~~Decay mode & Branching Ratio(\%)\\
 \hline
$\tau^-\to h^- \nut                      $ &  $( 11.571\pm0.120\pm0.114)   $ \\
$\tau^-\to h^-\pio\nut                   $ &  $( 25.740\pm0.201\pm0.138)   $ \\
$\tau^-\to h^- 2\pio\nut                 $ &  $(  9.498\pm0.320\pm0.275)   $ \\
$\tau^-\to h^-\geq 1\pio\nt              $ &  $( 36.641\pm0.155\pm0.127)   $ \\
$\tau^-\to h^-\geq 2\pio\nt              $ &  $( 10.901\pm0.173\pm0.118)  $ \\
$\tau^-\to h^-\geq 3\pio\nt              $ &  $(  1.403\pm0.214\pm0.224)   $ \\
 \hline
$\tau^-\to 3h^\pm \nut                   $ &  $(  9.317\pm0.090\pm0.082)   $ \\
$\tau^-\to 3h^\pm \pio\nut               $ &  $(  4.545\pm0.106\pm0.103)   $ \\
$\tau^-\to 3h^\pm\!\! \geq\!\! 1\pio\nut $ &  $(  5.106\pm0.083\pm0.103)   $ \\
$\tau^-\to 3h^\pm\!\! \geq\!\! 2\pio\nut $ &  $(  0.561\pm0.068\pm0.095)   $ \\
 \hline
$\tau^-\to 5h^\pm \nut                   $ &  $(  0.097\pm0.015\pm0.005)   $ \\
$\tau^-\to 5h^\pm\!\! \geq\!\! 1\pio\nut $ &  $(  0.016\pm0.012\pm0.006)   $ \\
 \hline
\end{tabular}
\caption{\it
Measured branching ratios in percent after subtraction of the contributions of channels including $\ko$.
The uncertainties are statistical followed by systematic.}
\label{tab:resnok}
\end{center}
\end{table}

The sum of the branching ratios of channels giving one-prong topologies, taking into account correlations and after
correcting for the decay modes not included in the 
analysis,$(0.266\pm0.027)$\%~\cite{pdg2002}, 
was $(85.417\pm0.094\pm0.075)$\%, consistent with the DELPHI 
topological one-prong branching ratio  measurement~\cite{delphitopologicalbr}
$\Bone=(85.316\pm0.093\pm0.049)$\%. Accounting for the strong correlation (0.80)
arising from the fact that the classification is very efficient and  
few events remain unclassified, these two numbers agree to 1.3 standard deviations.
%To compare these 
%numbers it is important to take into account the fact
%that the classification is very efficient and 
%few events remain unclassified. Therefore these
%results have a correlation of $0.80$.

These results are in good agreement with the current world
averages~\cite{pdg2002}.

\section{Conclusions}
\label{sec:conclusions}
The  measurement of $\tau$ exclusive branching ratio to final states 
containing up to five hadrons
has been performed with the DELPHI detector, with identification of neutral
pions. Different semi-exclusive branching ratios, with only a lower bound on the
number of $\pio$, were also measured for 
final states containing up to six hadrons.
A total of 134421 one-prong, 23847 three-prong and 112 five-prong candidate $\tau$-decays were identified.
Both sequential-cuts methods and neural networks have been used
in the selection of exclusive decay modes 
with different neutral pion multiplicity, giving compatible results.
The sum of the one-prong exclusive modes is 
consistent with our previous topological measurement. The good agreement in the
number of observed and expected events that are 
unclassified by the neural network shows no evidence of unexpected decays.

The branching ratios obtained are summarised in Table~\ref{tab:res}. 
Using the world-average measurements for channels involving
neutral kaons, this contribution was subtracted. 
The results are summarised in Table~\ref{tab:resnok}.
 
All the results are in good agreement with the current world
averages~\cite{pdg2002} and have similar
errors to the most precise single measurements. 

%         Modified on 04-06-1999 by dimartino
%-------------------------------------------------------------------
\subsection*{Acknowledgements}
\vskip 3 mm
 We are greatly indebted to our technical 
collaborators, to the members of the CERN-SL Division for the excellent 
performance of the LEP collider, and to the funding agencies for their
support in building and operating the DELPHI detector.\\
We acknowledge in particular the support of \\
Austrian Federal Ministry of Education, Science and Culture,
GZ 616.364/2-III/2a/98, \\
FNRS--FWO, Flanders Institute to encourage scientific and technological 
research in the industry (IWT), Belgium,  \\
FINEP, CNPq, CAPES, FUJB and FAPERJ, Brazil, \\
Czech Ministry of Industry and Trade, GA CR 202/99/1362,\\
Commission of the European Communities (DG XII), \\
Direction des Sciences de la Mati$\grave{\mbox{\rm e}}$re, CEA, France, \\
Bundesministerium f$\ddot{\mbox{\rm u}}$r Bildung, Wissenschaft, Forschung 
und Technologie, Germany,\\
General Secretariat for Research and Technology, Greece, \\
National Science Foundation (NWO) and Foundation for Research on Matter (FOM),
The Netherlands, \\
Norwegian Research Council,  \\
State Committee for Scientific Research, Poland, SPUB-M/CERN/PO3/DZ296/2000,
SPUB-M/CERN/PO3/DZ297/2000, 2P03B 104 19 and 2P03B 69 23(2002-2004)\\
FCT - Funda\c{c}\~ao para a Ci\^encia e Tecnologia, Portugal, \\
Vedecka grantova agentura MS SR, Slovakia, Nr. 95/5195/134, \\
Ministry of Science and Technology of the Republic of Slovenia, \\
CICYT, Spain, AEN99-0950 and AEN99-0761,  \\
The Swedish Natural Science Research Council,      \\
Particle Physics and Astronomy Research Council, UK, \\
Department of Energy, USA, DE-FG02-01ER41155. \\
EEC RTN contract HPRN-CT-00292-2002.\\

\clearpage
\clearpage

\clearpage
 
\clearpage

\clearpage

\clearpage

\clearpage

\clearpage

\clearpage

\clearpage

\clearpage

\clearpage

\clearpage

\clearpage

\clearpage

\clearpage

\clearpage

\clearpage

\clearpage

\end{document}